\documentclass[a4paper,11pt]{article}
\usepackage[margin=1.0in]{geometry}
\usepackage{graphics,graphicx}
\usepackage{longtable} 
\usepackage{arydshln}
\usepackage{amsmath,verbatim,amssymb,epsfig,wrapfig,color,lscape}
\usepackage{natbib}
\usepackage[table]{xcolor}
\definecolor{Gray}{gray}{0.9}  
\usepackage{booktabs}
\usepackage{lastpage}
\usepackage{authblk}
\usepackage{multirow} 
\newcommand{\hide}[1]{}  
\usepackage[normalem]{ulem}
\usepackage[ruled,vlined]{algorithm2e}
\usepackage{algpseudocode}
\usepackage{bm}
\usepackage{geometry}
\usepackage{tabularx}

\makeatletter

\usepackage[outerbars,color]{changebar}
\ifx\pdfoutput\undefined
\else\ifnum\pdfoutput>0
\usepackage{pdfcolmk}
\cbcolor{black}
\usepackage{array}
\newcommand{\PreserveBackslash}[1]{\let\temp=\\#1\let\\=\temp}
\newcolumntype{L}[1]{>{\raggedright\arraybackslash}m{#1}}
\newcolumntype{C}[1]{>{\centering\arraybackslash}m{#1}}
\newcolumntype{R}[1]{>{\raggedleft\arraybackslash}m{#1}}

\usepackage{tikz}
\usetikzlibrary{bayesnet}
\usetikzlibrary{fit,positioning}

\newfont{\rmm}{cmr10 at 11pt}
\rmm

\title{A Bayesian Precision Response-adaptive Phase II Clinical Trial Design for Radiotherapies with Competing Risk Survival Outcomes}
\author[1,2]{Jina Park}
\author[3]{Wenjing Hu}
\author[1,2]{Ick Hoon Jin}
\author[4]{Hao Liu}
\author[5]{Yong Zang}
\affil[1]{Department of Applied Statistics, Yonsei University, South Korea}
\affil[2]{Department of Statistics and Data Science, Yonsei University, South Korea}
\affil[3]{AT\&T Company, USA}
\affil[4]{Department of Biostatistics and Epidemiology, Cancer Institute of New Jersey, Rutgers University, USA}
\affil[5]{Department of Biostatistics and Health Data Sciences, Center of Computational Biology and Bioinformatics, Indiana University, USA}
\date{}

\begin{document}
	\maketitle
	
	\begin{abstract}
		Many phase II clinical trials have used survival outcomes as the primary endpoints in recent decades. Suppose the radiotherapy is evaluated in a phase II trial using survival outcomes. In that case, the competing risk issue often arises because the time to disease progression can be censored by the time to normal tissue complications, and vice versa. Besides, much literature has examined that patients receiving the same radiotherapy dose may yield distinct responses due to their heterogeneous radiation susceptibility statuses. Therefore, the ``one-dose-fit-all'' strategy often fails, and it is more relevant to evaluate the subgroup-specific treatment effect with the subgroup defined by the radiation susceptibility status. In this paper, we propose a Bayesian precision phase II trial design evaluating the subgroup-specific treatment effects of radiotherapy. We use the cause-specific hazard approach to model the competing risk survival outcomes. We propose restricting the candidate radiation doses based on each patient's radiation susceptibility status. Only the clinically feasible personalized dose will be considered, which enhances the benefit for the patients in the trial. In addition, we propose a stratified Bayesian adaptive randomization scheme such that more patients will be randomized to the dose reporting more favorable survival outcomes. Numerical studies have shown that the proposed design performed well and outperformed the conventional design ignoring the competing risk issue.     	  
		
	\end{abstract}
	
	\newpage
	\section{Introduction}
	A conventional phase II clinical trial tests whether the experimental drug has any anti-disease activity. The short-term efficacy outcome, such as the objective tumor response, is commonly used as the primary endpoint for a phase II clinical trial. Then, suppose the experimental drug shows sufficiently favorable short-term efficacy responses in a phase II trial. In that case, a large-scale phase III trial will be followed to test the long-term therapeutic effect using survival outcomes such as the overall survival (OS) or progression-free survival (PFS). This widespread clinical practice assumes that the short-term efficacy outcome is an excellent surrogate marker for the long-term survival outcome. This assumption, however, does not always hold. For example, complete remission (CR) is the most desirable short-term efficacy outcome. However, achieving CR is necessary but not sufficient for prolonging survival because many patients may relapse shortly after achieving CR. Indeed, many cytotoxic agents report favorable CR rates in phase II trials. However, only a few of them can transform the improvement of CR rates into a substantial survival benefit in the following phase III trials \citep{Kola2004}. Hence, to resolve this issue, in recent decades, there has been a growing trend to use the survival outcome as the primary endpoint for phase II clinical trials \citep{Iten2007, Jarnagin2009, Liu2019, Pimentel2019, Kudo2021}. This paper studies the phase II clinical trial design using survival outcomes, focusing on radiotherapies (RT).       
	
	The RT is a ``double-edged sword'' for cancer patients. On the one hand, the X-ray on tumor cells can prevent disease progression; on the other hand, the X-ray on normal cells can induce normal tissue complications such as severe and irreversible organ damage (fibrosis, vascular damage, atrophy, etc.) \citep{Hendry2006, Barnett2009}. Therefore, although the dose-limiting toxicity (DLT) has already been evaluated in the phase I dose-finding trial, the normal tissue complications still needs to be monitored in the phase II trial because (1) the normal tissue complications can be fatal, (2) DLT is typically evaluated within a short period whereas RT induced normal tissue complications may happen long after the follow-up (e.g., late-onset toxicity) and (3) the limited sample size (10$\sim$30) for phase I trial may be insufficient to provide an accurate estimate for toxicity. Consequently, for a phase II trial for RT using survival outcome, it is reasonable to treat time to disease progression and time to normal tissue complications as co-primary endpoints in a single trial. Moreover, for most phase II cancer oncology trials, if a patient experiences either disease progression or normal tissue complications, he/she should be treated off the protocol for ethical consideration. Since only the first event is observable, the competing risk issue arises.    
	
	Most phase II trial designs assume population homogeneity and either assign patients to a single treatment arm (Phase IIA) or randomize them to receive different treatments. The randomization scheme is typically independent of patients' personalized information (Phase IIB), which is disconnected from clinical practice. For RT, recent research has revealed that patients' responses can be remarkably different due to heterogeneous radiation susceptibility status \citep{Schipper2014}. Specifically, while some radiation-sensitive (SE) patients may yield desirable performances at a relatively low RT dose, some radiation-resistant (RE) patients require a very high RT dose to control disease \citep{Busch1994, Chistiakov2008, Kleinerman2009}. Studies in stereotactic body RT showed that a very high dose is required to reach at least 90\% tumor control for RE patients with stage I non-small cell lung cancers (NSCLC) \citep{Wang2012}. Single-institution studies and secondary analysis of Radiation Therapy Oncology Group (RTOG) trials also showed that increasing the RT dose improved local control and survival for RE patients \citep{Carvalho2013}. However, as demonstrated in the RTOG 0617 trial where a high dose arm has poorer survival than the standard dose arm in treating the SE patients, a high dose will harm the SE patients because it can induce severe and irreversible normal tissue complications \citep{Bradley2015}. Hence, a precision design is needed to (1) handle the competing risk co-primary survival endpoints (time to disease progression and time to normal tissue complications) and (2) incorporate each patient's radiation susceptibility status (RE and SE) into radiation dose assignment and evaluation procedures.   
	
	Our study is motivated by a phase II clinical trial conducted at the Department of Radiation Oncology, Indiana University Melvin and Bren Simon Comprehensive Cancer Center. The purpose of this trial is to evaluate the PFS and monitor the normal tissue complications for stage-III NSCLC patients receiving different doses of stereotactic body RT. A total of 92 patients will be enrolled and randomized into the trial. Patients will be classified into SE and RE subgroups, using a well-established ERCC1/2 SNP signature\citep{Wang2012, Carvalho2013}. ERCC1/2 genes are well known for repairing the ultraviolet-induced DNA damage through the nucleotide excision repair pathway \citep{Sinha2002}. Studies also showed that they are involved in DNA repairs for ionizing radiation-induced damage \citep{Santivasi2014}. There are three RT doses for consideration, referred to as the low dose (62 Gy in 2 Gy/fraction), the standard dose (74 Gy in 2 Gy/fraction), and the high dose (82 Gy in 2 Gy/fraction). Only the first two will be considered for the SE patients, and the last two will be considered for the RE patients. Each patient will be followed for three months to assess PFS. If any patient in the trial has experienced either disease progression or normal tissue complications, he/she will be treated by a second-line treatment off the protocol. 
	
	In this paper, we develop a Bayesian precision response-adaptive phase II clinical trial design fitting the requirement of the motivating trial. As illustrated in Figure 1, we use a proportional hazard regression model to characterize the association between the time-to-event and the RT dose and radiation susceptibility status. We treat disease progression and normal tissue complications as two cause-specific events and use the cause-specific hazard competing risk model to link these two events. We construct a utility function to measure the risk-benefit tradeoff between the competing risk outcomes. Stratified by the radiation susceptibility status, we develop a response-adaptive randomization scheme. More patients will be randomized to the RT dose reporting more favorable response outcomes in the posterior mean utility estimates. A subgroup-specific RT dose will be selected for SE and RE patients separately at the end of the trial.
	
	\begin{figure}[htbp]
		\centering
		\includegraphics[width=0.5\textwidth]{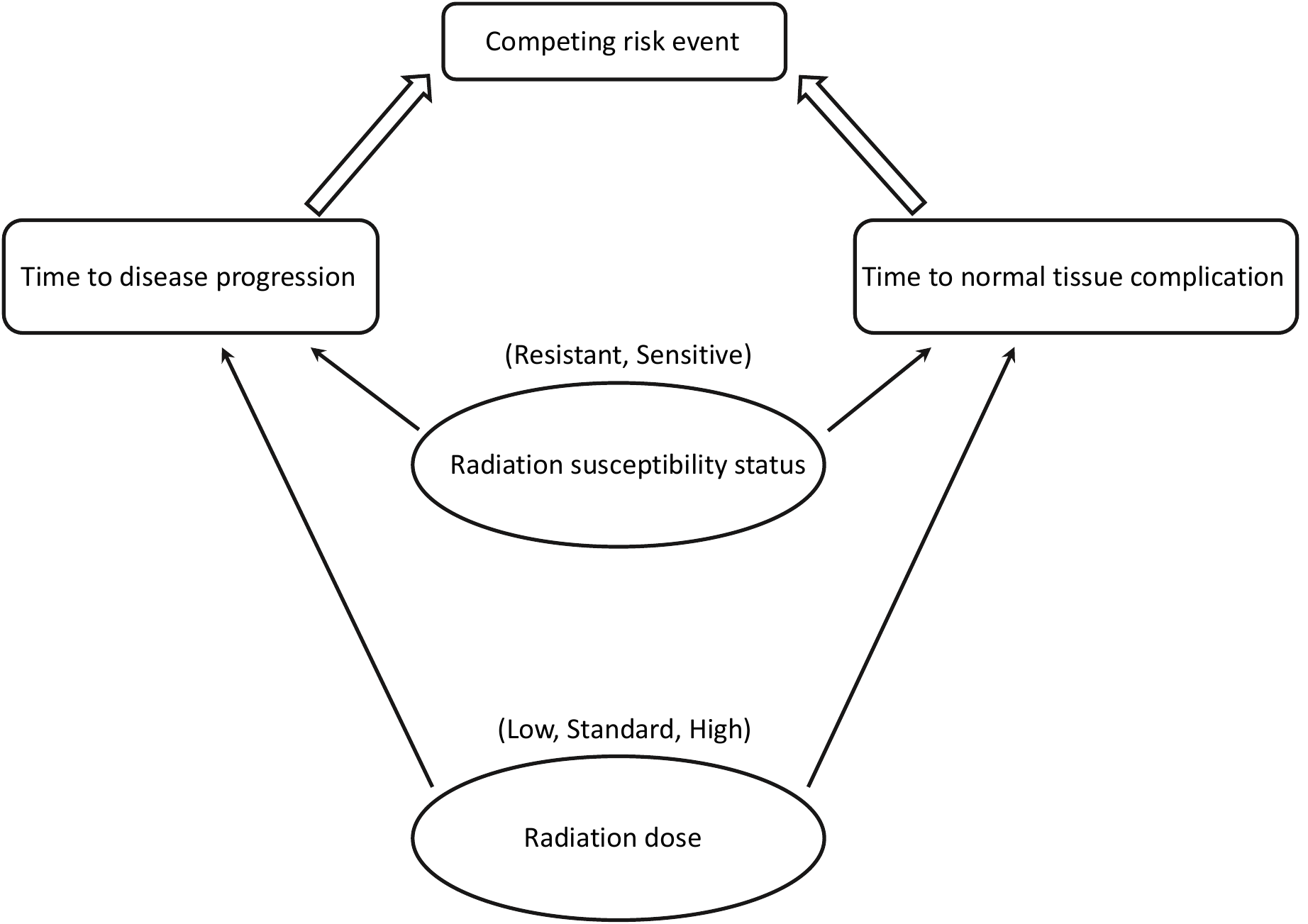}
		\caption{Illustration of the proposed competing risk model for cause-specific events.}
		\label{fig:crm}
	\end{figure}
	
	Numerous phase II clinical trial designs have been proposed. Frequentist designs includes the Simon's two-stage design \citep{Simon1989} and its extensions \citep{Ensign1994, Chen1997, Hanfelt1999, Jung2001, Shuster2002, Lin2004}. A lot of Bayesian adaptive phase II designs have also been developed using posterior probability, predictive probabilities, and Bayes factors, for both single-arm trials \citep{Thall1994, Thall1995, Heitjan1997, Lee2008, Johnson2009} and randomized trials \citep{Huang2009, Yin2012, Yuan2016, Guo2019, Guo2020}. There are also adaptive designs developed for biomarker-guided phase II clinical trials, such as the tandem two-stage design \citep{Pusztai2007}, sequential enrichment design \citep{Zang2017}, parallel two-stage design \citep{Jones2007, Parashar2016} and its extension \citep{Dutton2018}. However, all the existing biomarker-guided designs are for short-term binary efficacy outcomes only. To the best of our knowledge, the design proposed in this paper represents the first precision phase II clinical trial design dealing with competing risk survival outcomes.     
	
	We have proposed a Bayesian adaptive phase I/II design for competing risk outcomes \citep{Zhang2021}. The differences between the \citet{Zhang2021}'s design and this new design are (1) the previous design is for phase I dose-finding trials whereas the new design is for randomized phase II trials; (2) the previous design treats the competing risk data as an ordinal outcome and develops a Bayesian data augmentation method to impute the late-onset outcomes whereas this new design treats the competing risk data as survival outcome and uses the cause-specific hazard approach to model the competing risk survival data; (3) the previous design assumes population homogeneity whereas this new design incorporates each patient's biomarker information; and (4) the previous design is mainly used for immunotherapies and targeted therapies whereas the new design is more suitable for RT. In addition, \citet{Biard2021} recently proposed another phase I/II design dealing with competing risk outcomes. However, similar to our previous phase I/II design, this design is also for dose-finding trials only. It cannot be directly used for a phase II trial, nor can it incorporate biomarker information.  
	
	The remainder of this paper is organized as follows. In Section 2, we describe the probability model. In Section 3, we present the precision response-adaptive randomization design. In Section 4, we investigate the operating characteristics of the proposed design through numerical studies. We provide concluding remarks in Section 5.          
	
	\section{Probability Model}
	
	We first develop the competing risk probability model for survival outcomes. For the $i$-th patient in the trial, we define $Y_{ki}$ as the event happening time for the cause-specific events $k$ with $k=1$ representing disease progression and $k=2$ representing normal tissue complications. Due to the competing risk issue, we can only obtain the first event happening time $T_i = \min(Y_{1i}, Y_{2i})$. Let $C_i$ be the censoring time due to incomplete follow-up at any interim analysis stage or administrative censoring at the end of the follow-up. We have $X_i={\rm min}(T_i, C_i)$ as the observation time.  
	
	We use the cause-specific hazard approach to model the competing risk outcomes. Let $W_i$ be the radiation susceptibility status with $W_i=0,1$ representing the RE status and SE status. Let $D_i$ be the RT dose with $D_i=0,1,2$ representing the low RT dose, standard RT dose and high RT dose. We use $\lambda_{k}(X_i \mid W_i, D_i)$ to denote the cause-specific hazard function for the $i$-th patient with event $k$. We assume that the baseline hazard function follows the Weibull distribution due to its flexibility and generality (satisfying both the proportional hazard and accelerated failure time model assumptions). Then, $\lambda_{k}(X_i \mid W_i, D_i)$ can be expressed as $\lambda_k(X_i\mid W_i,D_i)=\alpha_k \beta_k X_i^{\alpha_k -1}{\rm exp}\Big(h_k(W_i,D_i)\Big)$ with $h_k(W_i,D_i)$ representing the logarithm of cause-specific hazard ratio for event $k$.
	
	There are many ways to specify $h_k(W_i, D_i)$, and we propose a specification following the motivating trial and clinical practice. As illustrated in the introduction and the motivating trial, for clinical practice of RT, a low dose is rarely considered for RE patients due to the lack of capability to control the disease. Along the same line, a high dose is, in general, not an option for SE patients due to the unacceptable normal tissue complications. That is, although we consider three doses in the trial, for a RE patient ($W_i=0$), the treatment comparison is restricted to the standard and high RT doses ($D_i=1,2$). For a SE patient ($W_i=1$), the treatment comparison is restricted to the low and standard RT doses ($D_i=0,1$). Then for event $k$, by treating the low dose as the reference level, for SE patients ($W_i=1$), we use $\gamma_{k1}$ to denote the treatment effect for the standard dose ($D_i=1$); for RE patients ($W_i=0$), we use $\gamma_{k2}$ to denote the treatment effect for the standard dose ($D_i=1$) and $\gamma_{k3}$ to denote the treatment effect for the high dose ($D_i=2$). Finally, we can write $h_k(W_i,D_i)$ as:
	\begin{equation}\label{eq:hr}
		h_k(W_i,D_i)=\gamma_{k1}D_iW_i+\Big( \gamma_{k2}{\rm I}(D_i=1)+\gamma_{k3}{\rm I}(D_i=2)  \Big)(1-W_i),
	\end{equation}
	with ${\rm I}(\cdot)$ representing the indicator function. Hence, by utilizing the inherent RT dose restrictions for different groups of patients to limit the model space, we develop a parsimonious yet flexible model with little parametric model assumption. We have considered two alternative model specifications such as  $h_k(W_i,D_i)=\gamma_{k1}W_i+\gamma_{k2}{\rm I}(D_i=1)+\gamma_{k3}{\rm I}(D_i=2)$ and $h_k(W_i,D_i)=\gamma_{k1}W_i+\gamma_{k2}{\rm I}(D_i=1)+\gamma_{k3}{\rm I}(D_i=2)+\gamma_{k4}W_i{\rm I}(D_i=1)+\gamma_{k5}W_i{\rm I}(D_i=2)$. However, none of them yields even comparable results as using the proposed model (\ref{eq:hr}) (results not shown). Indeed, the first alternative model makes strong additive model assumption and is sensitive to the model mis-specification, and the second alternative model is too flexible and hard to be fitted using the observed data because we do not have any observations with $W_i=1$ and $D_i=2$ or $W_i=0$ and $D_i=0$.      
	
	After specifying $h_k(W_i,D_i)$, we further denote $\delta_{ki} = 1$ if the $i$-th patient experiences the $k$-th cause-specific event as the first event and $\delta_{ki} = 0$, otherwise. Let $S_{k}(X_i|W_i,D_i) = \exp\Big\{-\int_0^{X_i} \lambda_{k}(x|W_i,D_i)dx \Big\}$ be the survival function. Then, the likelihood function for all the $n$ patients is expressed as following:
	
	\begin{equation}\label{eq:likelihood}
		L\Big({\cal M}|\Theta\Big)=\prod_{i=1}^n\prod_{k=1}^2 \lambda_{k}\Big(X_i \mid W_i, D_i\Big)^{\delta_{ki}} S_{k}\Big(X_i \mid W_i, D_i\Big), 
	\end{equation}
	where ${\cal M}$ represents the data, and $\Theta$ represents all the parameters of interest. 
	
	We propose to estimate $\Theta$ under the Bayesian framework. Prior distributions for $\Theta$ are given as:
	\begin{equation}\label{eq:prior}
		\pi\big(\alpha_k\big) \sim \mbox{Gamma}\big(a, b\big), \quad \pi\big(\beta_k\big) \sim \mbox{Gamma}\big(a, b\big), \quad \mbox{and} \quad \pi\big(\gamma_{kl}\big) \sim \mbox{Normal}\big(0, c^2\big),
	\end{equation}
	where $k = 1, 2$, $l = 1, 2, 3$, Gamma$(a, b)$ is the Gamma distribution with mean $ab$ and variance $ab^2$, and $\pi(\cdot)$ is the density function for prior distribution. Then, the posterior distribution of the proposed model is given as
	\begin{equation}\label{eq:posterior}
		\pi\Big(\Theta \mid {\cal M}\Big) \propto  \prod_{i=1}^n\prod_{k=1}^2 \lambda_{k}\Big(X_i \mid W_i,D_i\Big)^{\delta_{ki}} S_{k}\Big(X_i \mid W_i,D_i\Big)\prod_{k=1}^2 \Big\{\pi\big(\alpha_k\big) \pi\big(\beta_k\big) \prod_{l=1}^3 \pi\big(\gamma_{kl}\big)\Big\}.
	\end{equation}
	We derive the full conditional distribution for each parameter from the formula (\ref{eq:posterior}) and use the Metropolis-within-Gibbs-Sampler algorithm to draw posterior samples of $\Theta$ sequentially from the full conditional distribution. These posterior distributions will be used to guide the patients' randomization and treatment evaluation. 
	
	\section{Precision Response-adaptive Randomization Design} 
	
	We now propose the phase II clinical trial design based on the aforementioned probability model (\ref{eq:posterior}). The purpose of the proposed design is to evaluate the overall risk-benefit profile for each RT dose; randomize patients to receive more desirable RT doses based on patients' radiation susceptibility statuses, and select the best subgroup-specific RT dose. Towards these goals, we need a tradeoff measurement to compromise two cause-specific events (disease progression and normal tissue complications). We propose to use a utility function to measure each patient's survival benefit, which is a function of the RT dose given the radiation susceptibility status. The utility function should consider the event-happening time point because the later event is preferable. 
	
	Assuming that each patient will be followed in a time interval $[0, \nu]$ with $0$ denoting the beginning of randomization and $\nu$ denoting the end of follow-up. We equally partition $[0, \nu]$ into two sub-intervals and define five response-specific events referred to as $E_1$ to $E_5$. $E_1$ and $E_2$ are the events that disease progression ($k=1$) or normal tissue complications ($k=2$) occur between times 0 and $\nu /2$, respectively. Similarly, $E_3$ and $E_4$ are the events that disease progression or normal tissue complications occur between time $\nu /2$ and $\nu$, respectively.  $E_5$ is the best event that neither disease progression nor normal tissue complications occur during the whole follow-up $[0, \nu]$. After consulting the clinicians of the trial, we can assign different desirability weights to each response-specific event. The desirability weight ranges from 0 to 100, with a larger value representing higher desirability. We denote the desirability weights as $O_1$ to $O_5$. In Table \ref{tab:weight} we provide an example of the desirability weights. Conceptually, we can partition $\nu$ into more sub-intervals rather than two and correspondingly redefine the desirability weight. However, assigning an appropriate weight for each sub-interval is not straightforward for clinicians when the number of sub-intervals becomes large.
	\begin{table}[htbp]   
		\centering
		
		\scalebox{0.8}{			
			\begin{tabular}{c|ccccc}
				\hline\hline 
				& $E_1$ & $E_2$ & $E_3$ & $E_4$ & $E_5$  \\ 
				\multirow{-2}{*}{Response-specific event} & $X_i\leq \nu/2, k=1$; & $X_i\leq \nu/2, k=2$; & $\nu/2<X_i\leq \nu, k=1$;  & $\nu<X_i\leq \nu, k=2$; & $X_i>\nu$ \\ \hline 
				Weight & 0 & 5 & 10 & 20 & 100 \\
				\hline\hline 
			\end{tabular}
		}
		\caption{\label{tab:weight} 
			An example of the desirability weights for the utility function.
		} 
	\end{table}	 
	
	We note that the utility function is very general, and the desirability weights can be easily tailored to each trial's specific requirement. For example, if the time point of the event happening is not important, we can specify $E_1=E_3$ and $E_2=E_4$. If a trial is only interested in disease progression, we can specify $E_2=E_4=100$. Finally, for a patient with the radiation susceptibility status $W$, we construct the true utility function as $U(D \mid W, \Theta) = \sum_{s=1}^5 P(E_s \mid \Theta) O_s$ to measure his/her survival benefit at RT dose $D$ for $D=0,1,2$ by jointly considering disease progression and normal tissue complications. The true utility function contains unknown parameters $\Theta$, estimated through the proposed probability model. Under the competing risk model (4), given the current data ${\cal M}$ and the radiation susceptibility status $W$, we can derive the posterior mean utility function at RT dose $D$ as:    
	\begin{equation}
		\begin{split}
			\widetilde{U}(D \mid W,{\cal M} ) &= \int \sum_{s=1}^2 \bigg[\int_0^{\nu/2} \Big\{ \prod_{k=1}^2 S_k\big(x \mid W, D, \Theta \big) \Big\} \lambda_s\big(x \mid W, D, \Theta\big) dx \bigg] O_{s} \pi(\Theta|{\cal M})d\Theta\\ 
			&+ \int\sum_{s=1}^2 \bigg[\int_{\nu/2}^{\nu} \Big\{ \prod_{k=1}^2 S_k \big(x \mid W, D, \Theta\big)  \Big\} \lambda_s \big(x \mid W, D, \Theta\big)dx \bigg] O_{s+2}\pi(\Theta|{\cal M})d\Theta\\ 
			&+ \int\sum_{s=1}^2 \bigg[\int_{\nu}^{\infty} \Big\{ \prod_{k=1}^2 S_k \big(x \mid W, D, \Theta \big)  \Big\} \lambda_s (x \mid W, D, \Theta\big) dx \bigg]O_5\pi(\Theta|{\cal M})d\Theta.
		\end{split}
		\label{equ:utility}
	\end{equation}
	$\widetilde{U}(D \mid W,{\cal M} )$ integrates $\Theta$ over its posterior distribution $\pi(\Theta|{\cal M})$ and therefore is a function of only the RT dose $D$ so it can be used to conduct the trial. 
	
	In addition to the utility function, we also construct two admissible sets to safeguard the patients in the trial. The purpose of developing these admissible sets is to avoid treating patients at overly toxic or less efficacious RT doses. To achieve this goal, we propose continuously monitoring the cumulative incidence rates for disease progression and normal tissue complications events, respectively. The cumulative incidence rate is the probability that a cause-specific event occurs first within $[0, \nu]$, which can be written as: 
	\begin{equation}
		\begin{split}
			P_{k}(D| W, \Theta) &= {\rm Pr}(X\leq \nu, \mbox{cause}=k \mid W, D, \Theta)\\
			&=\int_{0}^{\nu} \Big\{\prod_{k=1}^2 S_k (x \mid W, D, \Theta)\Big\} \lambda_k (x \mid W, D, \Theta) dx.
		\end{split}   
		\label{equ:cir}
	\end{equation}  
	
	Then, let $\tau_k$ be the highest acceptable cumulative incidence rate for the cause-specific event $k$. For RE patients, the admissible set is constructed as
	$$
	{\cal A}_0=\Big\{D \in (1,2):  \cap_{k=1}^2 {\rm Pr}( P_k(D|W=0, \Theta)>\tau_k| {\cal M}  )<q_k       \Big\}; 
	$$ 
	and for SE patients, the admissible set is constructed as
	$$
	{\cal A}_1=\Big\{D \in (0,1):  \cap_{k=1}^2 {\rm Pr}( P_k(D|W=1, \Theta)>\tau_k| {\cal M}  )<q_k       \Big\} 
	$$   
	with ${\rm Pr}( P_k(D|W, \Theta)>\tau_k| {\cal M}  )=\int {\rm I}(P_k(D|W, \Theta)>\tau_k)\pi(\Theta|{\cal M})d\Theta$. $q_k$ is the pre-determined cut-off value, which is typically calibrated through simulation studies to yield good operating characteristics. During each interim analysis of the trial, we restrict the randomization scheme within the admissible sets to strengthen patients' benefit.  
	
	Our proposed precision response-adaptive randomization design starts by equally randomizing the first $n_1$ cohorts of patients to different RT doses based on patients' radiation susceptibility statuses. That is, we first measure each patient's radiation susceptibility status. Then, we equally randomize a RE patient to receive either the high RT dose or standard RT dose and equally randomize a SE patient to receive the low RT dose or standard RT dose. Then, starting from the $n_1+1$th cohort of the patient, we measure the radiation susceptibility statuses for the current cohort of patients and use the following response-adaptive randomization scheme for RT dose assignment based on all the observable data ${\cal M}$:
	\begin{enumerate}
		\item Construct the admissible sets ${\cal A}_0$ for RE patients, and ${\cal A}_1$ for SE patients, and restrict the randomization within the admissible sets.
		\item If ${\cal A}_0$ is empty, early terminate the enrollment for RE patients and claim no promising RT dose for RE patients. If ${\cal A}_0$ contains only one RT dose, assign all the RE patients in the current cohort to that RT dose. A similar RT dose assignment procedure is followed for the SE patients with different RT dose options.  
		\item If ${\cal A}_0$ contains two RT doses, randomize the RE patients to receive either RT dose $D=1$ or $D=2$ with the randomization ratios proportional to the posterior mean utility $\widetilde{U}(D=1|W=0, {\cal M})$ and $\widetilde{U}(D=2|W=0, {\cal M})$. If ${\cal A}_1$ contains two RT doses, a similar response-adaptive randomization procedure is followed for SE patients with different RT dose options, and $W=1$ is used in calculating the posterior mean utility.
		\item We repeat steps 1-3 until the trial is early terminated or the maximum sample size is reached. 
	\end{enumerate}
	
	At the end of the trial, for either RE or SE patients, if there is only one RT dose remaining in the admissible set, that RT dose is recommended as the subgroup-specific RT dose for the corresponding radiation susceptibility status subgroup. Otherwise, let us define $\mu_0$ and $\mu_1$ as the pre-determined cut-off values for final RT dose selection. For RE patients, we recommend the high RT dose if ${\rm Pr}\Big( U(D=2|W=0, \Theta)>U(D=1|W=0, \Theta) | {\cal M} \Big)>\mu_0$, and recommend the standard RT dose otherwise. For SE patients, we recommend the low RT dose if ${\rm Pr}\Big( U(D=0|W=1, \Theta)>U(D=1|W=1, \Theta) | {\cal M} \Big)>\mu_1$, and recommend the standard RT dose otherwise. Similar to $q_0$ and $q_1$, $\mu_0$ and $\mu_1$ can be calibrated through simulation studies, but the pre-preference of RT doses in clinical practice also needs to be considered. For example, if the standard RT dose is commonly used for the RE patients in the trial, then a large value of $\mu_0$ should be used to indicate that we will consider the high RT dose for RE patients only if the data strongly support that selection.      
	
	\section{Numerical Studies}
	
	As an essential step to apply the proposed design to the motivating NSCLC trial, we evaluate the operating characteristics of the design through numerical studies. Reporting operating characteristics is often required in trial protocols when a new design is involved.
	
	We specified that each cohort consisted of 5 patients and enrolled 4 cohorts of patients for equal randomization and 16 additional cohorts of patients for response-adaptive randomization. So the maximum sample size of the trial was 100. The radiation susceptibility status $W_i$ was generated from a Bernoulli distribution with a probability of 0.5. We used the Weibull distribution to
	generate the cause-specific hazards. The Weibull distributions were specified in a way that $50\%$ of the cause-specific events would occur within the first half of the follow-up sub-interval $[0, \nu/2]$. We set the highest acceptable disease progression rate and normal tissue complications rate at 0.4 ($\tau_1 = \tau_2 = 0.4$) and the cut-off values for admissible sets at 0.95 ($q_1 = q_2 = 0.95$). For the final subgroup-specific RT dose selection, we considered the general setting of no pre-preference of RT doses for both the RE and SE patients and set $\mu_0=\mu_1=0.5$. We used the desirability weights as those given in Table 1.
	
	We compared the proposed design with two alternative designs. The first design ignored the competing risk issue and modeled the time to disease progression and time to normal tissue complications events separately. We refer to this design as the ``separate'' design. The second design used equal randomization instead of the response-adaptive randomization, and we refer to this design as the ``ER'' design. We refer to the proposed design as the ``AR'' (adaptive randomization) design. We consider seven scenarios for the numerical studies. Under each scenario, we simulated 5,000 trials. In Scenarios 1-2, all doses are admissible. Details of Scenarios 1-2 are given as follows:
	\begin{itemize}
		\item In Scenario 1, the amount of decrease in cumulative incidence rate (CIR) for disease progression is equal to the increase in CIR for normal tissue complications.  
		\item In Scenario 2, the amount of CIR decrease for disease progression is higher than the amount increase in CIR for normal tissue complications.
	\end{itemize}
	
	In Scenario 3-7, at least one of the RT doses is not admissible. Details of Scenarios 3-7 are given as follows:
	\begin{itemize}
		\item In Scenario 3, only the standard RT dose is admissible for the RE patients, and none of the RT doses are admissible for the SE patients. 
		\item In Scenario 4, only the standard RT dose is admissible for the RE patients, and only the low RT dose is admissible for the SE patients. 
		\item In Scenario 5, both the standard and high RT doses are admissible for the RE patients. The amount of decrease in CIR for disease progression is larger than the amount of increase in CIR for normal tissue complications. Only the standard RT dose is admissible for the SE patients. 
		\item In Scenario 6, only the high RT dose is admissible for the RE patients. The low and standard RT doses are admissible for the SE patients, and the amount of CIR decrease for disease progression is less than the amount of CIR increase for normal tissue complications. 
		\item In Scenario 7, both the standard and high RT doses are admissible for the RE patients. The decrease in cumulative incidence rate (CIR) for disease progression is equal to the increase in CIR for normal tissue complications. Only the standard RT dose is admissible for the SE patients. 
	\end{itemize} 
	
	\begin{table}[htbp]
		\begin{minipage}{\linewidth}
			\begin{center}
				\scalebox{.5}{
					\begin{tabular}{ccccccccccc ccccccccc}
						\rowcolor{Gray} & & & \multicolumn{4}{c}{RE} & \multicolumn{4}{c}{SE} & &  \multicolumn{4}{c}{RE} & \multicolumn{4}{c}{SE} \\ 
						\rowcolor{Gray} & & &\multicolumn{2}{c}{STD} & \multicolumn{2}{c}{HIGH} & \multicolumn{2}{c}{LOW} & \multicolumn{2}{c}{STD} & & \multicolumn{2}{c}{STD} & \multicolumn{2}{c}{HIGH} & \multicolumn{2}{c}{LOW} & \multicolumn{2}{c}{STD} \\ 
						\rowcolor{Gray} \multirow{-3}{*}{Design} & & \multirow{-3}{*}{Scenario}& DP & NC & DP & NC & DP & NC & DP & NC & \multirow{-3}{*}{Scenario} & DP & NC & DP & NC & DP & NC & DP & NC \\  \hline
						
						& CIR & &
						0.2 & 0.2 & 0.1 & 0.3 & 0.3 & 0.1 & 0.2 & 0.2 & &
						0.3 & 0.1 & 0.05 & 0.2 & 0.6 & 0.1 & 0.2 & 0.2 \\ 
						& Utility & &
						\multicolumn{2}{c}{63.5} & \multicolumn{2}{c}{64.3} & \multicolumn{2}{c}{62.8} & \multicolumn{2}{c}{63.6} & & 
						\multicolumn{2}{c}{62.6} & \multicolumn{2}{c}{\textbf{77.7}} & \multicolumn{2}{c}{35.2} & \multicolumn{2}{c}{\textbf{63.4}} \\ 
						
						\cdashline{1-2} \cdashline{4-11} \cdashline{13-20}
						
						\multirow{4}{*}{AR} & $\#$ of patient treated & &
						\multicolumn{2}{c}{24.97} & \multicolumn{2}{c}{25.12} &
						\multicolumn{2}{c}{23.45} & \multicolumn{2}{c}{26.45} & &
						\multicolumn{2}{c}{18.38} & \multicolumn{2}{c}{31.76} &
						\multicolumn{2}{c}{13.19} & \multicolumn{2}{c}{36.6} \\
						& $\#$ DP, $\#$ NC & &
						5.03 & 5.03 & 2.47 & 7.58 & 7.14 & 2.32 & 5.3 & 5.28 & &
						5.53 & 1.83 & 1.68 & 6.25 & 8.02 & 1.29 & 7.36 & 7.27 \\
						& Selection probability & &
						\multicolumn{2}{c}{47.7} & \multicolumn{2}{c}{\textbf{52.3}} &
						\multicolumn{2}{c}{45.4} & \multicolumn{2}{c}{\textbf{54.6}} & &
						\multicolumn{2}{c}{23.5} & \multicolumn{2}{c}{\textbf{76.5}} &
						\multicolumn{2}{c}{8.8} & \multicolumn{2}{c}{\textbf{90.6}} \\ 
						& Early stop probability & &
						\multicolumn{4}{c}{0}  & \multicolumn{4}{c}{0} & &
						\multicolumn{4}{c}{0} & \multicolumn{4}{c}{0.6} \\ 
						
						\cdashline{1-2} \cdashline{4-11} \cdashline{13-20}
						
						\multirow{4}{*}{Separate} & $\#$ of patient treated  & &
						\multicolumn{2}{c}{25.72} & \multicolumn{2}{c}{24.62} &
						\multicolumn{2}{c}{24.59} & \multicolumn{2}{c}{25.08} & & 
						\multicolumn{2}{c}{22.46} & \multicolumn{2}{c}{27.5} &
						\multicolumn{2}{c}{13.24} & \multicolumn{2}{c}{36.78} \\
						& $\#$ DP, $\#$ NC & &
						5.12 & 5.18 & 2.49 & 7.31 & 7.36 & 2.52 & 4.99 & 4.94 & &
						6.75 & 2.28 & 1.38 & 5.52 & 8.02 & 1.28 & 7.29 & 7.48 \\
						& Selection probability ($\%$) & &
						\multicolumn{2}{c}{\textbf{54}} & \multicolumn{2}{c}{46} &
						\multicolumn{2}{c}{44.2} & \multicolumn{2}{c}{\textbf{55.8}} & &
						\multicolumn{2}{c}{44.3} & \multicolumn{2}{c}{\textbf{55.7}} &
						\multicolumn{2}{c}{7.1} & \multicolumn{2}{c}{\textbf{92.8}} \\ 
						& Early stop probability ($\%$) & &
						\multicolumn{4}{c}{0} & \multicolumn{4}{c}{0} & &
						\multicolumn{4}{c}{0} & \multicolumn{4}{c}{0.5} \\ 
						
						\cdashline{1-2} \cdashline{4-11} \cdashline{13-20}
						
						\multirow{4}{*}{ER} & $\#$ of patient treated  & &
						\multicolumn{2}{c}{25.26} & \multicolumn{2}{c}{24.96} &
						\multicolumn{2}{c}{24.96} & \multicolumn{2}{c}{24.82} & &
						\multicolumn{2}{c}{24.75} & \multicolumn{2}{c}{25.42} &
						\multicolumn{2}{c}{21.55} & \multicolumn{2}{c}{28.15} \\
						& $\#$ DP, $\#$ NC & &
						5.14 & 4.99 & 2.5 & 7.54 & 7.59 & 2.51 & 4.98 & 4.97 & &
						7.34 & 2.42 & 1.28 & 4.98 & 12.92 & 2.21 & 5.7 & 5.7 \\ 
						& Selection probability ($\%$) & &
						\multicolumn{2}{c}{48.4} & \multicolumn{2}{c}{\textbf{51.6}} &
						\multicolumn{2}{c}{42.9} & \multicolumn{2}{c}{\textbf{57.1}} & &
						\multicolumn{2}{c}{21.9} & \multicolumn{2}{c}{\textbf{78.1}} &
						\multicolumn{2}{c}{8} & \multicolumn{2}{c}{\textbf{91.5}} \\ 
						& Early stop probability ($\%$) &
						\multirow{-13}{*}{1} & \multicolumn{4}{c}{0} &  \multicolumn{4}{c}{0} &
						\multirow{-14}{*}{5} & \multicolumn{4}{c}{0} & \multicolumn{4}{c}{0.5} \\ \hline
						
						
						\rowcolor{Gray} & CIR & &
						0.3 & 0.1 & 0.05 & 0.2 & 0.25 & 0.2 & 0.1 & 0.3 & &
						0.5 & 0.1 & 0.1 & 0.15 & 0.1 & 0.05 & 0.08 & 0.35 \\ 
						\rowcolor{Gray} &  Utility & &
						\multicolumn{2}{c}{62.8} & \multicolumn{2}{c}{\textbf{77.7}} &
						\multicolumn{2}{c}{58.8} & \multicolumn{2}{c}{\textbf{64.2}} & &
						\multicolumn{2}{c}{44.3} & \multicolumn{2}{c}{\textbf{77.3}} &
						\multicolumn{2}{c}{\textbf{86.0}} & \multicolumn{2}{c}{61.8} \\ 
						
						\cdashline{1-2} \cdashline{4-11} \cdashline{13-20}
						
						\rowcolor{Gray} & $\#$ of patient treated & &
						\multicolumn{2}{c}{16.33} & \multicolumn{2}{c}{33.95} & 
						\multicolumn{2}{c}{22.44} & \multicolumn{2}{c}{27.65} & &
						\multicolumn{2}{c}{10.57} & \multicolumn{2}{c}{39.52} &
						\multicolumn{2}{c}{36.11} & \multicolumn{2}{c}{13.8} \\
						\rowcolor{Gray} & $\#$ DP, $\#$ NC & &
						4.92 & 1.59 & 1.73 & 6.76 & 5.67 & 4.48 & 2.83 & 8.26 & &
						5.29 & 1.09 & 4.07 & 6 & 3.61 & 1.77 & 1.11 & 4.83 \\
						\rowcolor{Gray} & Selection probability & &
						\multicolumn{2}{c}{20.8} & \multicolumn{2}{c}{\textbf{79.2}} &
						\multicolumn{2}{c}{37.1} & \multicolumn{2}{c}{\textbf{62.9}} & &
						\multicolumn{2}{c}{5.3} & \multicolumn{2}{c}{\textbf{94.7}} &
						\multicolumn{2}{c}{\textbf{88.5}} & \multicolumn{2}{c}{11.5} \\ 
						\rowcolor{Gray} \multirow{-4}{*}{AR} & Early stop probability & &
						\multicolumn{4}{c}{0} &  \multicolumn{4}{c}{0} & &
						\multicolumn{4}{c}{0} & \multicolumn{4}{c}{0} \\ 
						
						\cdashline{1-2} \cdashline{4-11} \cdashline{13-20}
						
						\rowcolor{Gray} & $\#$ of patient treated & &
						\multicolumn{2}{c}{18.05} & \multicolumn{2}{c}{31.85} &
						\multicolumn{2}{c}{26.57} & \multicolumn{2}{c}{23.53} & &
						\multicolumn{2}{c}{13.12} & \multicolumn{2}{c}{36.75} &
						\multicolumn{2}{c}{37.38} & \multicolumn{2}{c}{12.75} \\
						\rowcolor{Gray} & $\#$ DP, $\#$ NC & &
						5.31 & 1.79 & 1.56 & 6.51 & 6.58 & 5.33 & 2.36 & 7.16 & &
						6.66 & 1.24 & 3.8 & 5.48 & 3.76 & 1.8 & 1.01 & 4.47\\
						\rowcolor{Gray} & Selection probability ($\%$) & &
						\multicolumn{2}{c}{29.8} & \multicolumn{2}{c}{\textbf{70.2}} &
						\multicolumn{2}{c}{\textbf{51.4}} & \multicolumn{2}{c}{48.6} & &
						\multicolumn{2}{c}{19.1} & \multicolumn{2}{c}{\textbf{80.9}} &
						\multicolumn{2}{c}{\textbf{90.6}} & \multicolumn{2}{c}{9.4} \\ 
						\rowcolor{Gray} \multirow{-4}{*}{Separate} & Early stop probability ($\%$) & &
						\multicolumn{4}{c}{0} & \multicolumn{4}{c}{0} & &
						\multicolumn{4}{c}{0} & \multicolumn{4}{c}{0} \\ 
						
						\cdashline{1-2} \cdashline{4-11} \cdashline{13-20}
						
						\rowcolor{Gray} & $\#$ of patient treated & &
						\multicolumn{2}{c}{24.95} & \multicolumn{2}{c}{25.11} &
						\multicolumn{2}{c}{25.35} & \multicolumn{2}{c}{24.59} & & 
						\multicolumn{2}{c}{22.11} & \multicolumn{2}{c}{28.06} &
						\multicolumn{2}{c}{25.53} & \multicolumn{2}{c}{24.29} \\
						\rowcolor{Gray} & $\#$ DP, $\#$ NC & &
						7.57 & 2.48 & 1.29 & 4.98 & 6.31 & 4.97 & 2.49 & 7.35 & &
						11.08 & 2.22 & 2.78 & 4.14 & 2.54 & 1.29 & 1.94 & 8.4 \\
						\rowcolor{Gray} & Selection probability ($\%$) & &
						\multicolumn{2}{c}{20} & \multicolumn{2}{c}{\textbf{80}} &
						\multicolumn{2}{c}{38.5} & \multicolumn{2}{c}{\textbf{61.5}} & & 
						\multicolumn{2}{c}{3.1} & \multicolumn{2}{c}{\textbf{96.9}} &
						\multicolumn{2}{c}{\textbf{88.8}} & \multicolumn{2}{c}{11.2} \\ 
						\rowcolor{Gray} \multirow{-4}{*}{ER} & Early stop probability ($\%$) &
						\multirow{-13}{*}{2} & \multicolumn{4}{c}{0} & \multicolumn{4}{c}{0} &
						\multirow{-13}{*}{6} & \multicolumn{4}{c}{0} & \multicolumn{4}{c}{0.3} \\ \hline
						
						& CIR & &
						0.3 & 0.1 & 0.2 & 0.6 & 0.6 & 0.1 & 0.2 & 0.6 & &
						0.15 & 0.1 & 0.1 & 0.15 & 0.5 & 0.05 & 0.1 & 0.35 \\ 
						& Utility & &
						\multicolumn{2}{c}{\textbf{62.3}} & \multicolumn{2}{c}{29.6} &
						\multicolumn{2}{c}{34.7} & \multicolumn{2}{c}{29.6} & &
						\multicolumn{2}{c}{77.3} & \multicolumn{2}{c}{77.0} &
						\multicolumn{2}{c}{48.4} & \multicolumn{2}{c}{\textbf{60.0}} \\
						
						\cdashline{1-2} \cdashline{4-11} \cdashline{13-20}
						
						\multirow{4}{*}{AR} & $\#$ of patient treated & &
						\multicolumn{2}{c}{39.46} & \multicolumn{2}{c}{10.25} &
						\multicolumn{2}{c}{26.18} & \multicolumn{2}{c}{19.47} & &
						\multicolumn{2}{c}{24.97} & \multicolumn{2}{c}{24.93} &
						\multicolumn{2}{c}{20.57} & \multicolumn{2}{c}{29.48} \\
						& $\#$ DP, $\#$ NC & &
						11.29 & 3.77 & 1.98 & 6.09 & 15.76 & 2.6 & 3.89 & 11.59 & &
						3.78 & 2.5 & 2.52 & 3.76 & 10.18 & 1.04 & 3.02 & 10.3 \\
						& Selection probability & &
						\multicolumn{2}{c}{\textbf{94.1}} & \multicolumn{2}{c}{4.9} &
						\multicolumn{2}{c}{34.5} & \multicolumn{2}{c}{32.2} & &
						\multicolumn{2}{c}{\textbf{51.8}} & \multicolumn{2}{c}{48.2} &
						\multicolumn{2}{c}{28.3} & \multicolumn{2}{c}{\textbf{71.4}}\\ 
						& Early stop probability & &
						\multicolumn{4}{c}{1} &  \multicolumn{4}{c}{33.3} & &
						\multicolumn{4}{c}{0} & \multicolumn{4}{c}{0.3} \\ 
						
						\cdashline{1-2} \cdashline{4-11} \cdashline{13-20}
						
						\multirow{4}{*}{Separate} & $\#$ of patient treated  & &
						\multicolumn{2}{c}{36.71} & \multicolumn{2}{c}{13.28} &
						\multicolumn{2}{c}{29.36} & \multicolumn{2}{c}{20.66} & & 
						\multicolumn{2}{c}{25.36} & \multicolumn{2}{c}{24.73} &
						\multicolumn{2}{c}{20.58} & \multicolumn{2}{c}{29.33} \\
						& $\#$ DP, $\#$ NC & &
						10.86 & 3.89 & 2.61 & 8.01 & 17.52 & 3.07 & 4.17 & 12.33 & &
						3.84 & 2.6 & 2.45 & 3.77 & 10.4 & 1 & 2.96 & 10.31  \\
						& Selection probability ($\%$) & &
						\multicolumn{2}{c}{\textbf{84.5}} & \multicolumn{2}{c}{15.5} &
						\multicolumn{2}{c}{60.6} & \multicolumn{2}{c}{39.4} & &
						\multicolumn{2}{c}{46.5} & \multicolumn{2}{c}{\textbf{53.5}} &
						\multicolumn{2}{c}{25.3} & \multicolumn{2}{c}{\textbf{74.7}} \\ 
						& Early stop probability ($\%$) & &
						\multicolumn{4}{c}{0} &  \multicolumn{4}{c}{0} & &
						\multicolumn{4}{c}{0} & \multicolumn{4}{c}{0} \\ 
						
						\cdashline{1-2} \cdashline{4-11} \cdashline{13-20}
						
						\multirow{4}{*}{ER} & $\#$ of patient treated  & &
						\multicolumn{2}{c}{30.18} & \multicolumn{2}{c}{19.7} &
						\multicolumn{2}{c}{24.46} & \multicolumn{2}{c}{21.3} & &
						\multicolumn{2}{c}{25.05} & \multicolumn{2}{c}{24.81} &
						\multicolumn{2}{c}{24.79} & \multicolumn{2}{c}{25.23}\\
						& $\#$ DP, $\#$ NC & &
						8.59 & 2.89 & 3.86 & 11.42 & 14.7 & 2.46 & 4.32 & 12.76 & &
						3.74 & 2.48 & 2.48 & 3.76 & 12.41 & 1.22 & 2.52 & 8.92\\ 
						& Selection probability ($\%$) & &
						\multicolumn{2}{c}{\textbf{96}} & \multicolumn{2}{c}{3.4}  &
						\multicolumn{2}{c}{38.3} & \multicolumn{2}{c}{27.3} & &
						\multicolumn{2}{c}{\textbf{50.2}} & \multicolumn{2}{c}{49.9} &
						\multicolumn{2}{c}{27} & \multicolumn{2}{c}{\textbf{72.3}}\\ 
						& Early stop probability ($\%$) &
						\multirow{-13}{*}{3} & \multicolumn{4}{c}{0.6} & \multicolumn{4}{c}{34.4} &
						\multirow{-14}{*}{7} & \multicolumn{4}{c}{0} & \multicolumn{4}{c}{0.7} \\ \hline
						
						\rowcolor{Gray} & CIR & &
						0.3 & 0.1 & 0.2 & 0.6 & 0.25 & 0.1 & 0.2 & 0.6 & &
						& & & & & & & \\ 
						\rowcolor{Gray} & Utility & &
						\multicolumn{2}{c}{\textbf{62.5}} & \multicolumn{2}{c}{29.8} &
						\multicolumn{2}{c}{\textbf{67.3}} & \multicolumn{2}{c}{29.8} & &
						& & & & & & & \\ 
						
						\cdashline{1-2} \cdashline{4-11} 
						
						\rowcolor{Gray} & $\#$ of patient treated & &
						\multicolumn{2}{c}{39.47} & \multicolumn{2}{c}{10.28} &
						\multicolumn{2}{c}{40.8} & \multicolumn{2}{c}{9.29} & &
						& & & & & & & \\
						\rowcolor{Gray} & $\#$ DP, $\#$ NC & &
						11.95 & 3.95 & 2.02 & 6.2 & 10.23 & 4.1 & 1.82 & 5.59 & &
						& & & & & & & \\
						\rowcolor{Gray} & Selection probability & &
						\multicolumn{2}{c}{\textbf{93.9}} & \multicolumn{2}{c}{5.3} &
						\multicolumn{2}{c}{\textbf{96.1}} & \multicolumn{2}{c}{3.9} & &
						& & & & & & &\\ 
						\rowcolor{Gray} \multirow{-4}{*}{AR} & Early stop probability & &
						\multicolumn{4}{c}{1.1} & \multicolumn{4}{c}{0} & &
						& & & & & & & \\ 
						
						\cdashline{1-2} \cdashline{4-11} 
						
						\rowcolor{Gray} & $\#$ of patient treated & &
						\multicolumn{2}{c}{35.57} & \multicolumn{2}{c}{14.33} &
						\multicolumn{2}{c}{40.99} & \multicolumn{2}{c}{9.11} & &
						& & & & & & &  \\
						\rowcolor{Gray} & $\#$ DP, $\#$ NC & &
						10.66 & 3.51 & 2.87 & 8.64 & 10.32 & 4.07 & 1.87 & 5.49 & &
						& & & & & & & \\
						\rowcolor{Gray} & Selection probability ($\%$) & &
						\multicolumn{2}{c}{\textbf{81.4}} & \multicolumn{2}{c}{18.6} &
						\multicolumn{2}{c}{\textbf{95.5}} & \multicolumn{2}{c}{4.5} & &
						& & & & & & & \\ 
						\rowcolor{Gray} \multirow{-4}{*}{Separate} & Early stop probability ($\%$) & &
						\multicolumn{4}{c}{0} & \multicolumn{4}{c}{0} & &
						& & & & & & & \\ 
						
						\cdashline{1-2} \cdashline{4-11} 
						
						\rowcolor{Gray} & $\#$ of patient treated & &
						\multicolumn{2}{c}{31.2} & \multicolumn{2}{c}{18.84} &
						\multicolumn{2}{c}{30.41} & \multicolumn{2}{c}{19.38} & & 
						& & & & & & & \\
						\rowcolor{Gray} & $\#$ DP, $\#$ NC & &
						9.46 & 3.06 & 3.71 & 11.41 & 7.71 & 3.06 & 3.98 & 11.45 & &
						& & & & & & & \\
						\rowcolor{Gray} & Selection probability ($\%$) & &
						\multicolumn{2}{c}{\textbf{96.4}} & \multicolumn{2}{c}{2.6} &
						\multicolumn{2}{c}{\textbf{98.2}} & \multicolumn{2}{c}{1.8} & & 
						& & & & & & & \\ 
						\rowcolor{Gray} \multirow{-4}{*}{ER} & Early stop probability ($\%$) &
						\multirow{-13}{*}{4} & \multicolumn{4}{c}{1} & \multicolumn{4}{c}{0} &
						& & & & & & & & \\ \hline
					\end{tabular}
				}
				\caption{ The results of numerical studies based on 5,000 simulated trials. CIR is the cumulative incidence rate, DP is disease progression, NC is normal tissue complications. AR is the proposed design; Separate is the conventional design ignoring the competing risk issue; ER is similar to AR but uses equal randomization.
				}
				\label{tab:result1}
			\end{center}
		\end{minipage}
	\end{table}
	
	Table \ref{tab:result1} summarized the operating characteristics of the three designs under investigation, including the RT dose selection probability, the number of patients treated, and early stopping percentages, all stratified by the radiation susceptibility status. The RT dose selection probability indicates the benefit for further patients outside the trial, and the number of patients treated indicates the benefit for current patients in the trial.
	
	In Scenario 1, the selection probabilities for the proposed AR and ER designs were comparable. The utility values of the RT doses were close within each subgroup under this scenario. Both designs yielded higher probabilities in selecting the RT doses with higher utility values. However, the Separate design, which ignored the competing risk issue, was preferable to the RT doses with fewer utility values. The total number of patients in the trial is identical for all the designs. However, due to the advantage of response-adaptive randomization, the AR design assigned more SE patients to the standard RT dose than the other two designs. The early stopping percentages were close to 0 for all the designs because all the RT doses were in the admissible set. 
	
	In Scenario 2, both the AR and ER designs exhibited desirable subgroup-specific RT dose selection probabilities, significantly outperforming the Separate design. In addition, the AR design was the most ethical design in terms of patients' allocation as it allocated the most number of patents to the true optimal RT dose, maximizing the survival benefit. Indeed, let us consider the ratio of patients assigned to the optimal RT dose to those assigned to the non-optimal RT dose (optimal/non-optimal ratio) as a measurement for the individual ethics of the design. The ratio was 2.07 for the AR design, 1.76 for the Separate design, and further dropped to 1.01 for ER because of equal randomization. In Scenario 3, all designs had overwhelmingly high correct subgroup-specific RT dose selection probabilities for the RE patients. The AR and Separate designs outperformed the ER designs in terms of patients' allocation. There were no admissible RT doses for the SE patients. Compared with the Separate design, the AR and ER designs yielded around $34\%$ higher early stopping percentage. 
	
	In Scenario 4, since only one RT dose is admissible for both subgroups, all three designs had high correct RT dose selection probabilities. However, the AR and ER designs yielded at least $10\%$ higher selection probability than the Separate design for the RE patients. Regarding optimal/non-optimal ratio, the AR and Separate designs reported similar values of 3.84 for the RE patients and 4.39 for the SE patients, which was significantly better than the ER design (1.66 for the RE patients and 1.56 for the SE patients). The results for Scenarios 5 to 7 were similar. 
	
	In summary, the proposed AR and ER designs surpassed the Separate design in terms of correct subgroup-specific RT dose selection, and the improvement can be substantial. For the individual ethics, the AR design generally performed best across all the scenarios due to the response-adaptive randomization scheme. Therefore, by jointly considering the RT dose selection and patients' allocation, we recommended the proposed AR design being used in practice. 
	
	We conducted additional sensitivity studies to investigate the robustness of the proposed AR designs by varying the sample size, proportion of RE patients, time-to-event data generation function, and the desirability weights. Specifically, we varied the sample size from 60 to 200 and the proportion from 0.1 to 0.9 and considered the Logistic survival distribution in addition to the Weibull distribution. We also considered different values of the desirability weights, as summarized in Table \ref{tab:sen_weight}. We summarized our sensitivity analysis results of Scenarios 1, 3, and 6 in Figures 2, 3, 4, and 5. Figures 2 and 3 depicted the dose selection probabilities for the RE and SE patients. Figures 4 and 5 show the proportions of RE and SE patients assigned at each RT dose. 
	
	\begin{table}[htbp]
		\centering
		
		\scalebox{0.9}{
			\begin{tabular}{*{6}{c}} 
				\hline\hline 
				& $E_1$ & $E_2$ & $E_3$ & $E_4$ & $E_5$   \\ 
				\multirow{-2}{*}{} & $X_i\leq \nu/2, k=1$; & $X_i\leq \nu/2, k=2$; & $\nu/2<X_i\leq \nu, k=1$;  & $\nu<X_i\leq \nu, k=2$; & $X_i>\nu$ \\\hline
				1 & 0 & 5 & 5 & 10 & 100 \\
				2 & 0 & 5 & 20 & 30 & 100 \\
				3 & 0 & 5 & 10 & 20 & 100 \\
				4 & 0 & 0 & 5 & 5 & 100 \\
				5 & 0 & 0 & 20 & 20 & 100 \\
				6 & 0 & 0 & 10 & 10 & 100 \\
				7 & 5 & 0 & 10 & 5 & 100 \\
				8 & 5 & 0 & 30 & 20 & 100 \\
				9 & 5 & 0 & 20 & 10 & 100 \\\hline
			\end{tabular}
		}
		\caption{\label{tab:sen_weight} 
			The desirability weights used in the sensitivity analysis
		}  
	\end{table}
	
	In general, the performance of the proposed design was improved with a larger sample size. The proportion of RE patients has mild impact on the performance of the proposed design and the trend depends on both the scenarios under investigation and patients' radiation susceptibility statuses. The data-generation function and the desirability weights did little impact on the proposed design.
	
	\begin{figure}[htbp]
		\centering
		\begin{tabular}{C{2.5cm}C{4cm}C{4cm}C{4cm}}
			& \multicolumn{3}{c}{Dose Selection Probability for RE patients} \\
			Sensitivity Analysis & Scenario 1  & Scenario 3 & Scenario 6  \\ \hline
			
			Number of total patients & \includegraphics[width=0.22\textwidth]{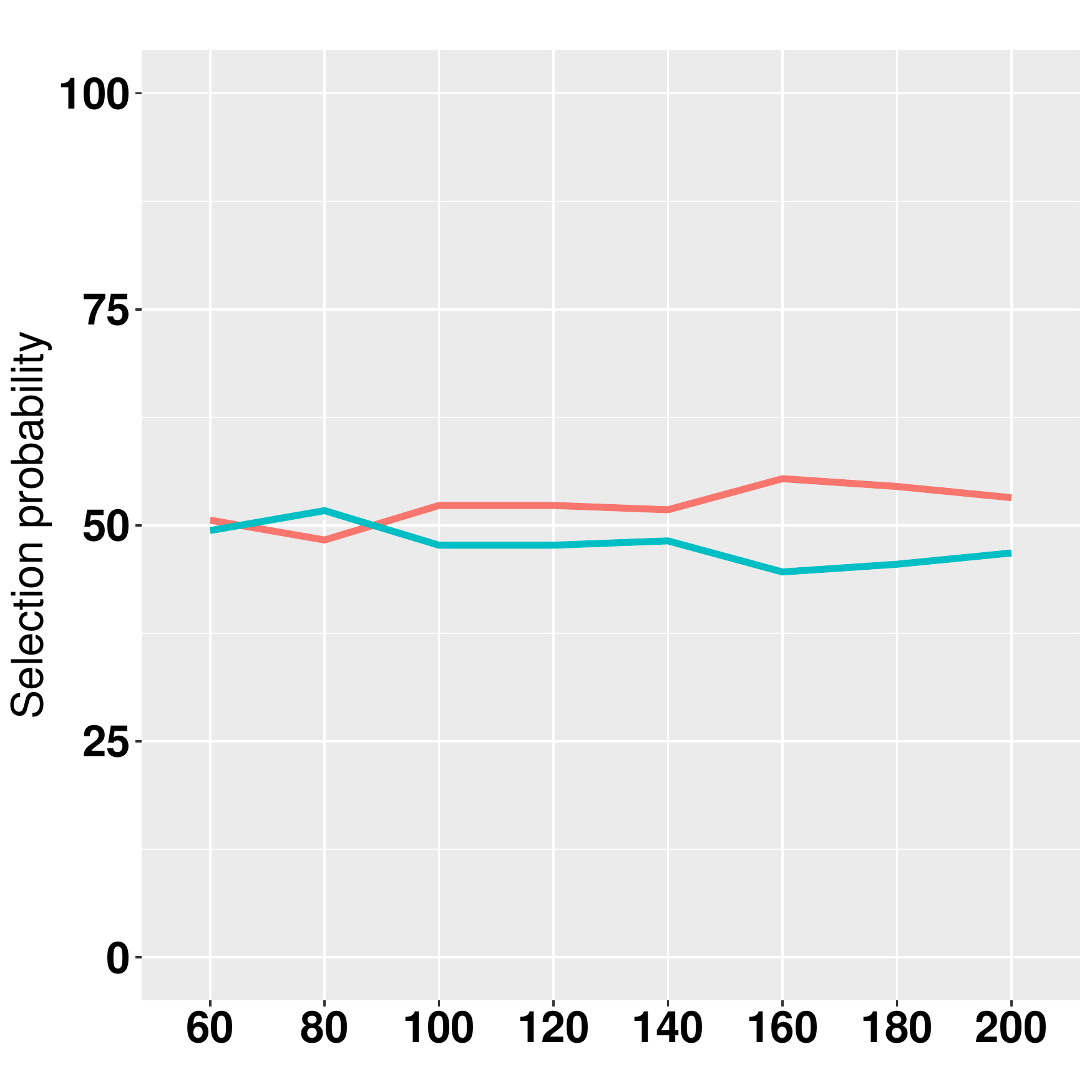}
			& \includegraphics[width=0.22\textwidth]{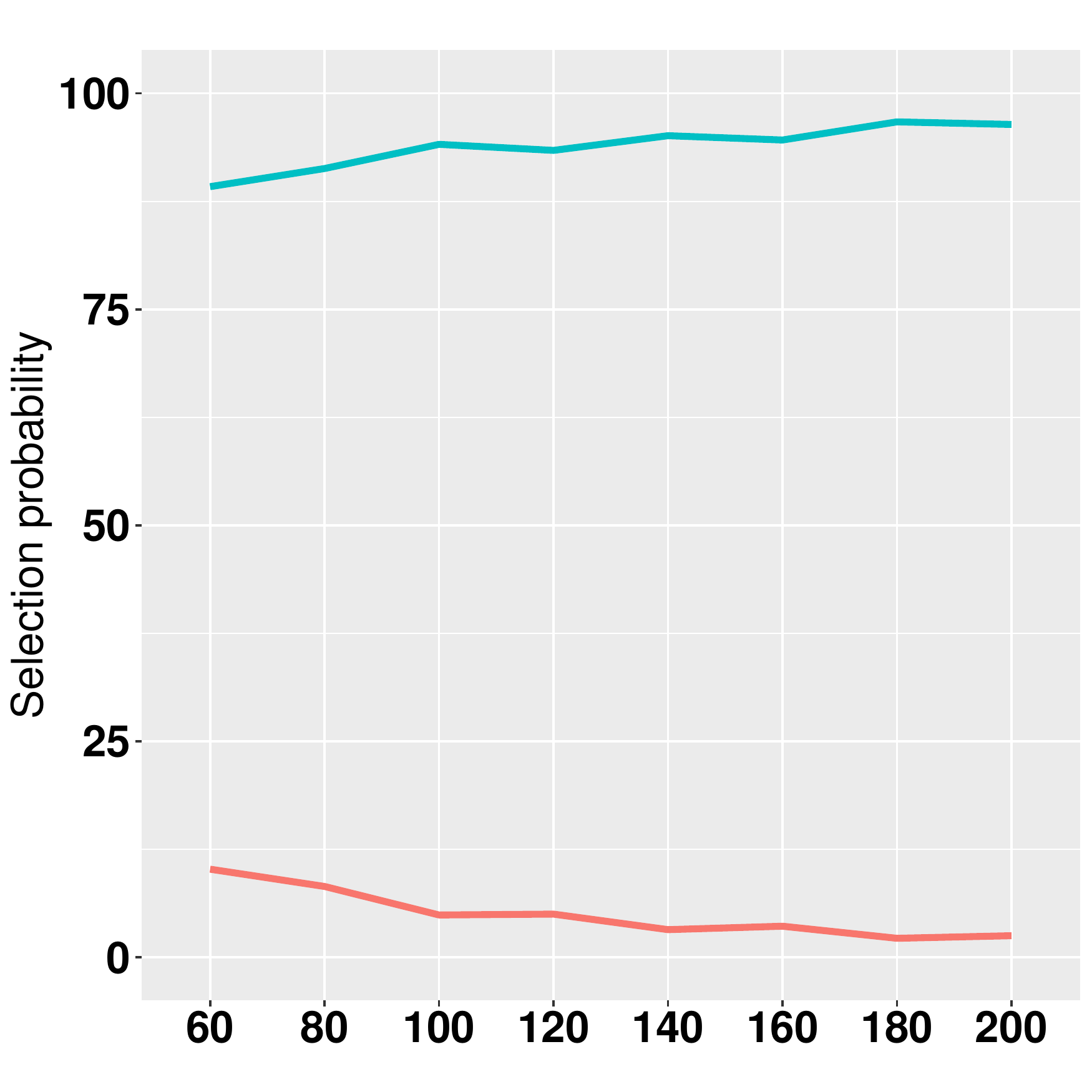}
			&\includegraphics[width=0.22\textwidth]{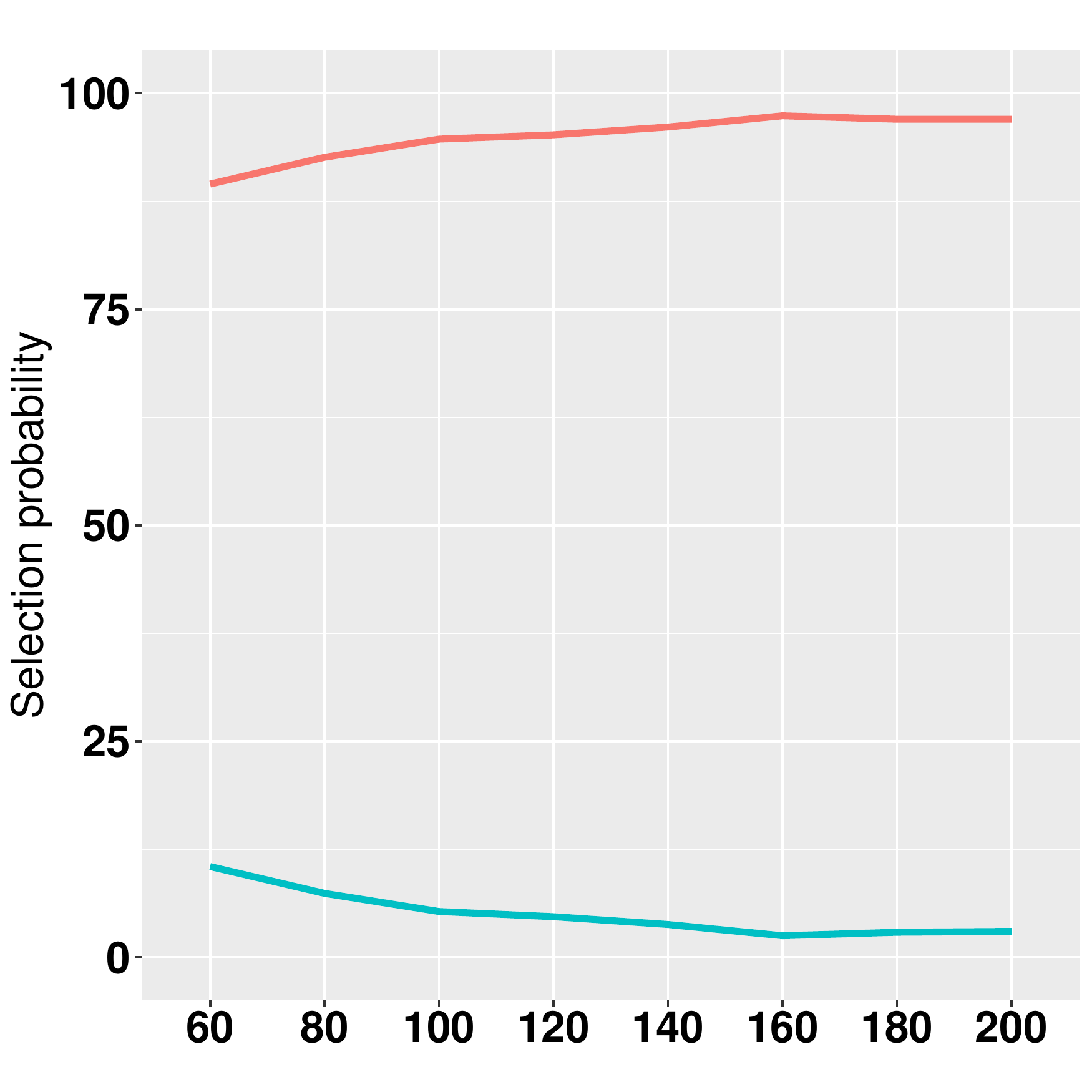}\\
			Proportion of the RE patients & \includegraphics[width=0.22\textwidth]{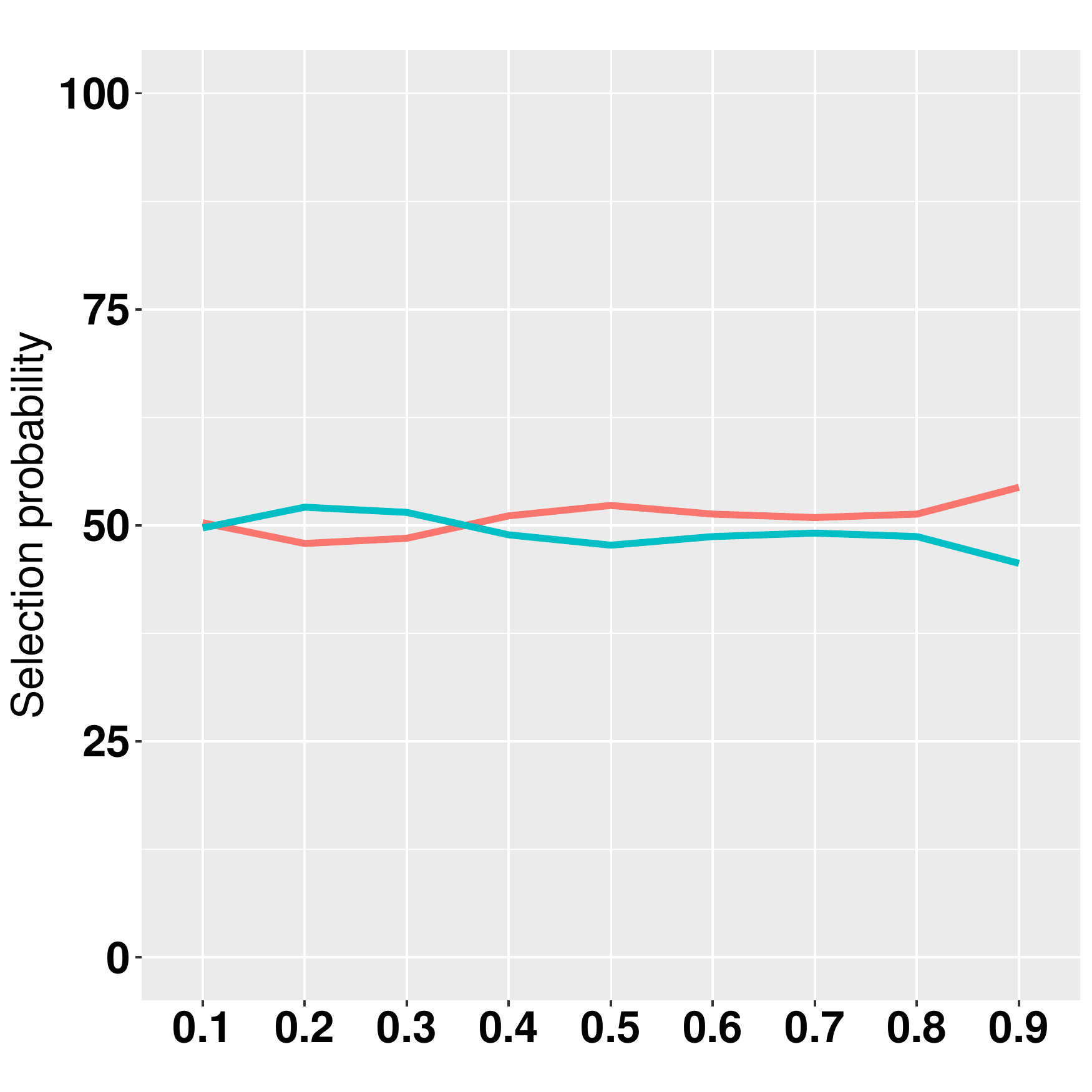}
			& \includegraphics[width=0.22\textwidth]{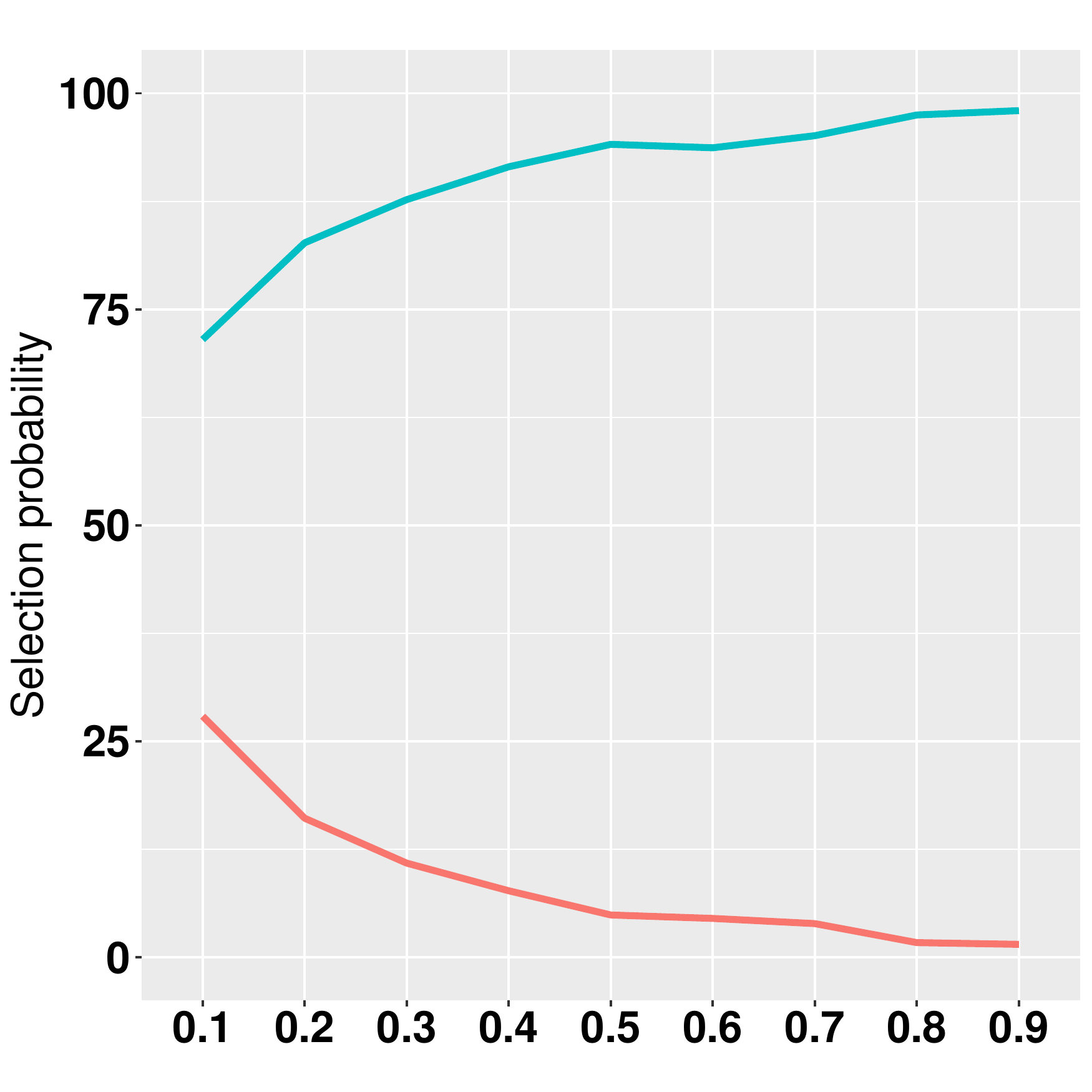}
			& \includegraphics[width=0.22\textwidth]{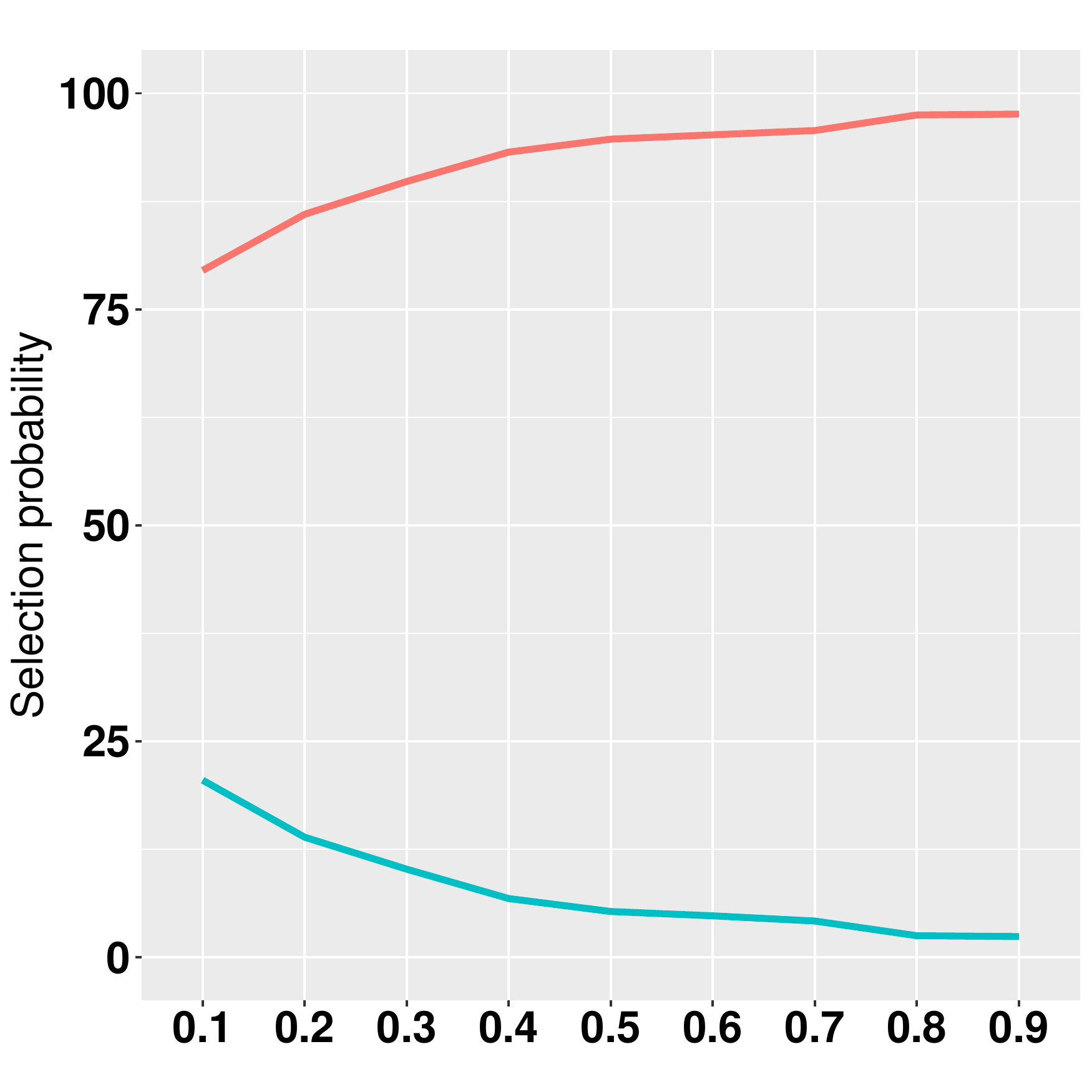}\\
			
			Distribution for the event time & \includegraphics[width=0.22\textwidth]{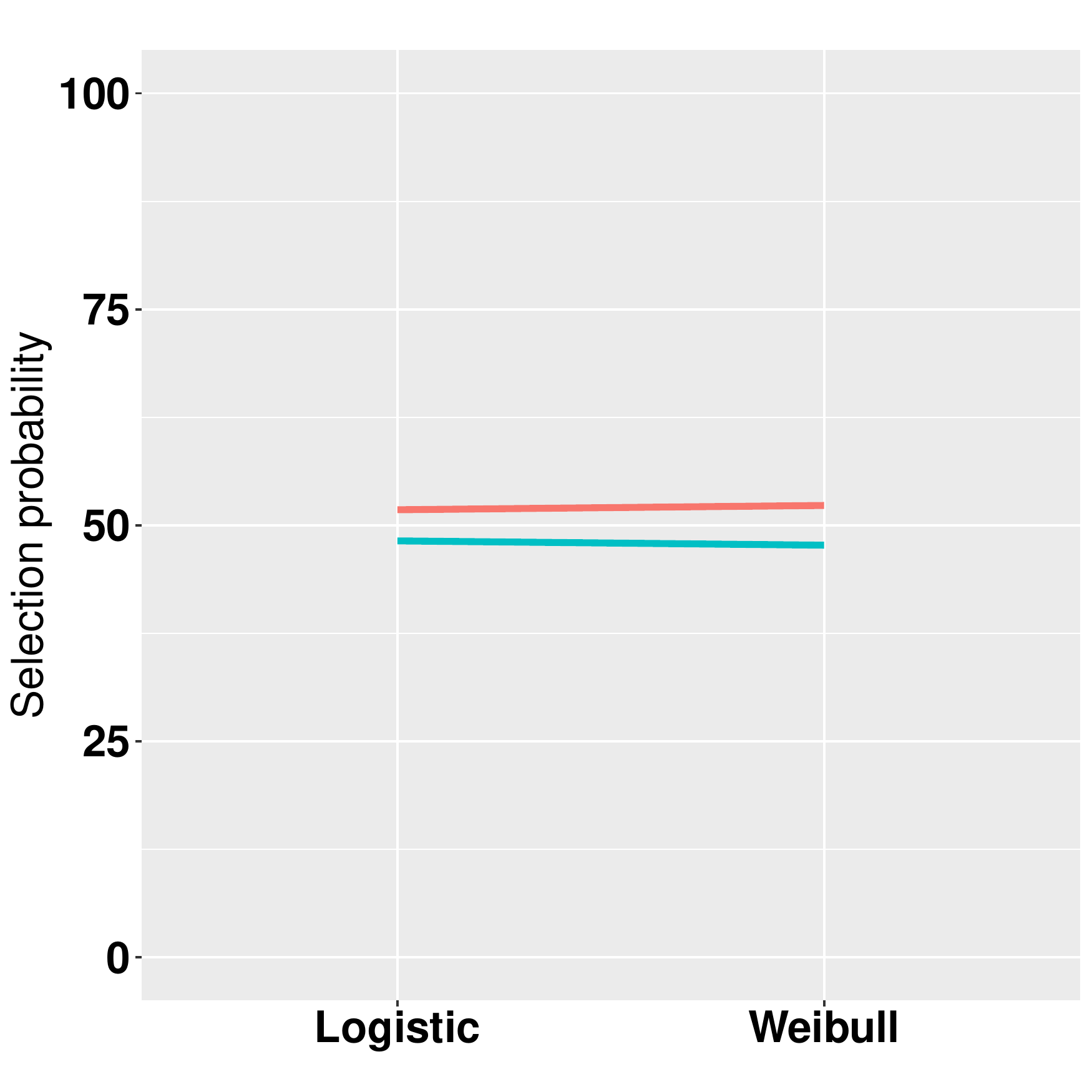}
			& \includegraphics[width=0.22\textwidth]{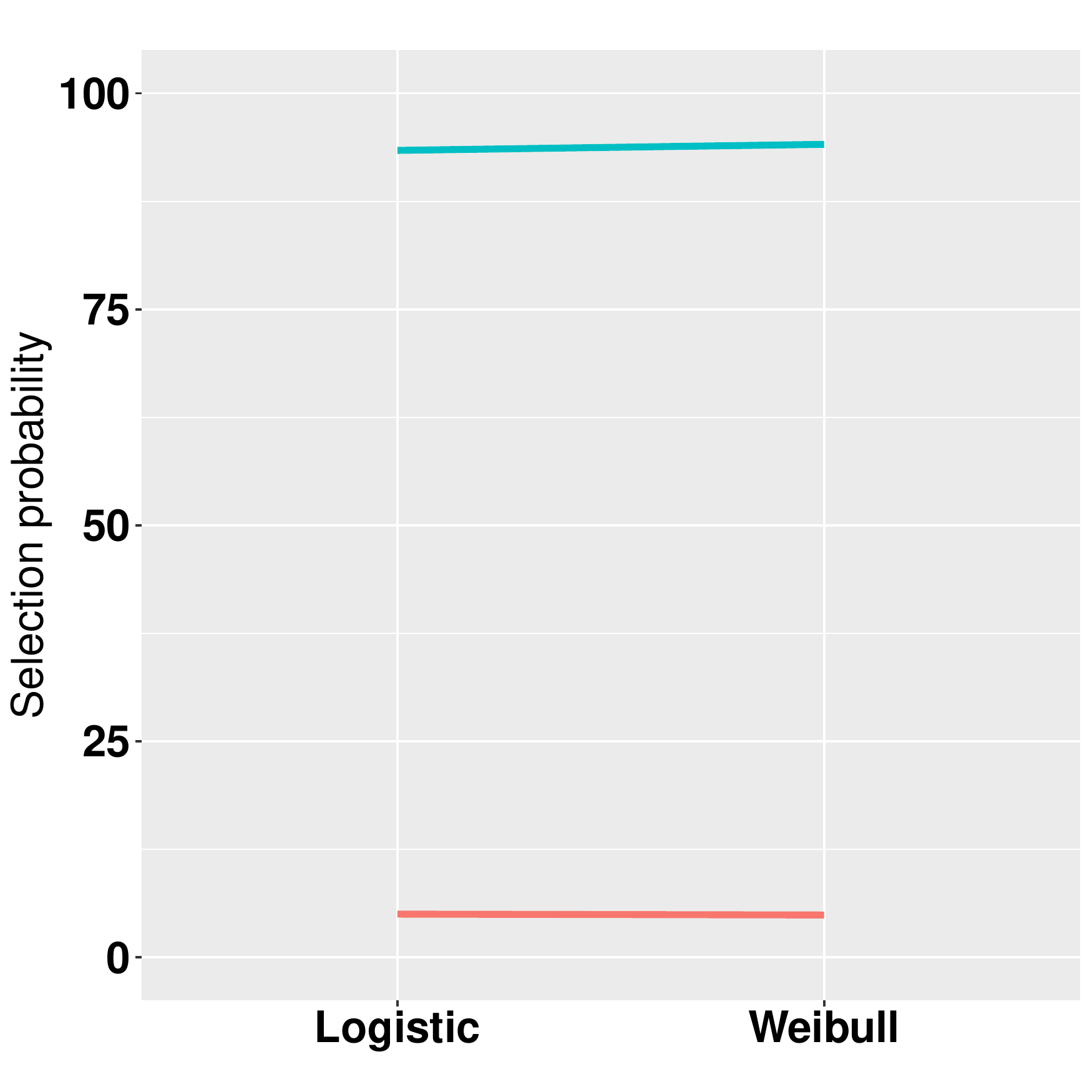}
			& \includegraphics[width=0.22\textwidth]{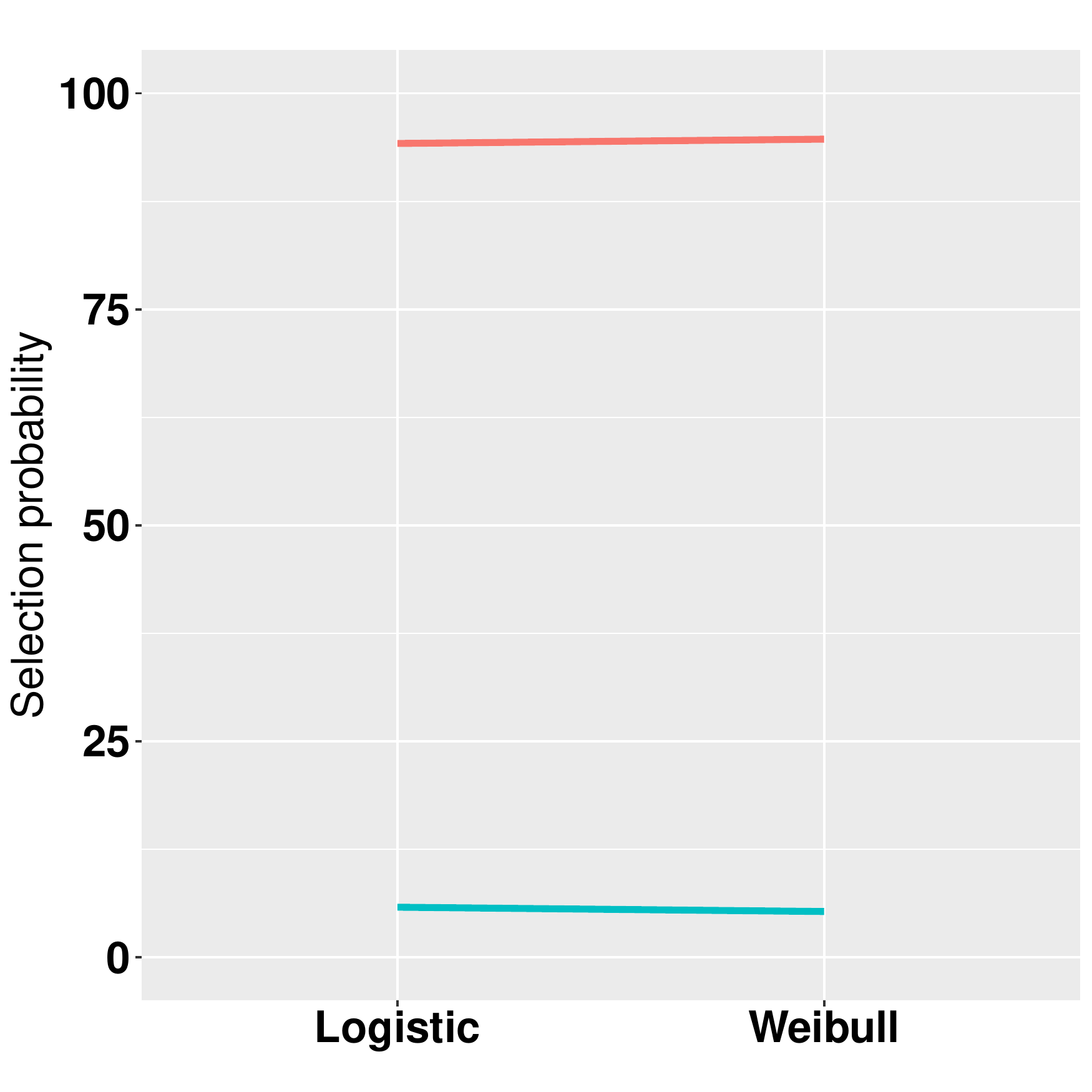}\\
			Utility function & \includegraphics[width=0.22\textwidth]{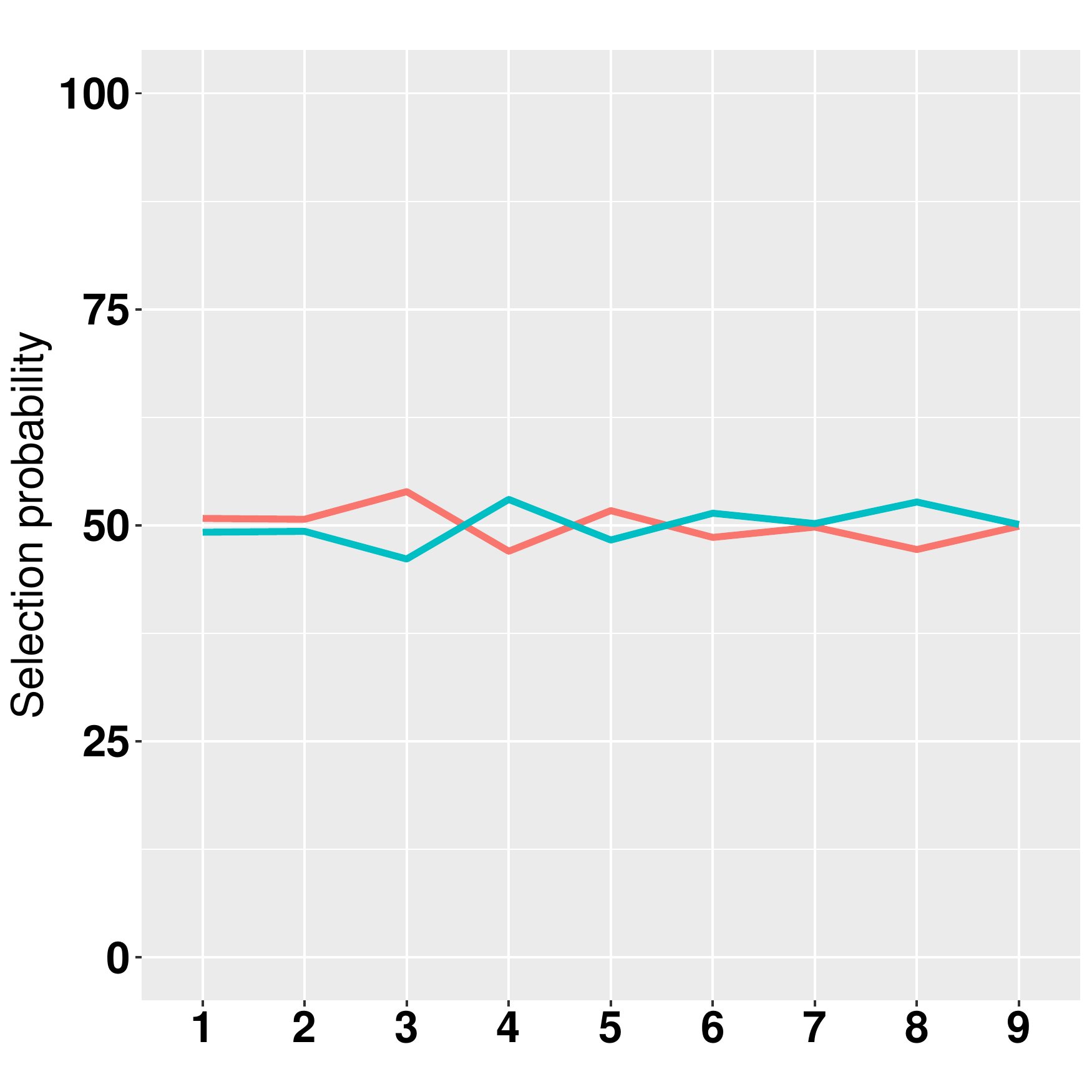}
			& \includegraphics[width=0.22\textwidth]{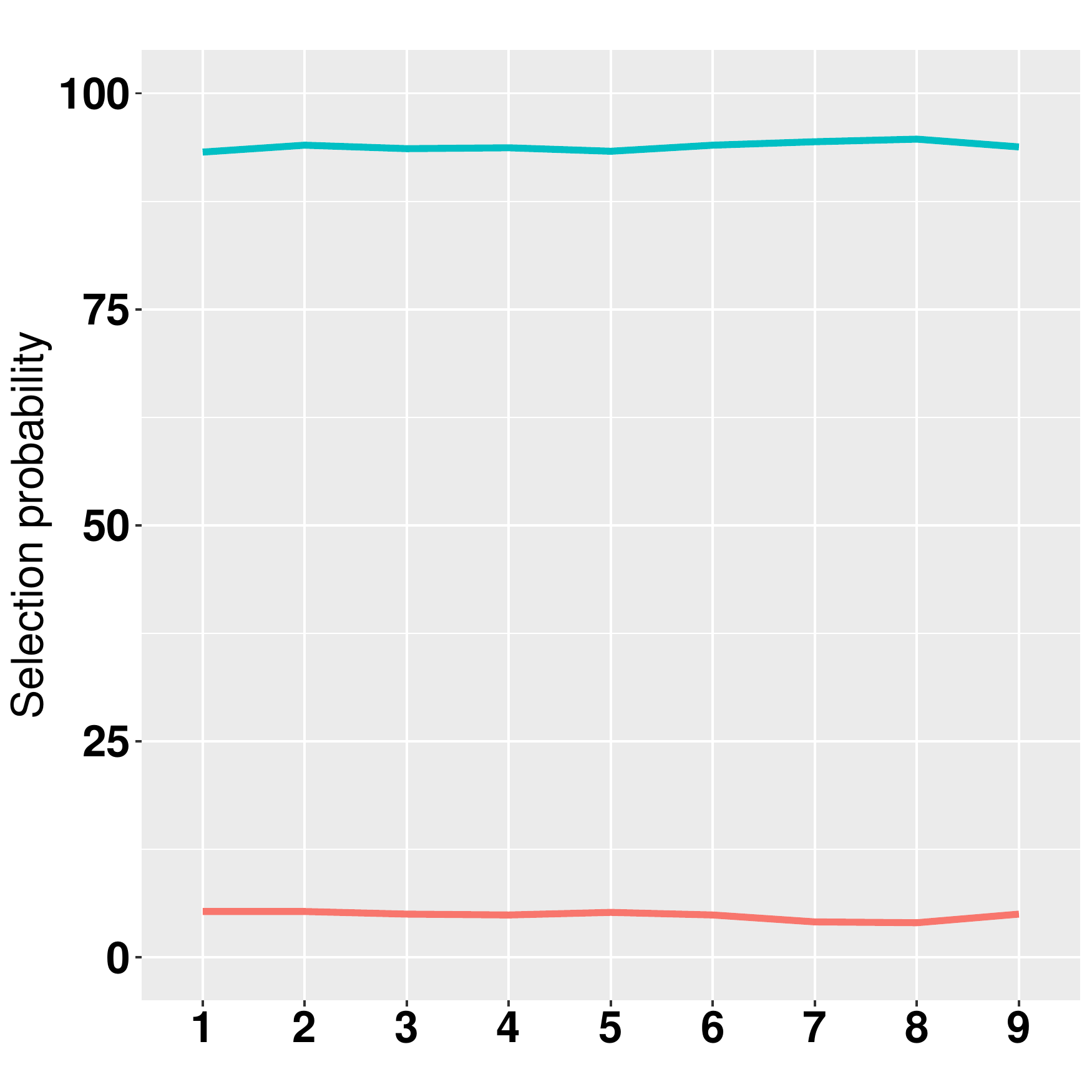}
			& \includegraphics[width=0.22\textwidth]{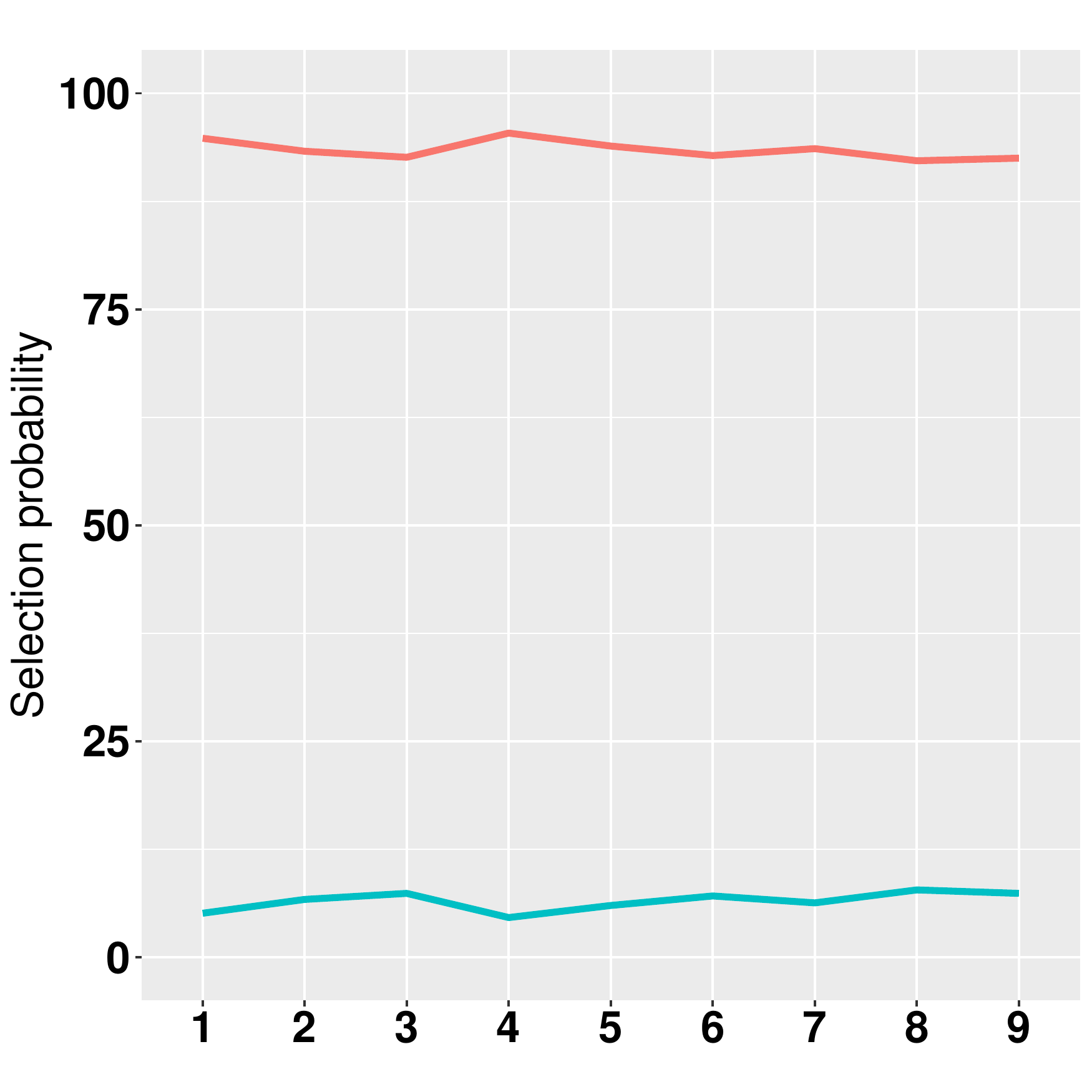}  
		\end{tabular}
		\caption{The sensitivity analysis of the RT dose selection probability for the RE patients. The blue and red lines represent high and standard RT doses, respectively.}
		\label{tab:sensitivity_result_resp}
	\end{figure}

	\begin{figure}[htbp]
		\centering
		\begin{tabular}{C{2.5cm}C{4cm}C{4cm}C{4cm}}
			& \multicolumn{3}{c}{Dose Selection Probability for SE patients} \\
			Sensitivity Analysis & Scenario 1  & Scenario 3 & Scenario 6  \\ \hline
			
			Number of total patients & \includegraphics[width=0.22\textwidth]{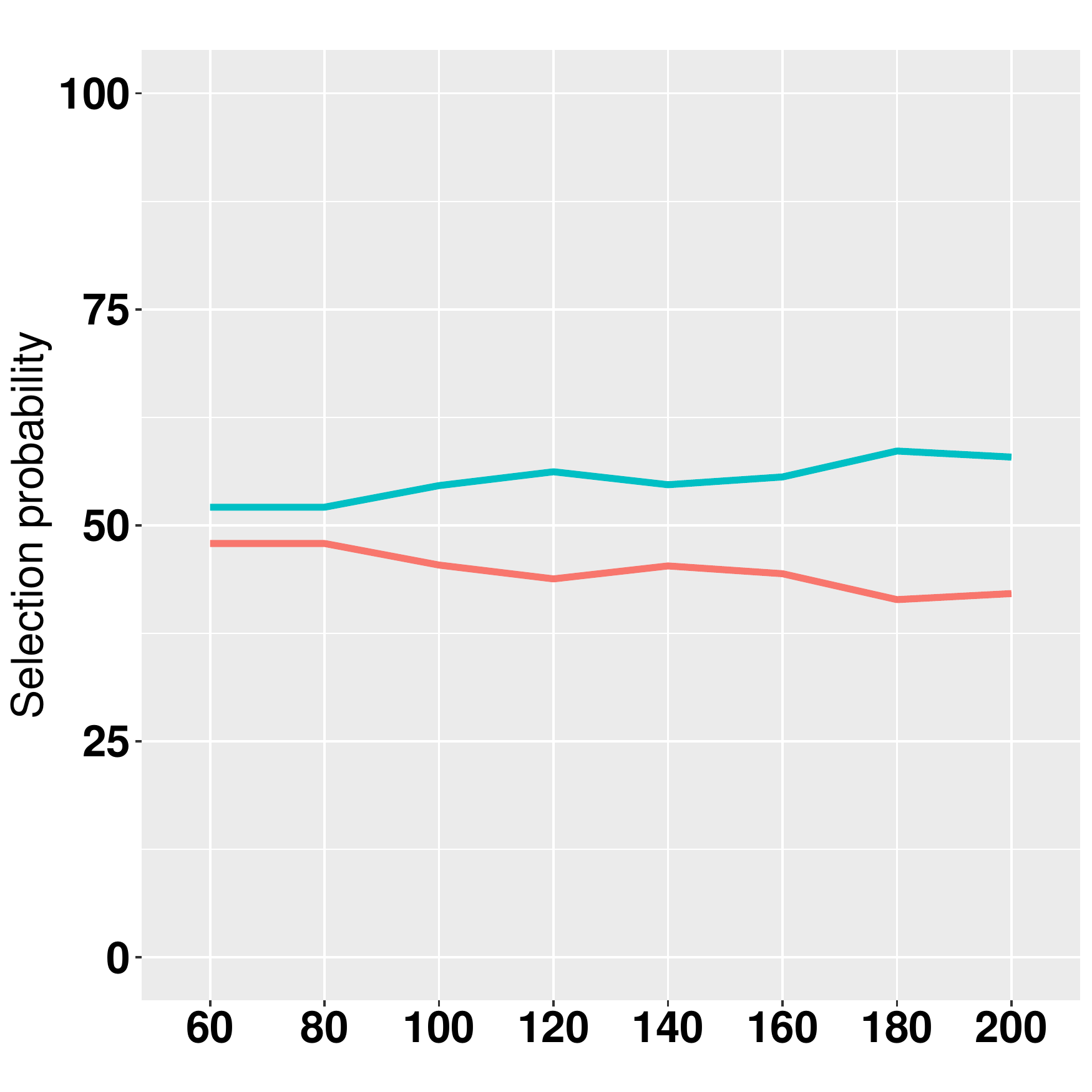}
			& \includegraphics[width=0.22\textwidth]{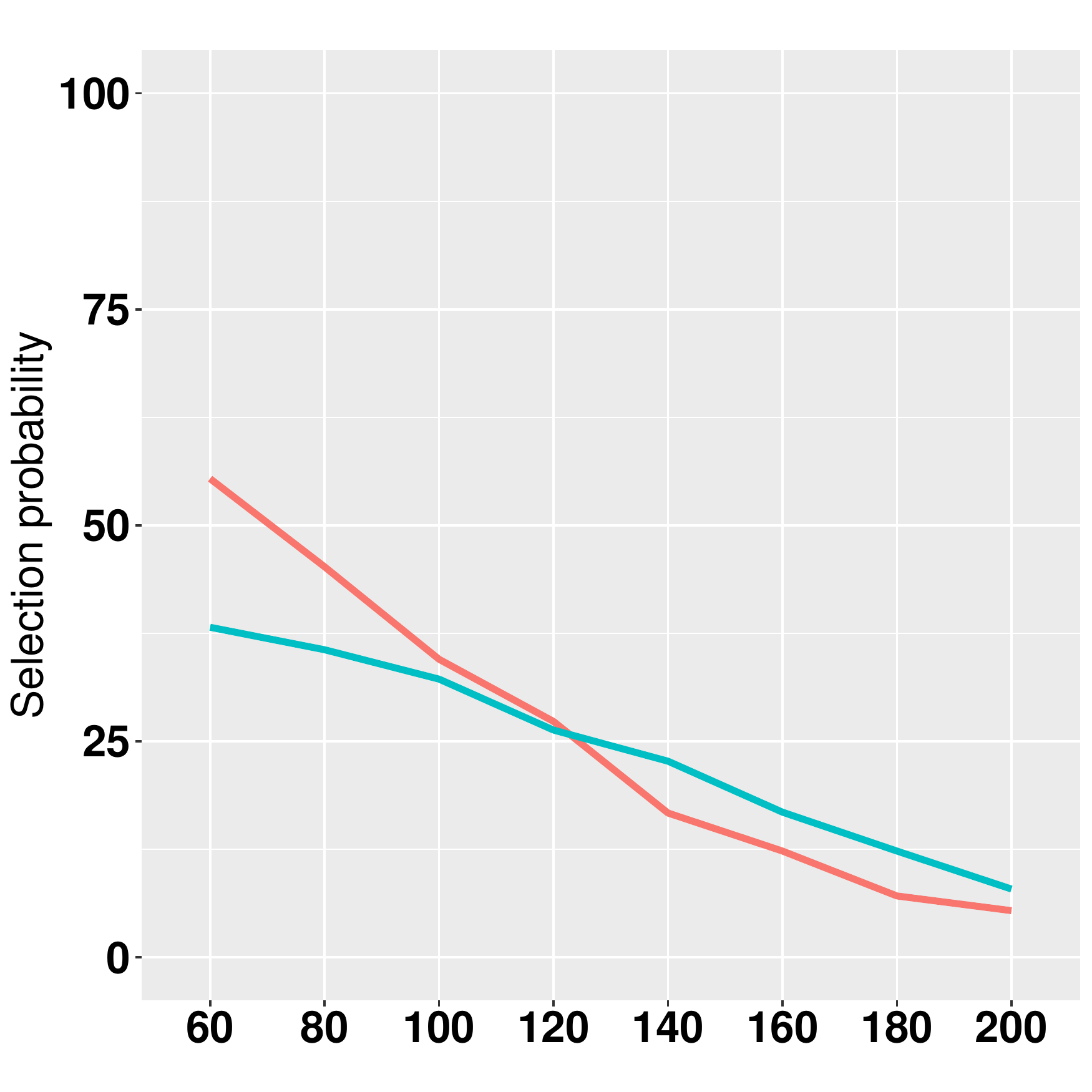}
			&\includegraphics[width=0.22\textwidth]{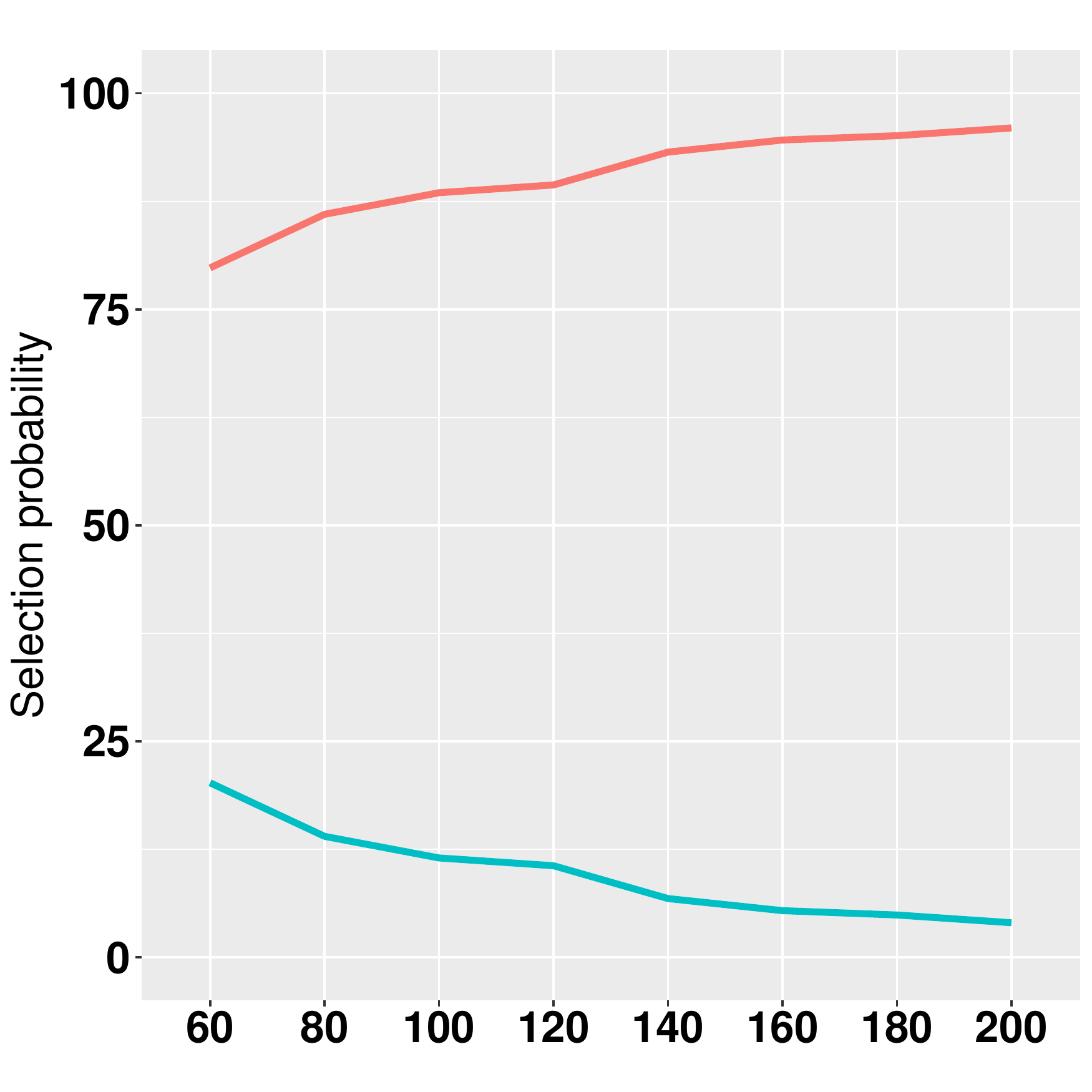}\\
			Proportion of the RE patients & \includegraphics[width=0.22\textwidth]{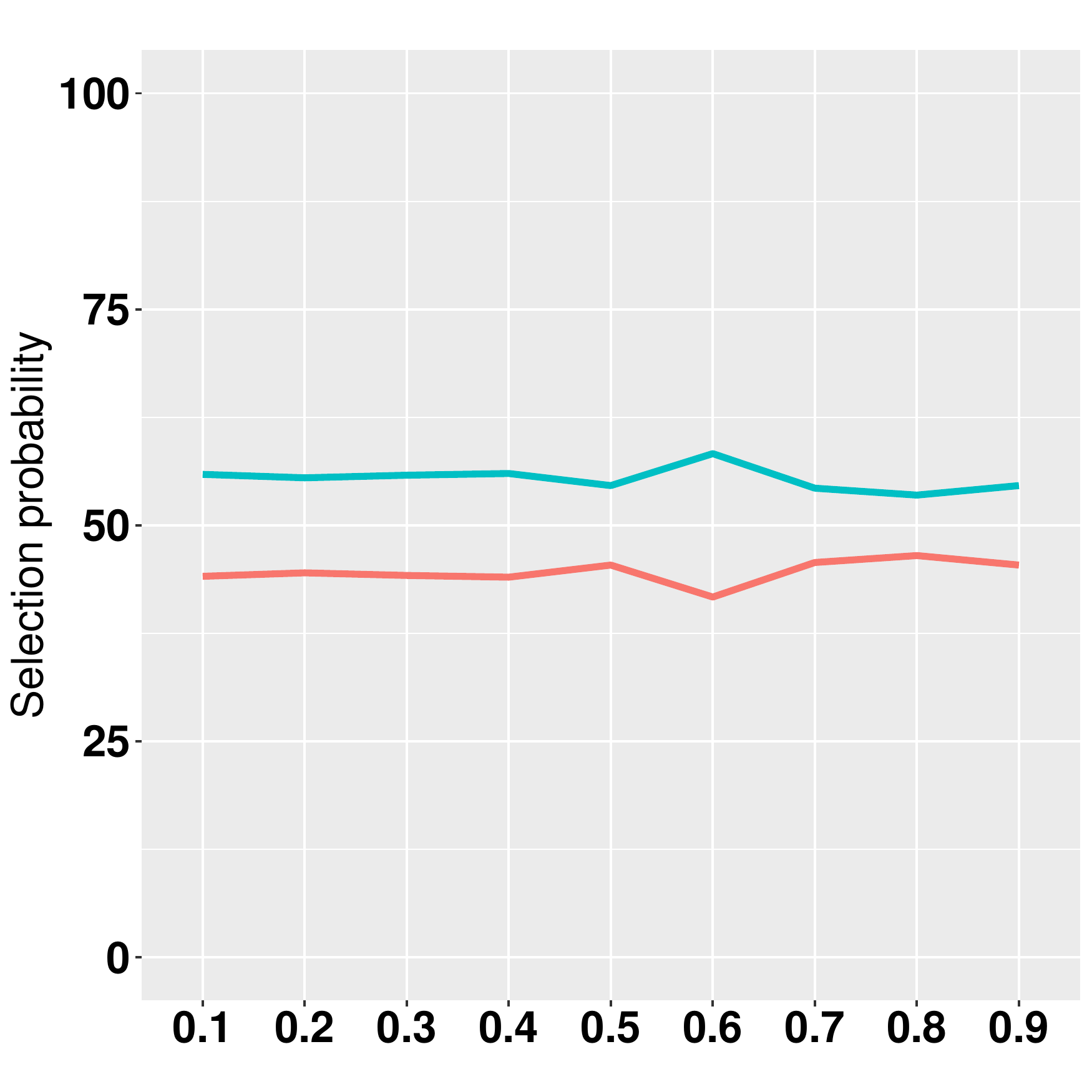}
			& \includegraphics[width=0.22\textwidth]{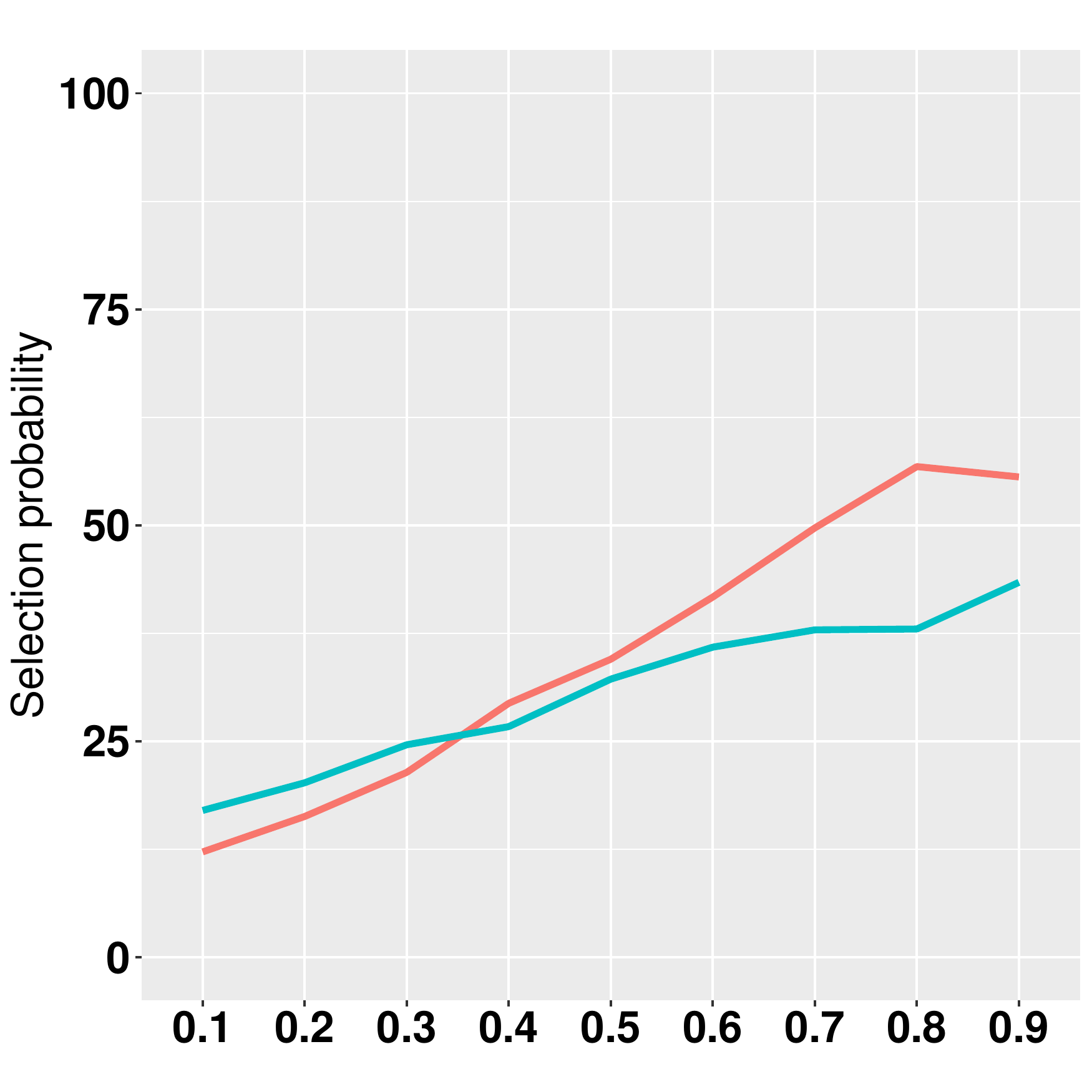}
			& \includegraphics[width=0.22\textwidth]{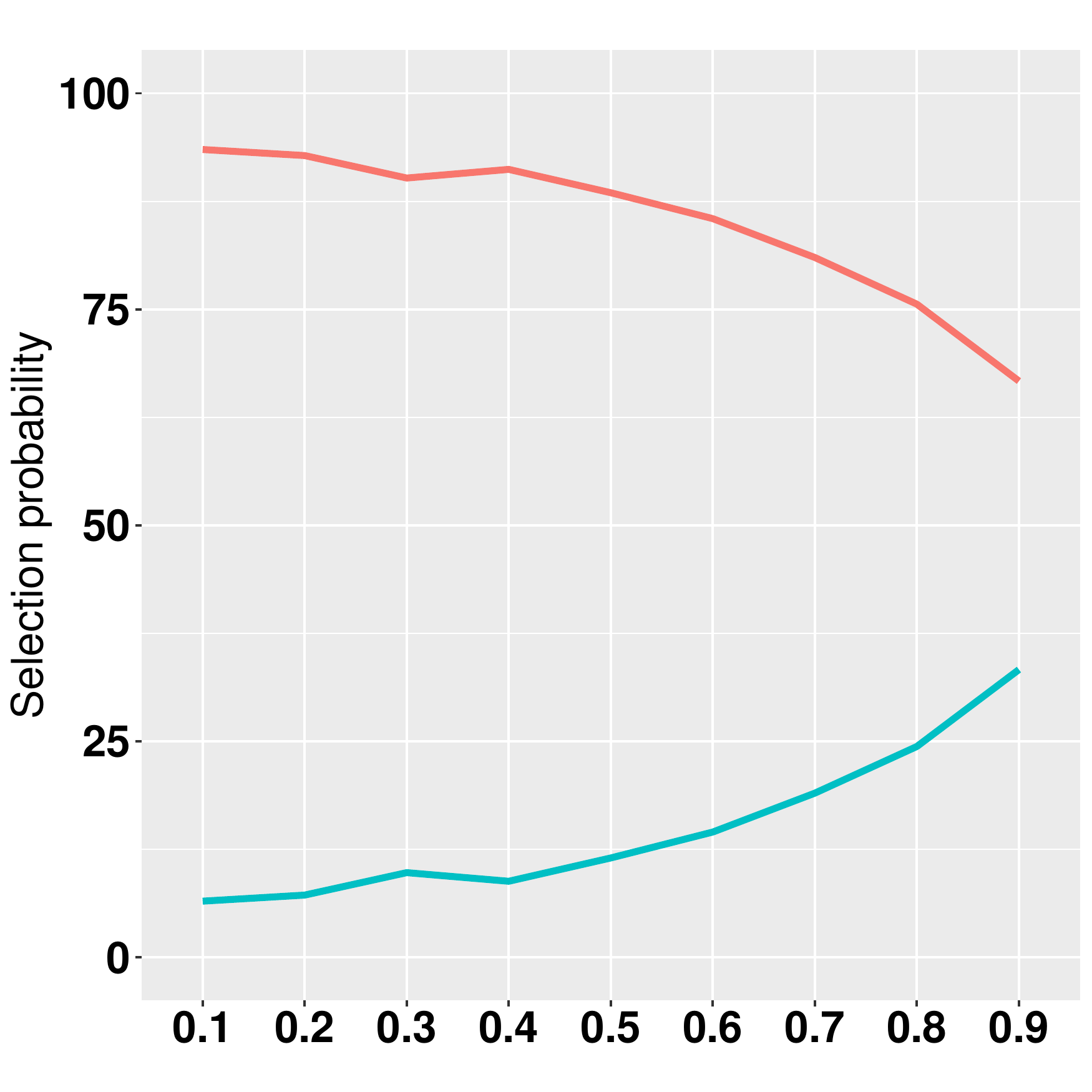}\\
			
			Distribution for the event time & \includegraphics[width=0.22\textwidth]{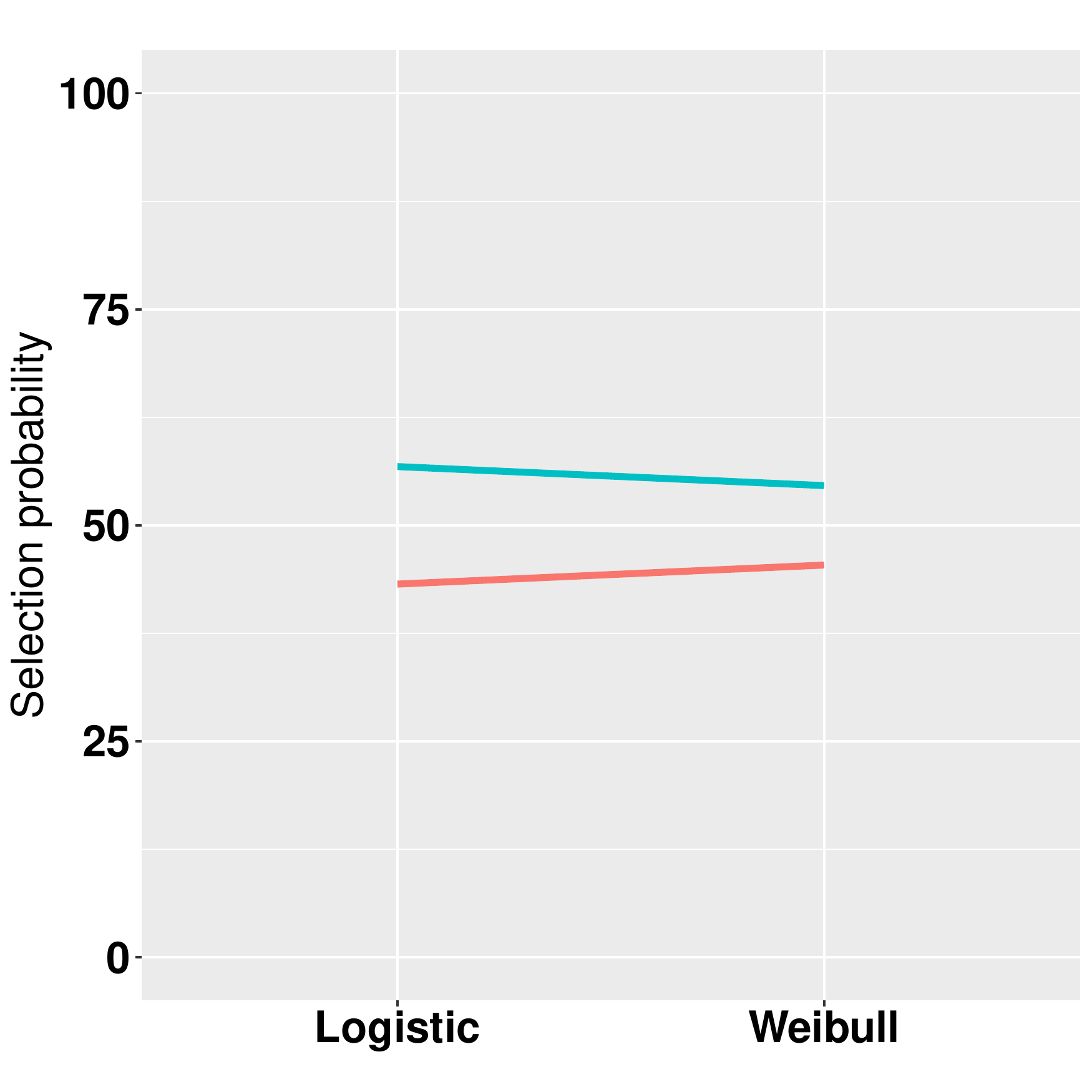}
			& \includegraphics[width=0.22\textwidth]{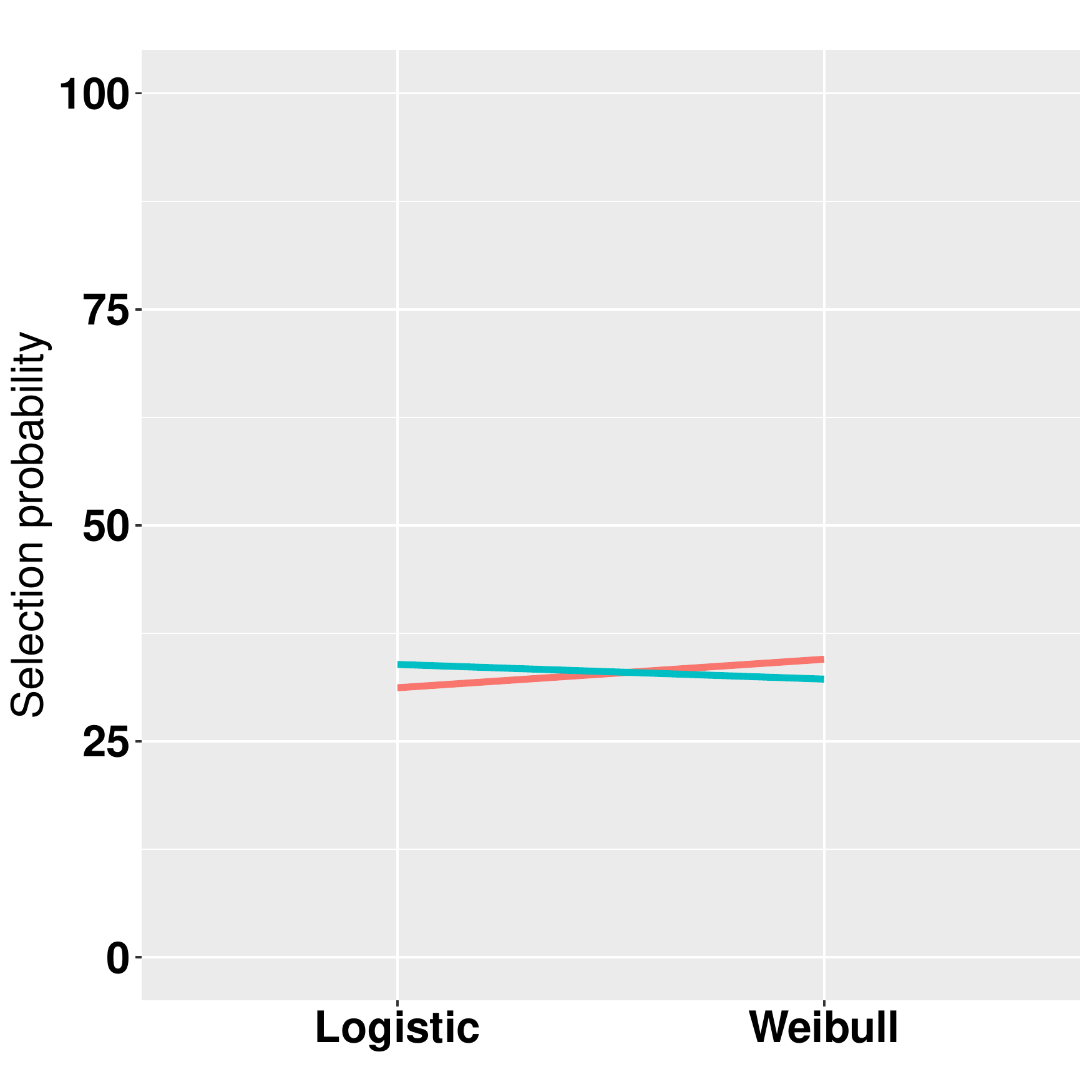}
			& \includegraphics[width=0.22\textwidth]{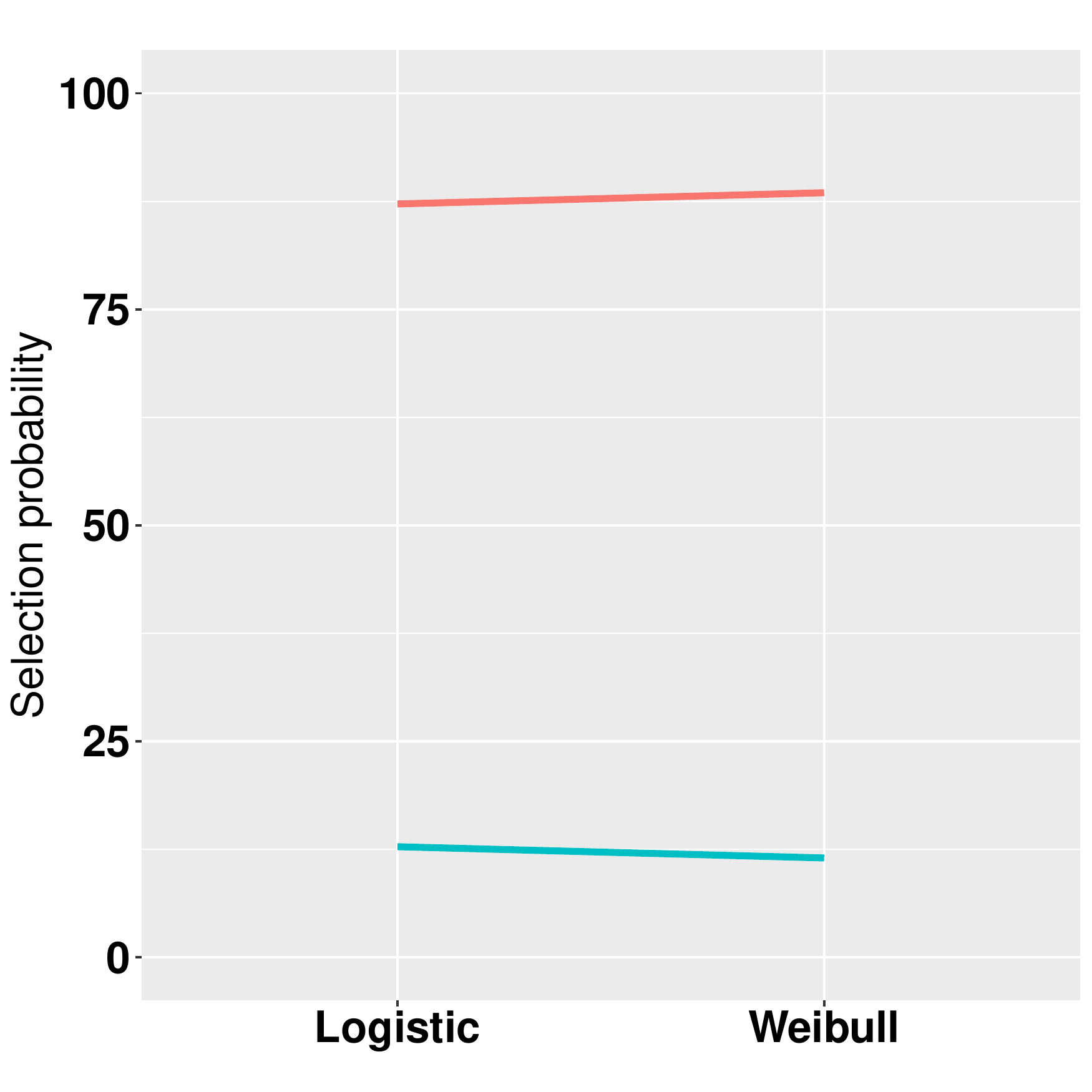}\\
			Utility function & \includegraphics[width=0.22\textwidth]{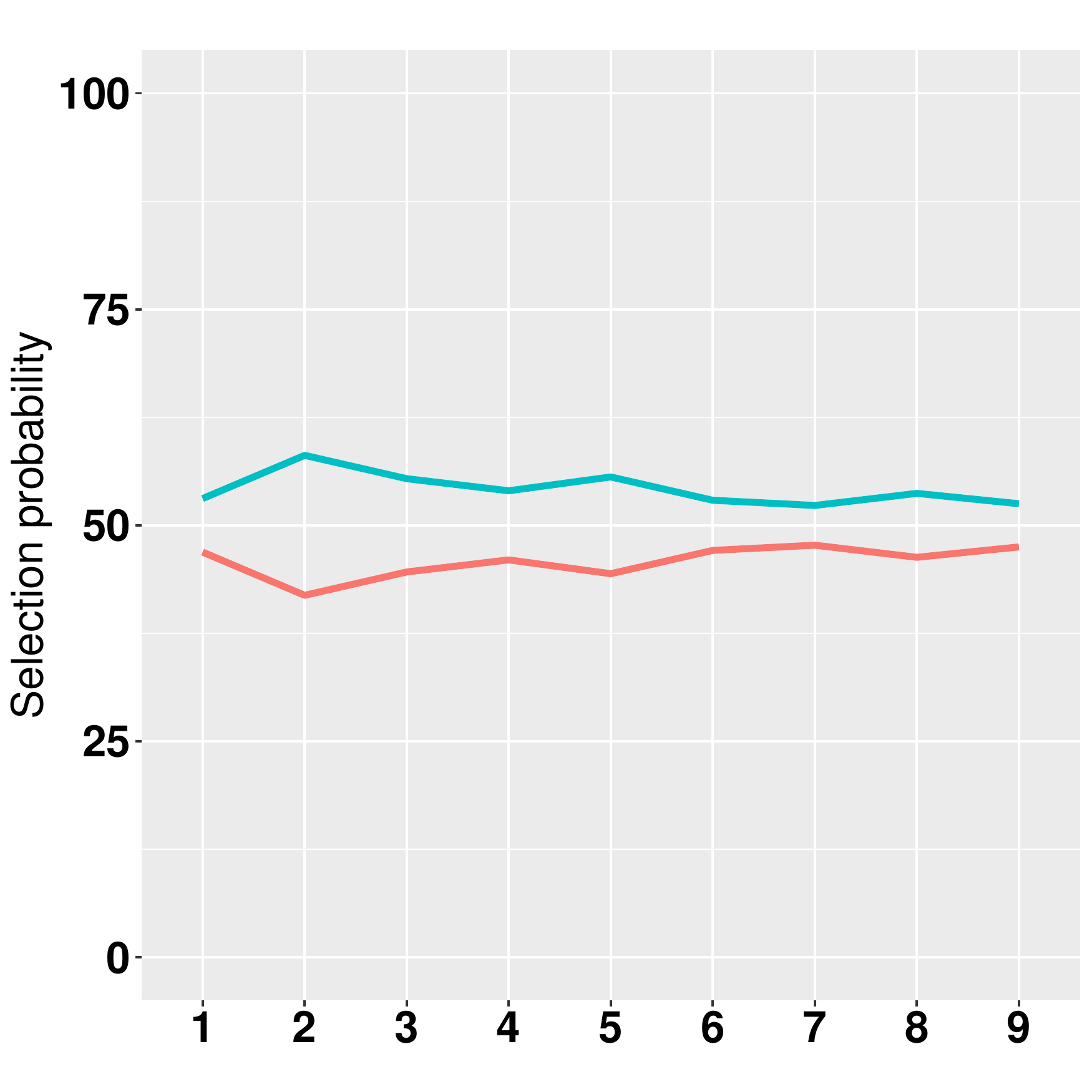}
			& \includegraphics[width=0.22\textwidth]{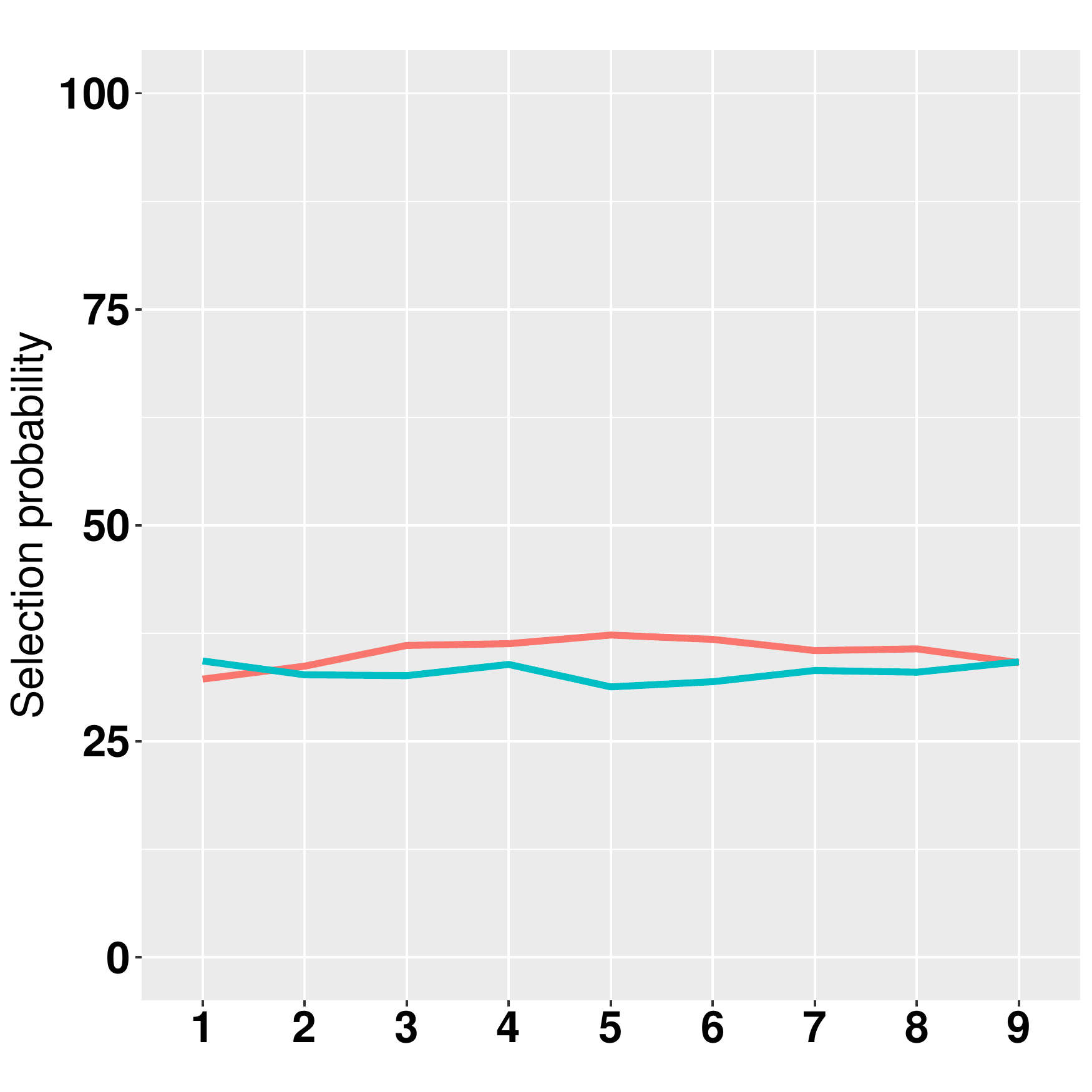}
			& \includegraphics[width=0.22\textwidth]{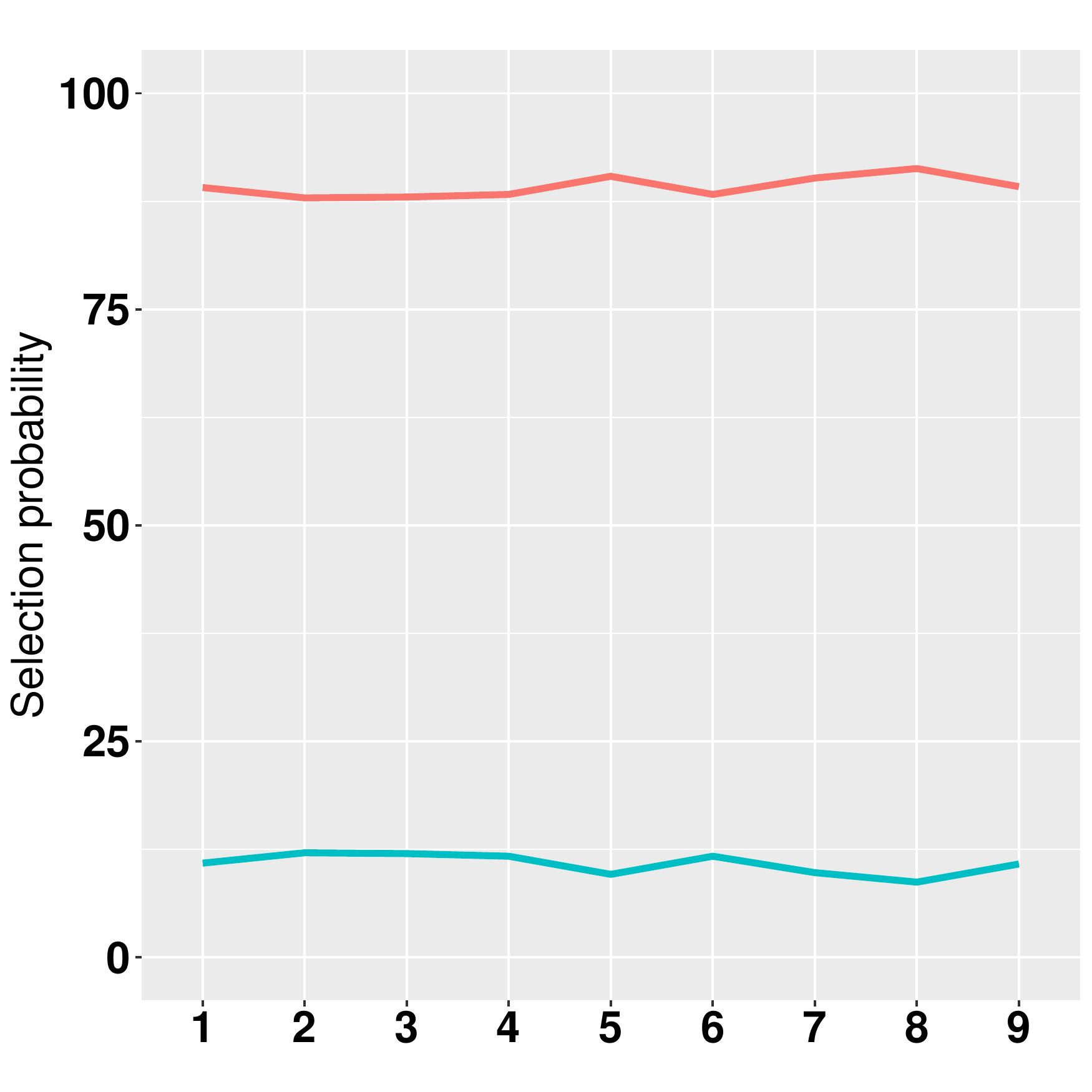}  
		\end{tabular}
		\caption{The sensitivity analysis of the RT dose selection probability for the SE patients. The blue and red lines represent low and standard RT doses, respectively.}
		\label{tab:sensitivity_result_sesp}
	\end{figure}

	\begin{figure}[htbp]
		\centering
		\begin{tabular}{C{2.5cm}C{4cm}C{4cm}C{4cm}}
			& \multicolumn{3}{c}{Proportion of RE patients treated} \\
			Sensitivity Analysis & Scenario 1  & Scenario 3 & Scenario 6  \\ \hline
			
			Number of total patients & \includegraphics[width=0.22\textwidth]{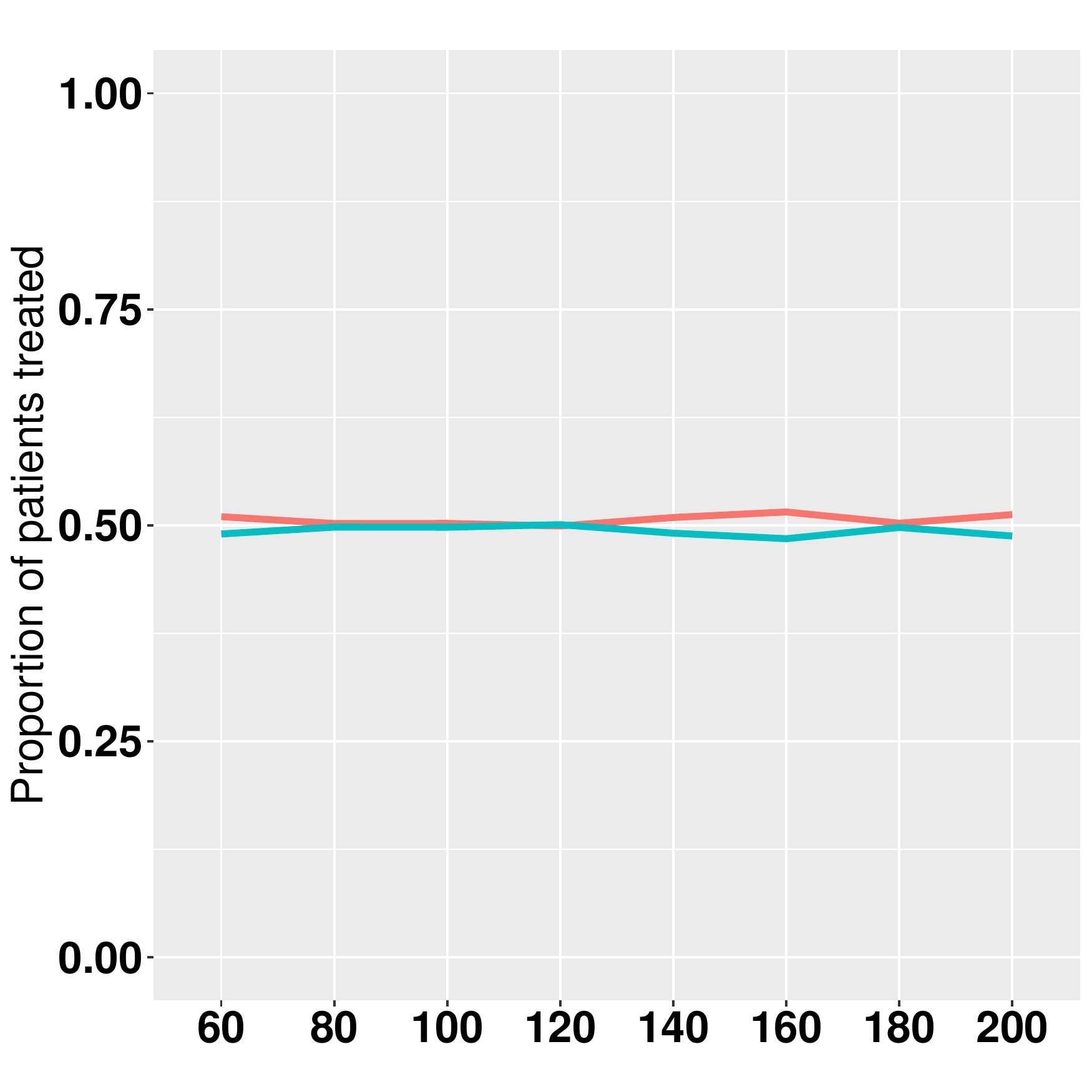}
			& \includegraphics[width=0.22\textwidth]{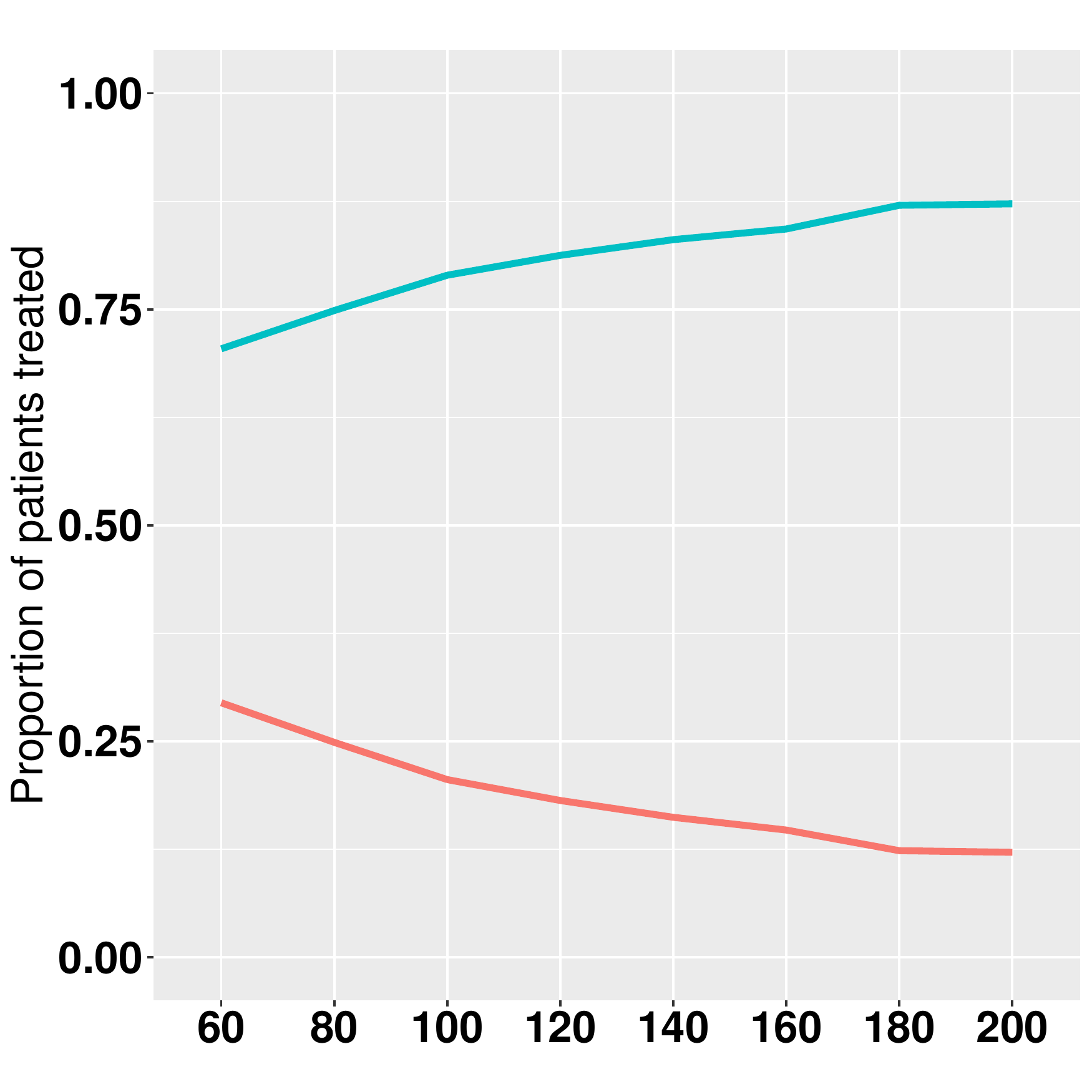}
			&\includegraphics[width=0.22\textwidth]{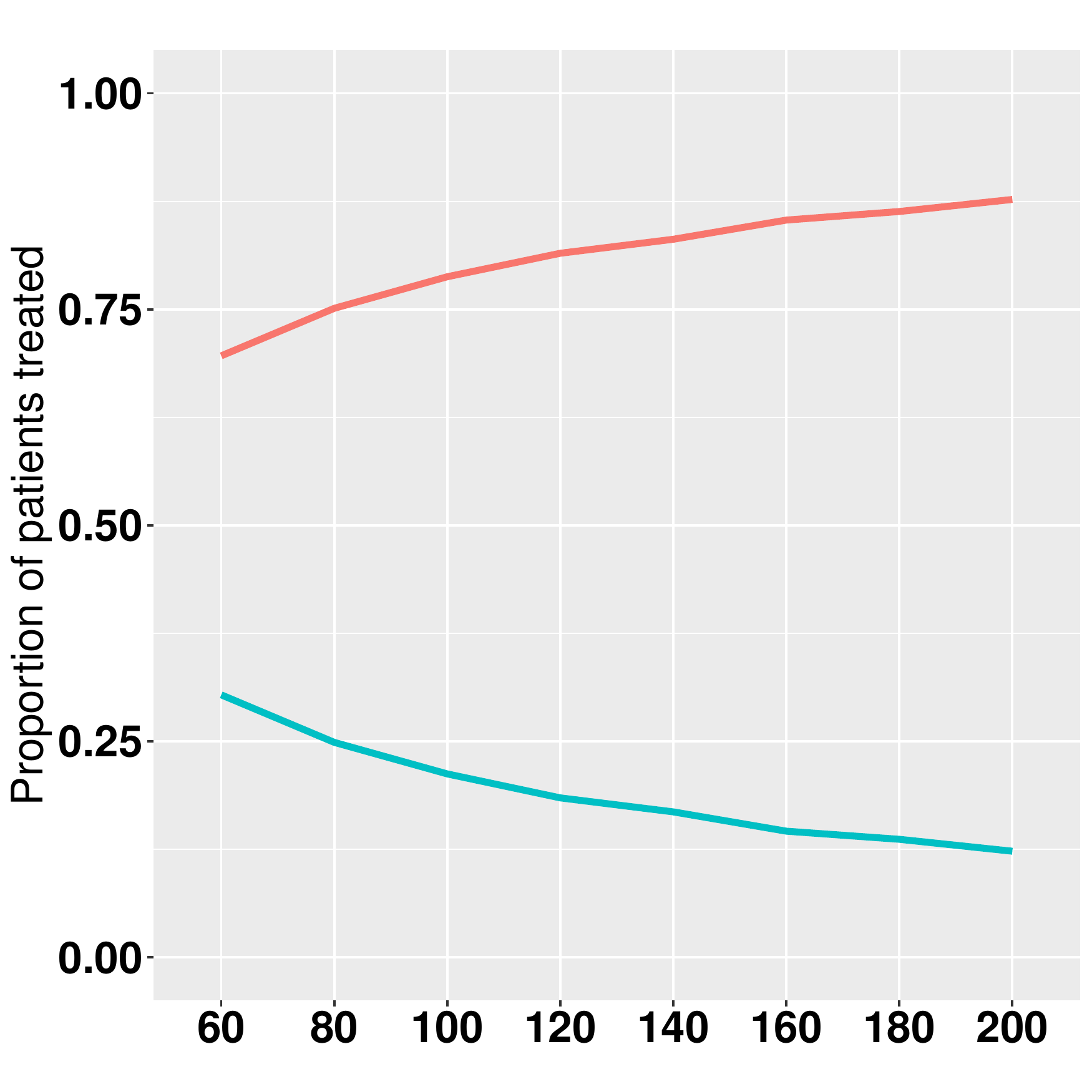}\\
			Proportion of the RE patients & \includegraphics[width=0.22\textwidth]{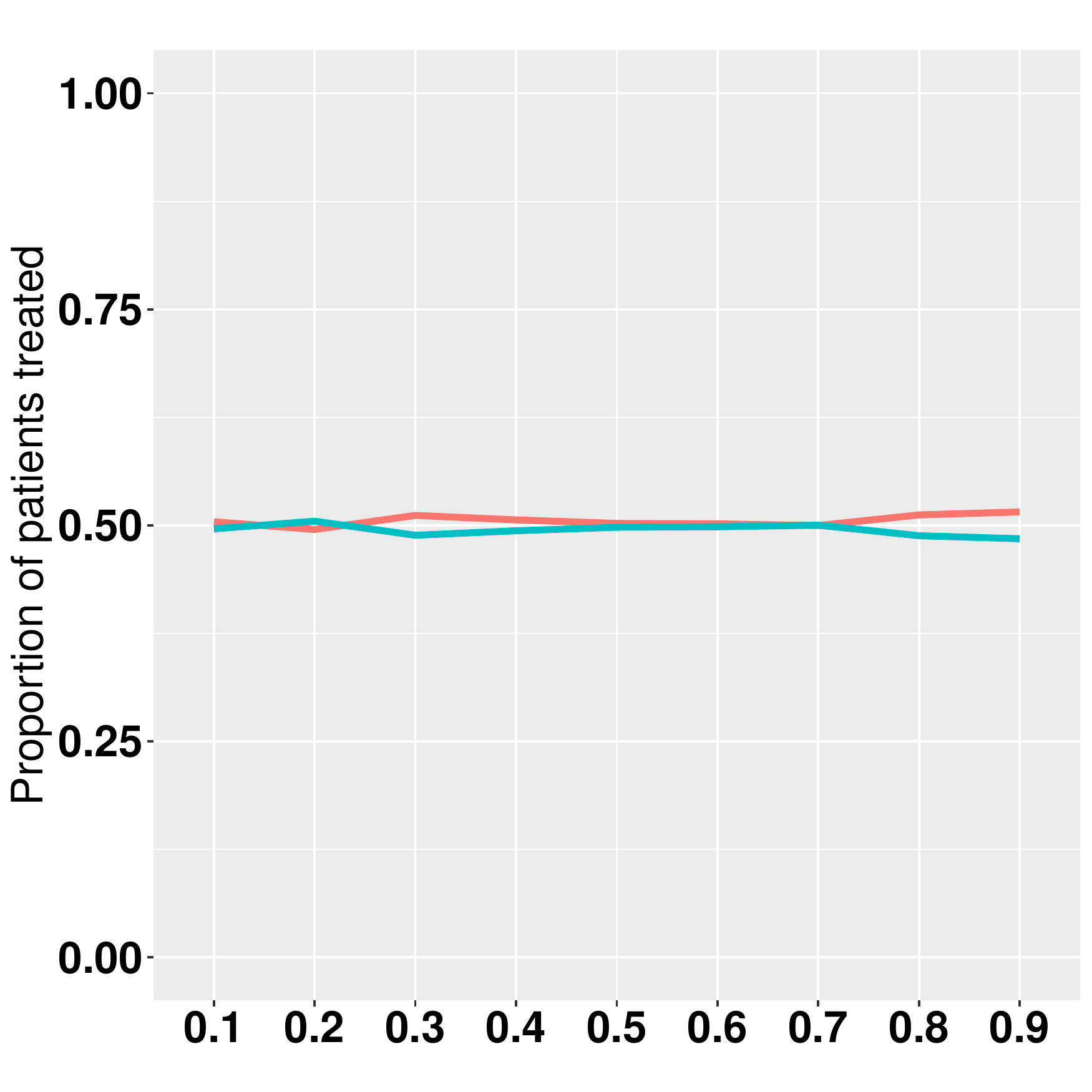}
			& \includegraphics[width=0.22\textwidth]{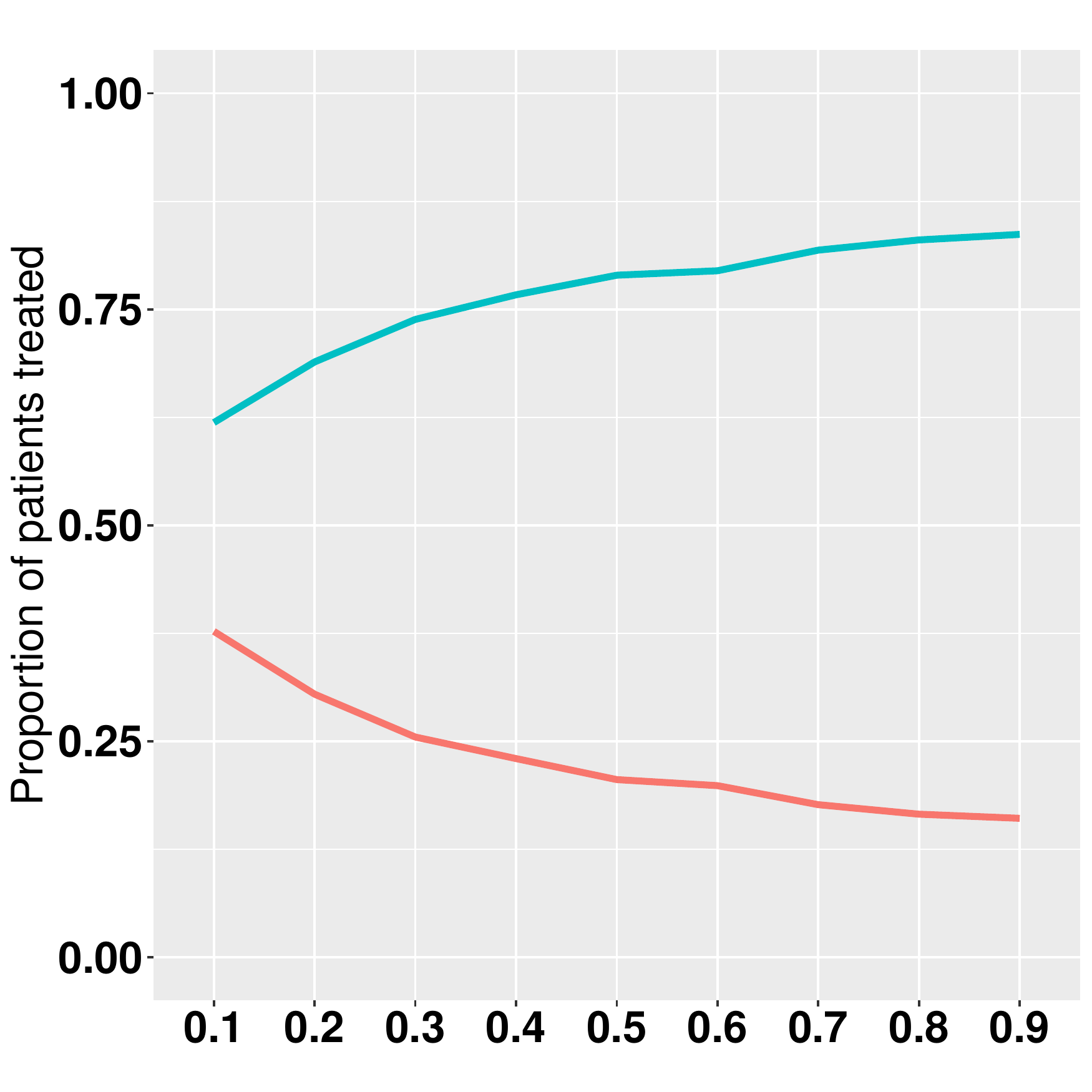}
			& \includegraphics[width=0.22\textwidth]{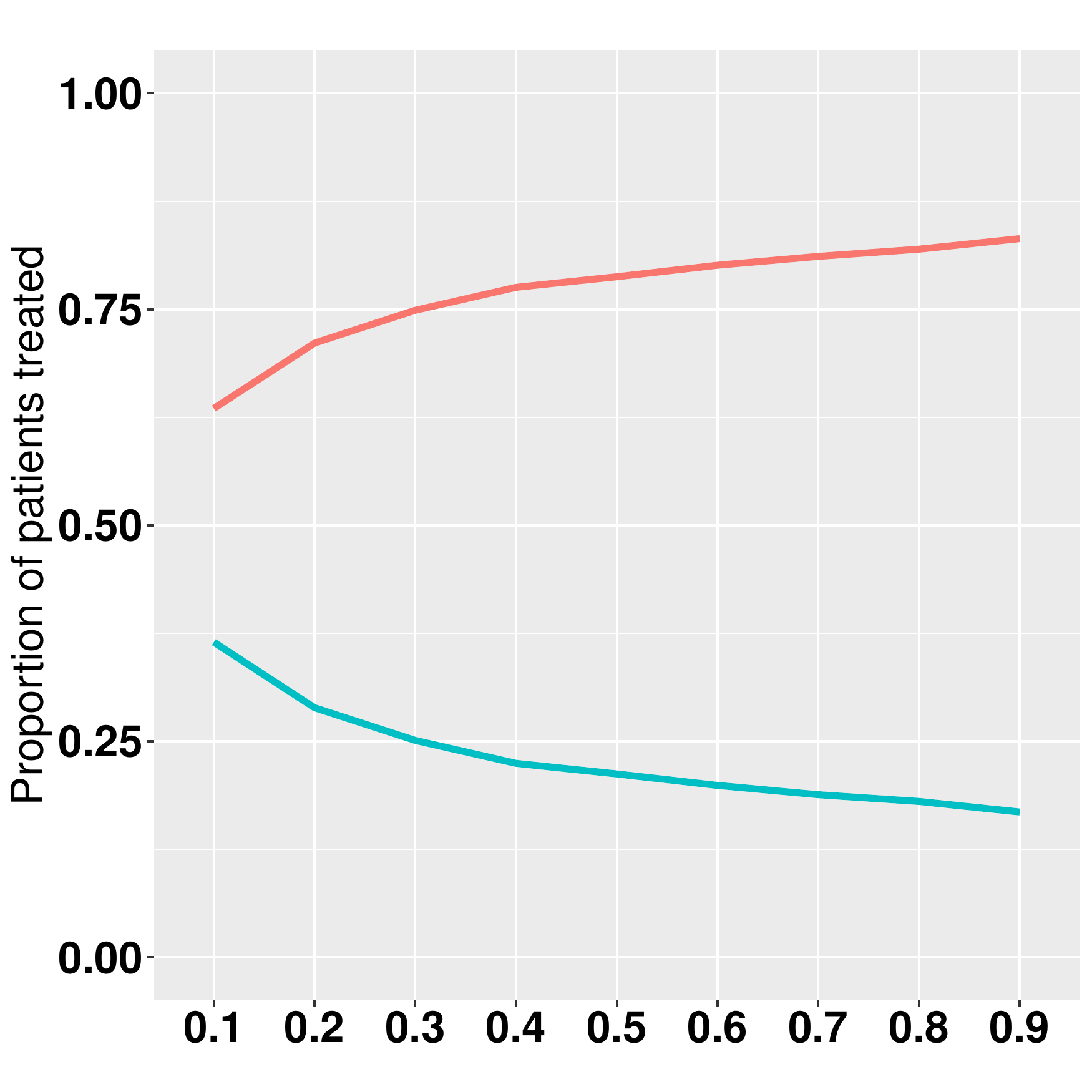}\\
			
			Distribution for the event time & \includegraphics[width=0.22\textwidth]{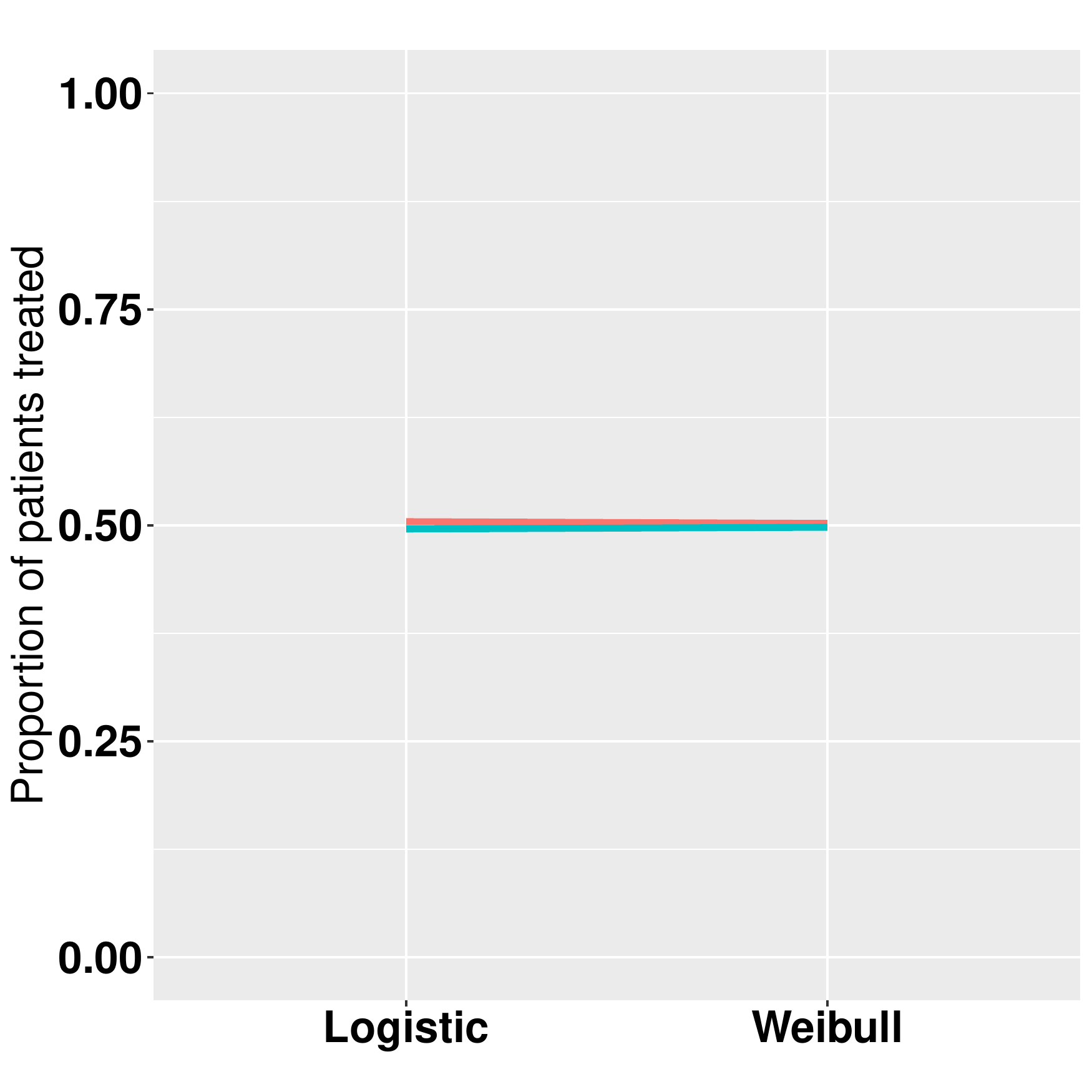}
			& \includegraphics[width=0.22\textwidth]{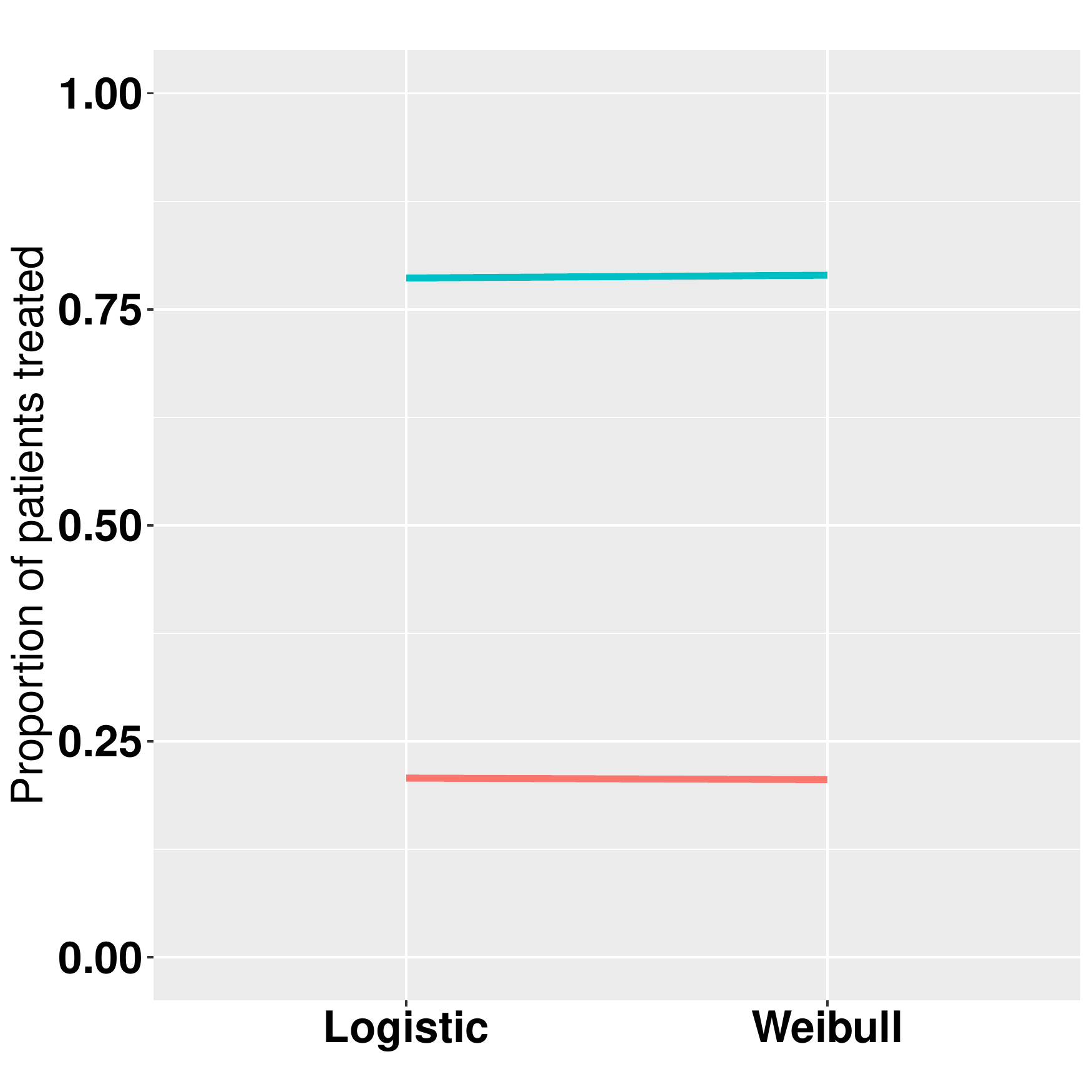}
			& \includegraphics[width=0.22\textwidth]{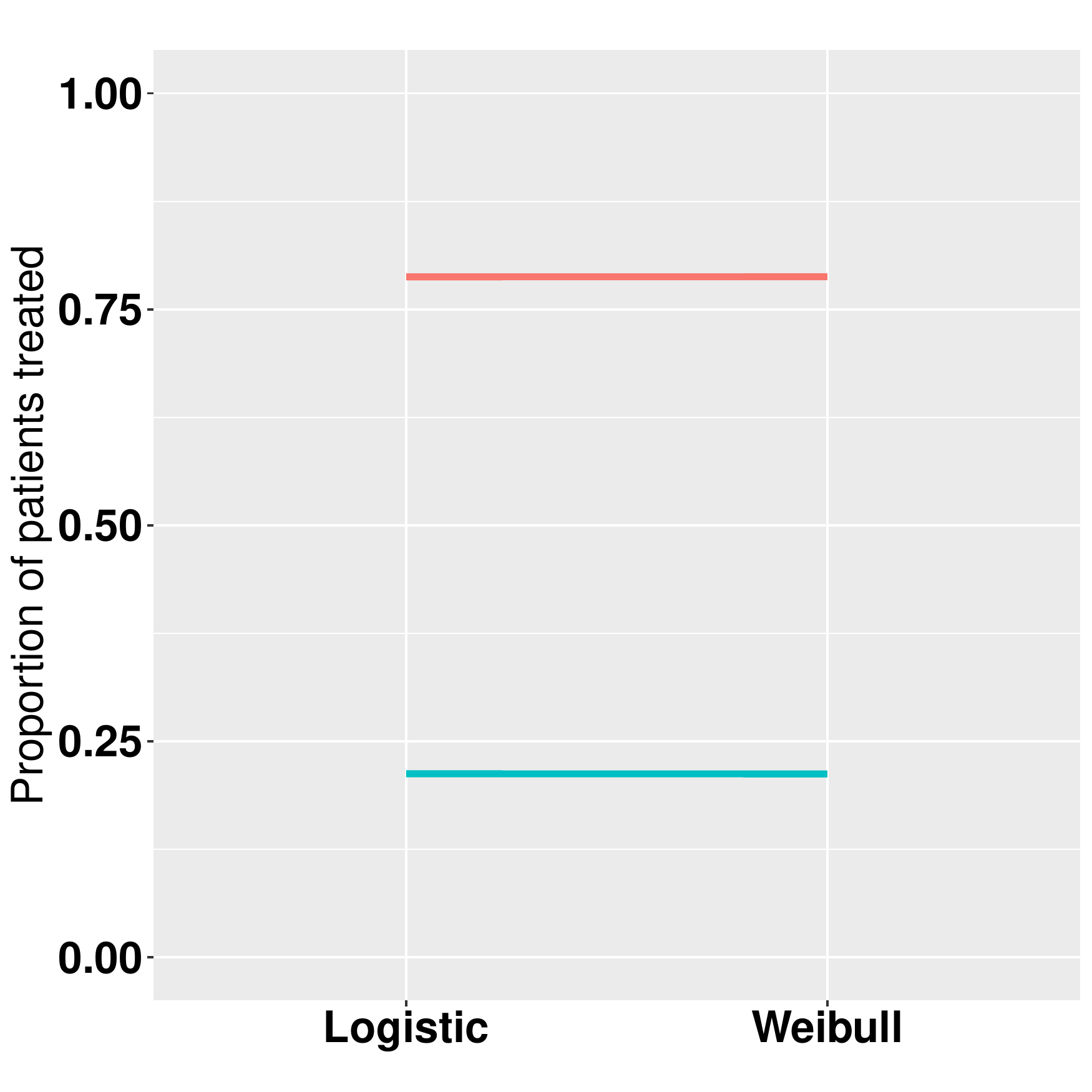}\\
			Utility function & \includegraphics[width=0.22\textwidth]{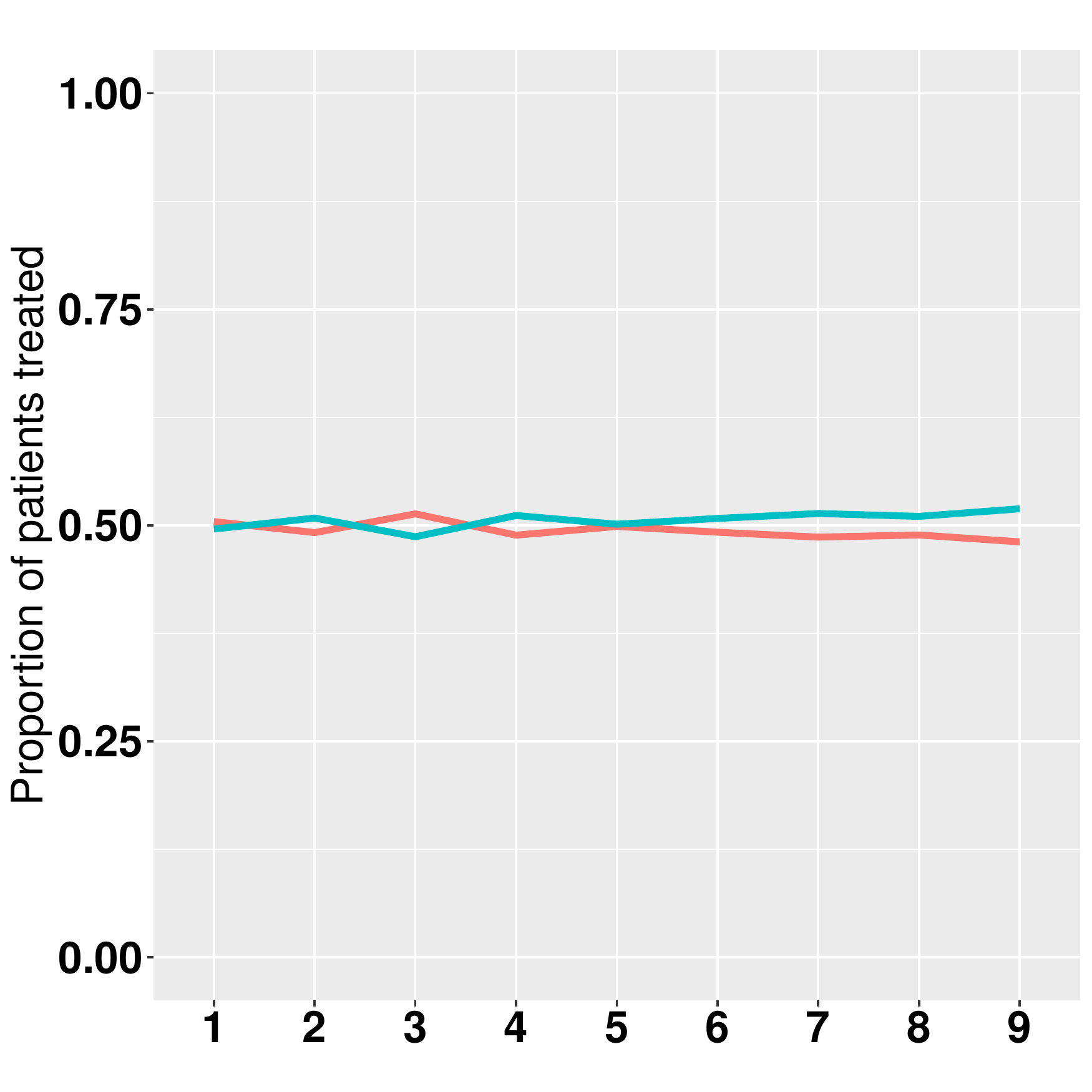}
			& \includegraphics[width=0.22\textwidth]{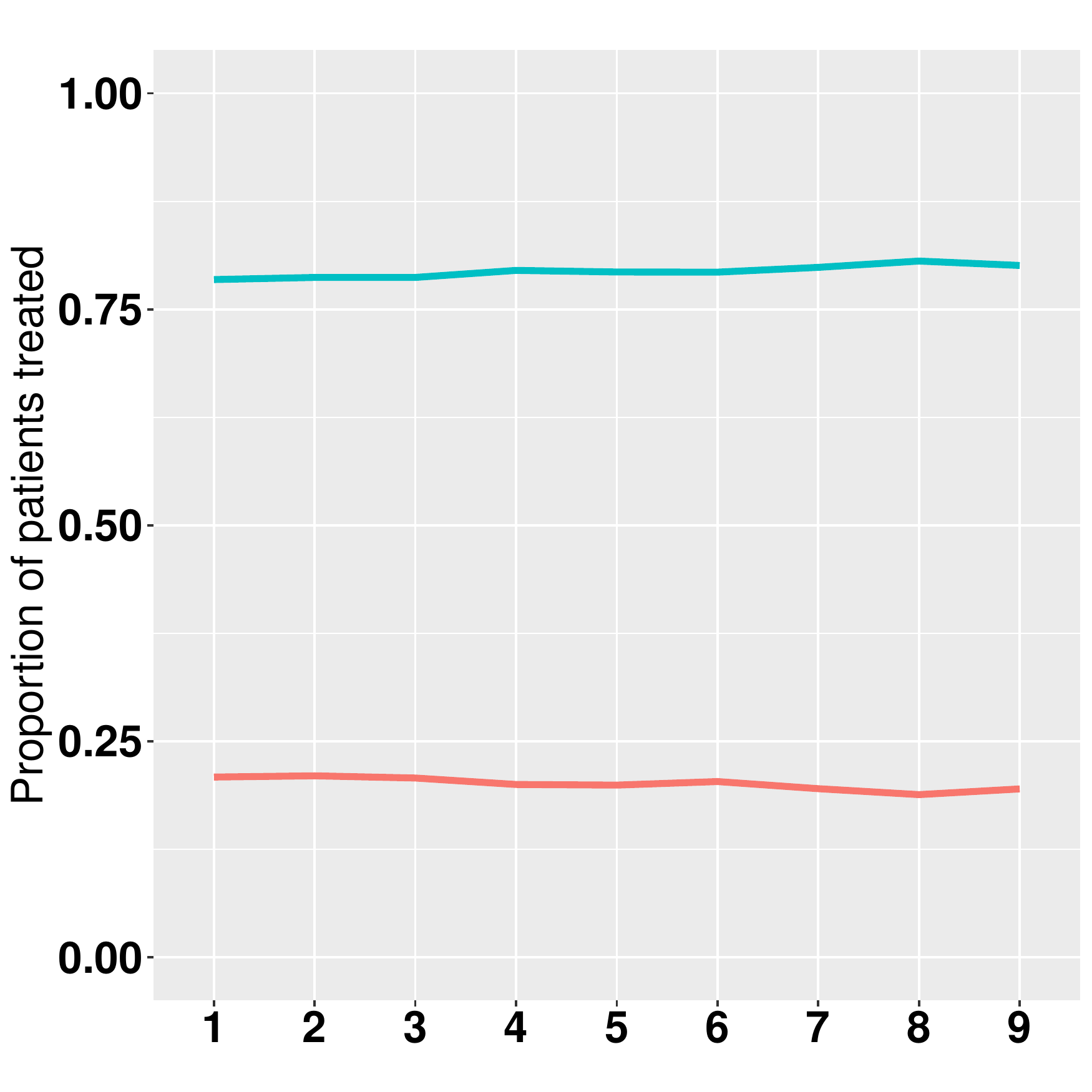}
			& \includegraphics[width=0.22\textwidth]{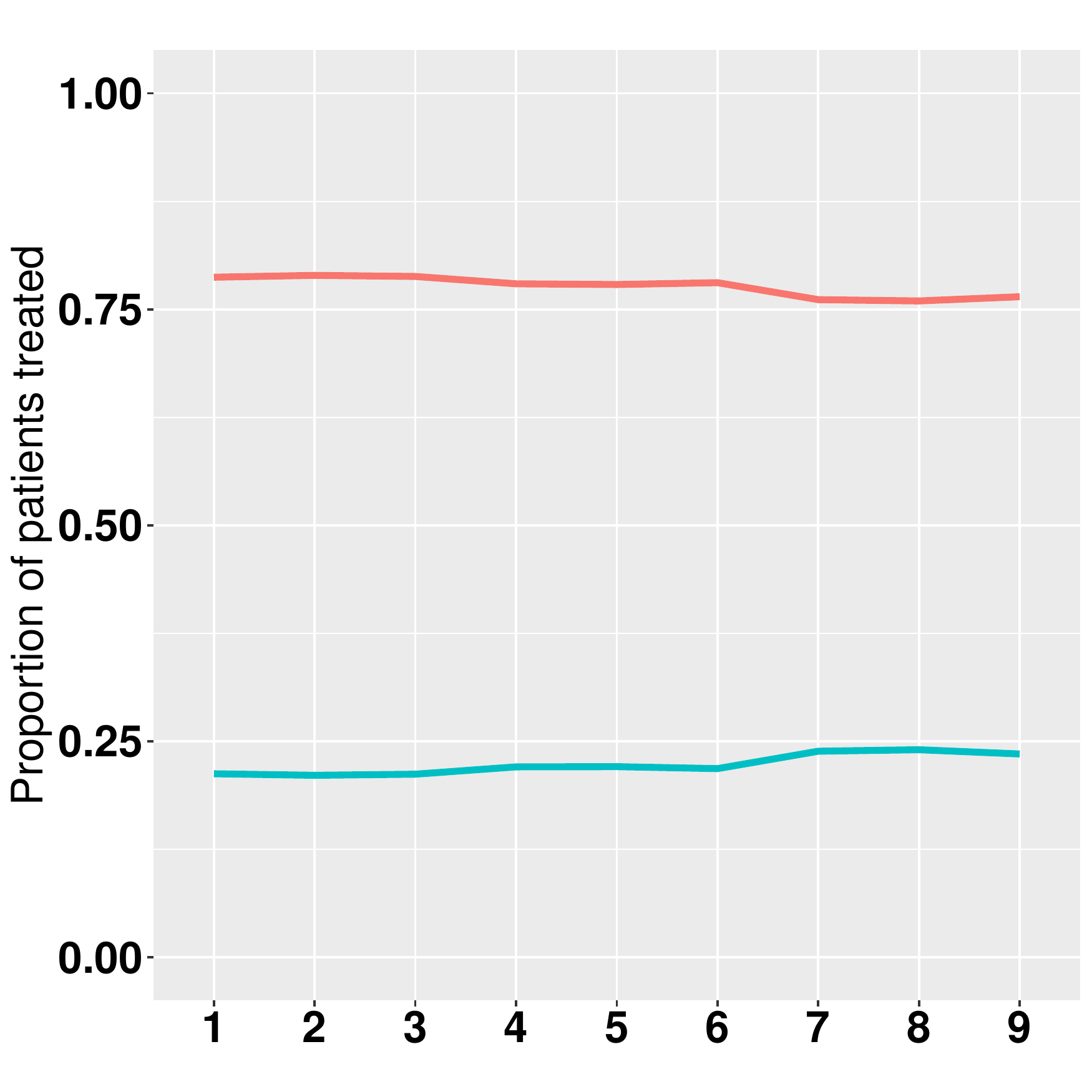}  
		\end{tabular}
		\caption{The sensitivity analysis of the proportion of the RE patients treated with each RT dose. The blue and red lines represent high and standard RT doses, respectively.}
		\label{tab:sensitivity_result_retp}
	\end{figure}

	\begin{figure}[htbp]
		\centering
		\begin{tabular}{C{2.5cm}C{4cm}C{4cm}C{4cm}}
			& \multicolumn{3}{c}{Proportion of SE patients treated} \\
			Sensitivity Analysis & Scenario 1  & Scenario 3 & Scenario 6  \\ \hline
			
			Number of total patients & \includegraphics[width=0.22\textwidth]{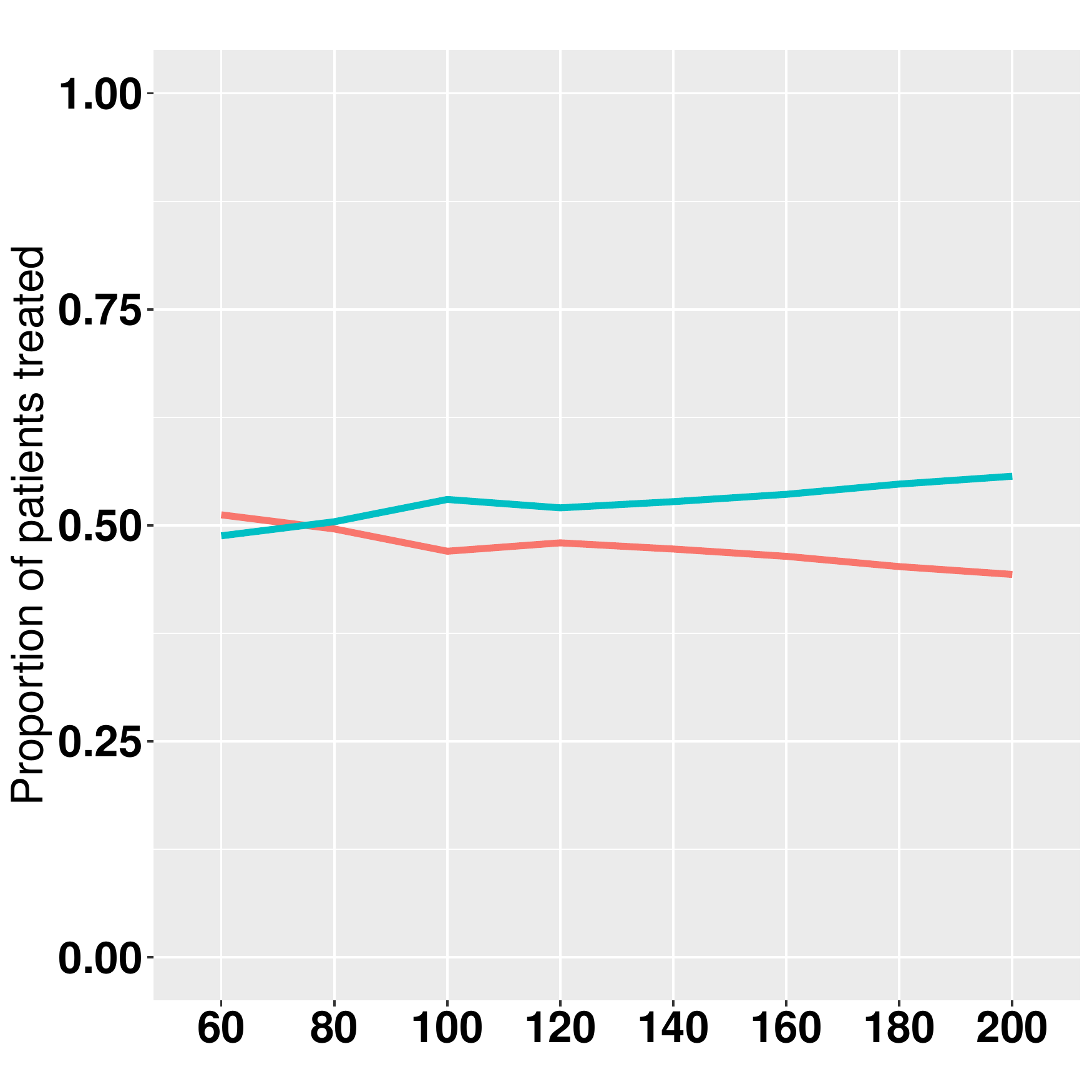}
			& \includegraphics[width=0.22\textwidth]{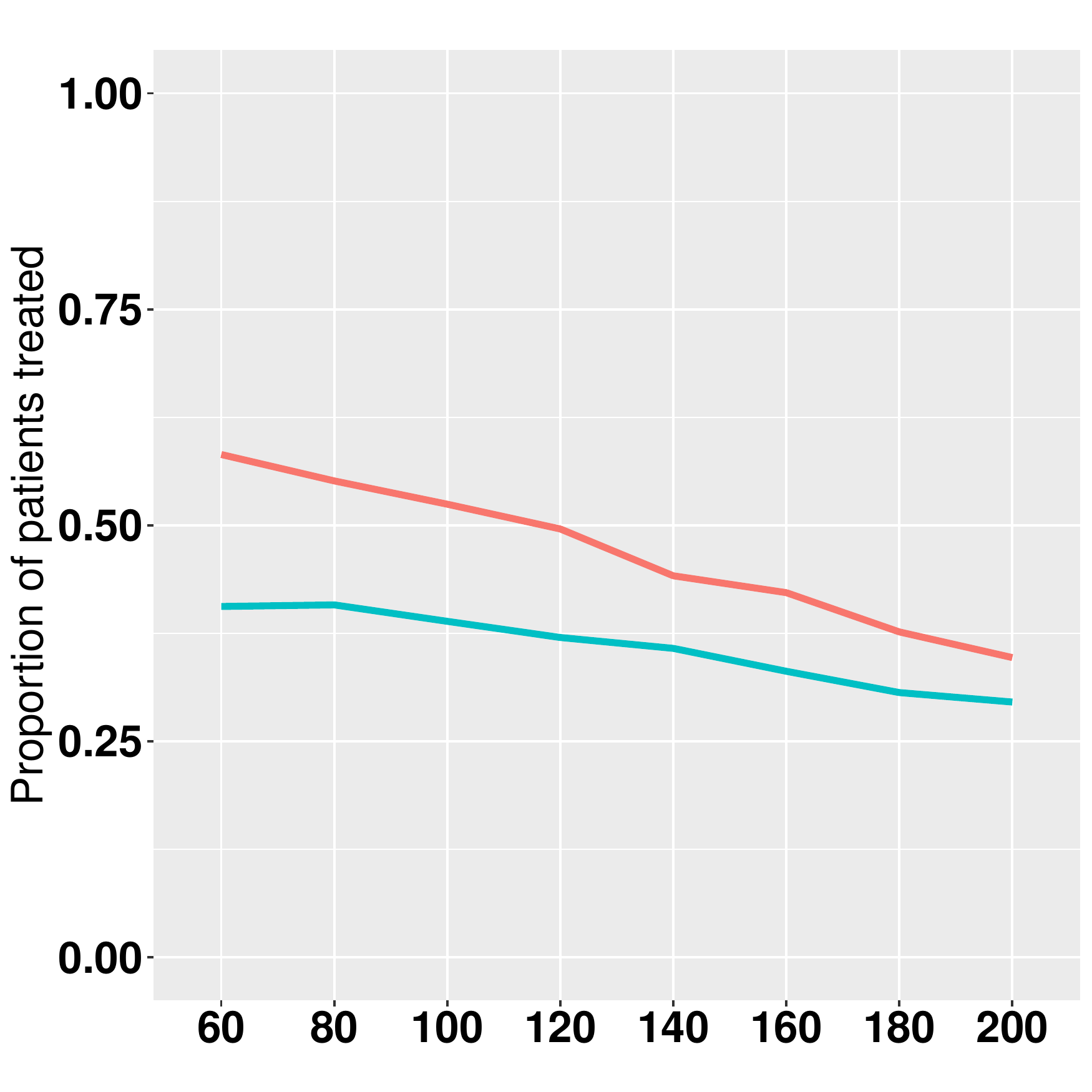}
			&\includegraphics[width=0.22\textwidth]{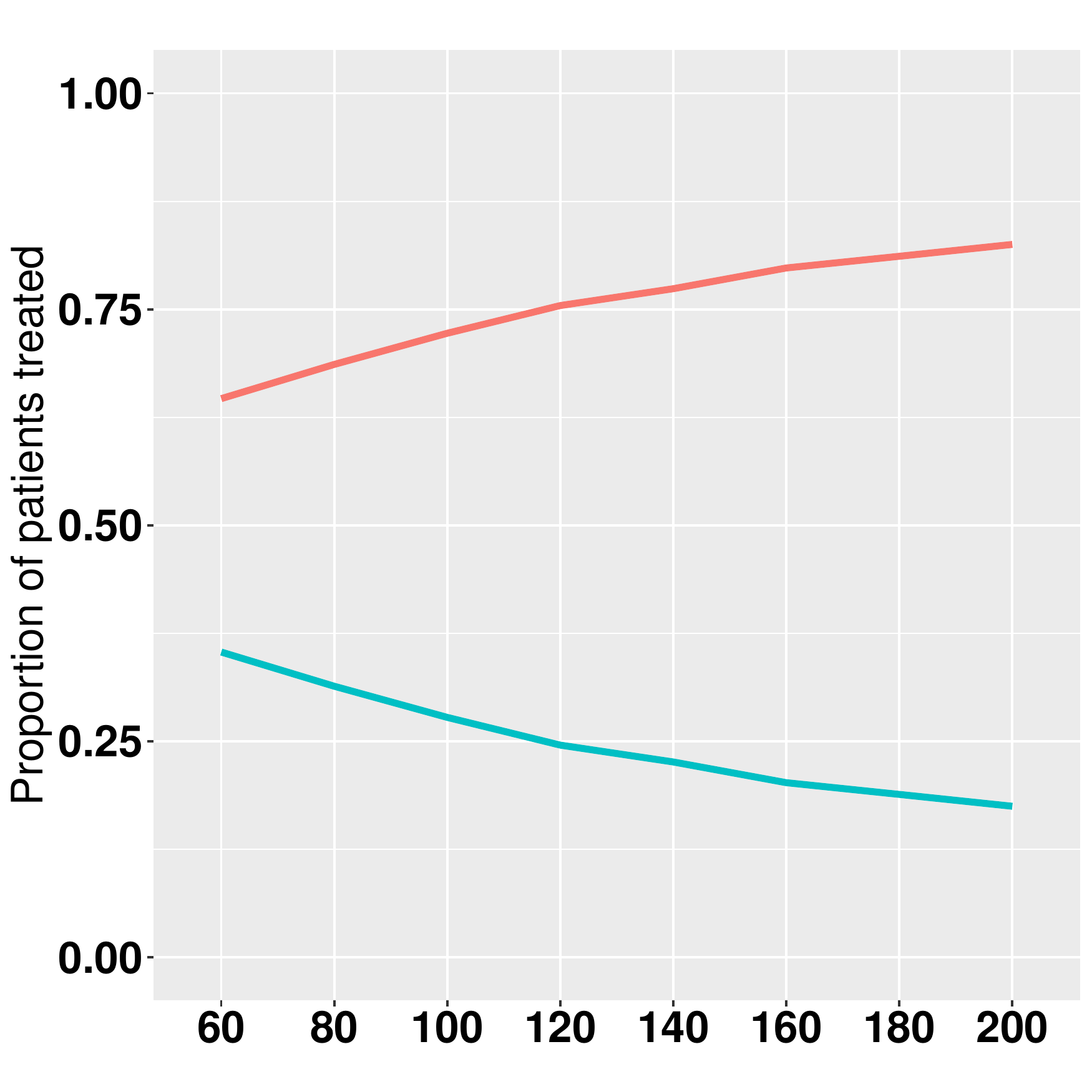}\\
			Proportion of the RE patients & \includegraphics[width=0.22\textwidth]{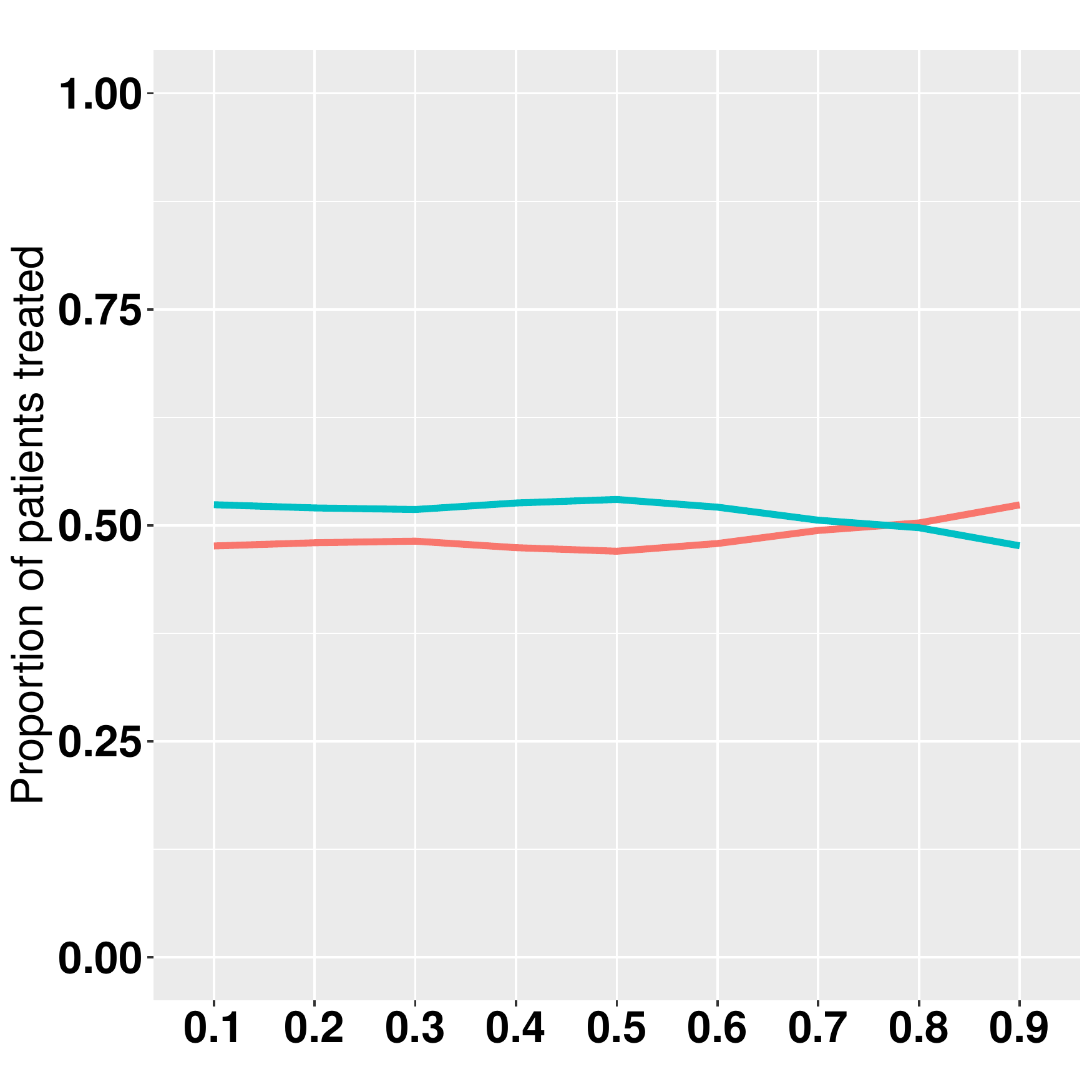}
			& \includegraphics[width=0.22\textwidth]{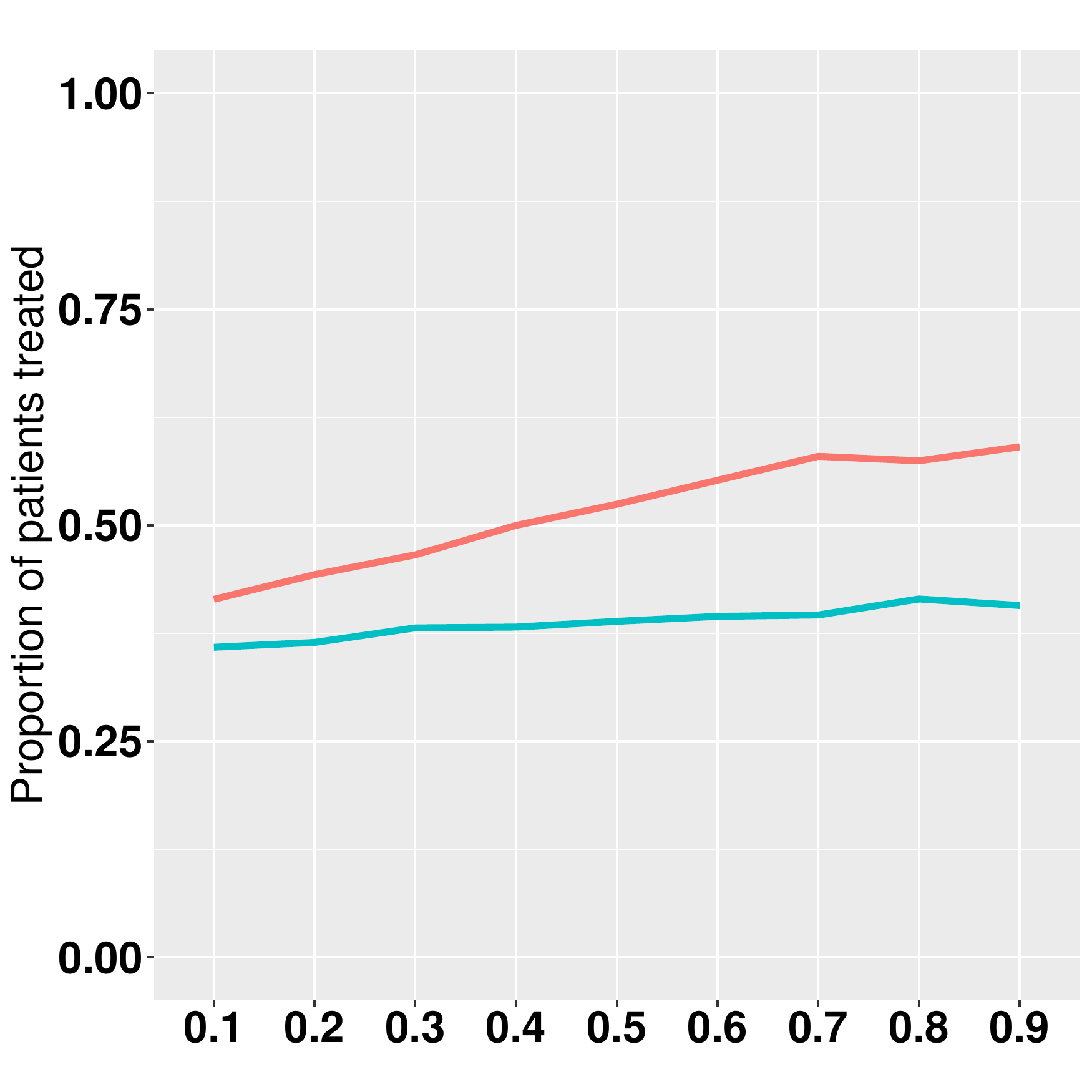}
			& \includegraphics[width=0.22\textwidth]{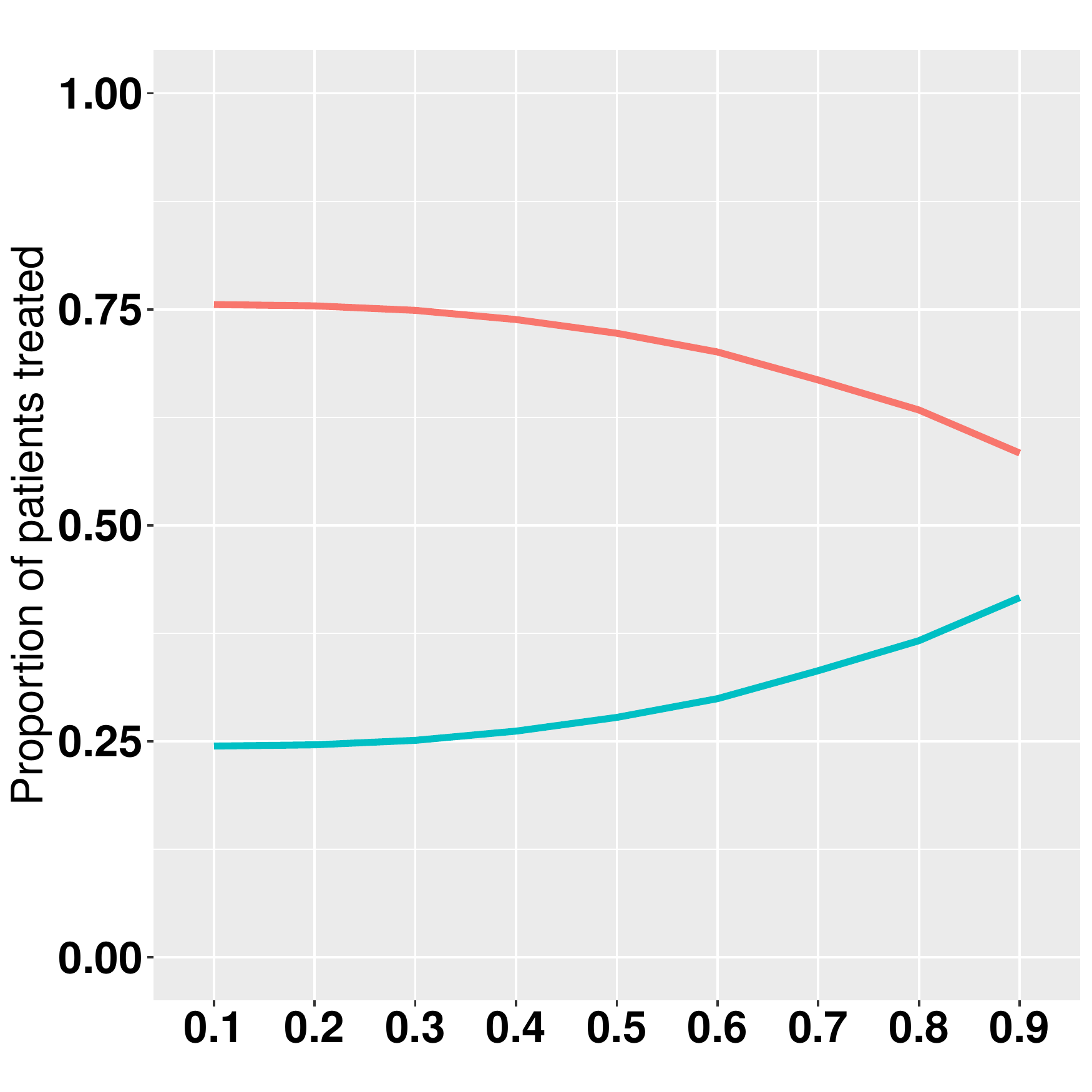}\\
			
			Distribution for the event time & \includegraphics[width=0.22\textwidth]{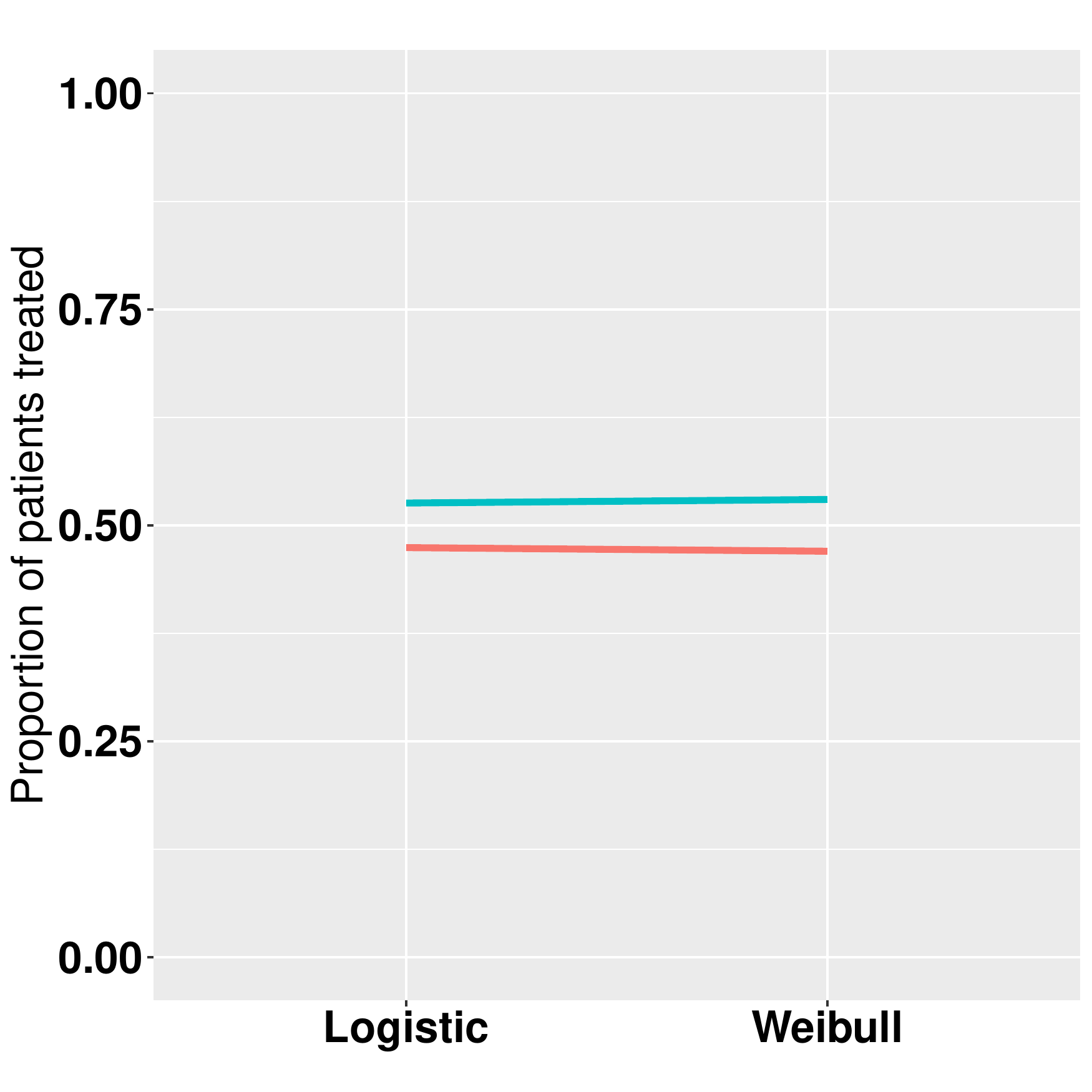}
			& \includegraphics[width=0.22\textwidth]{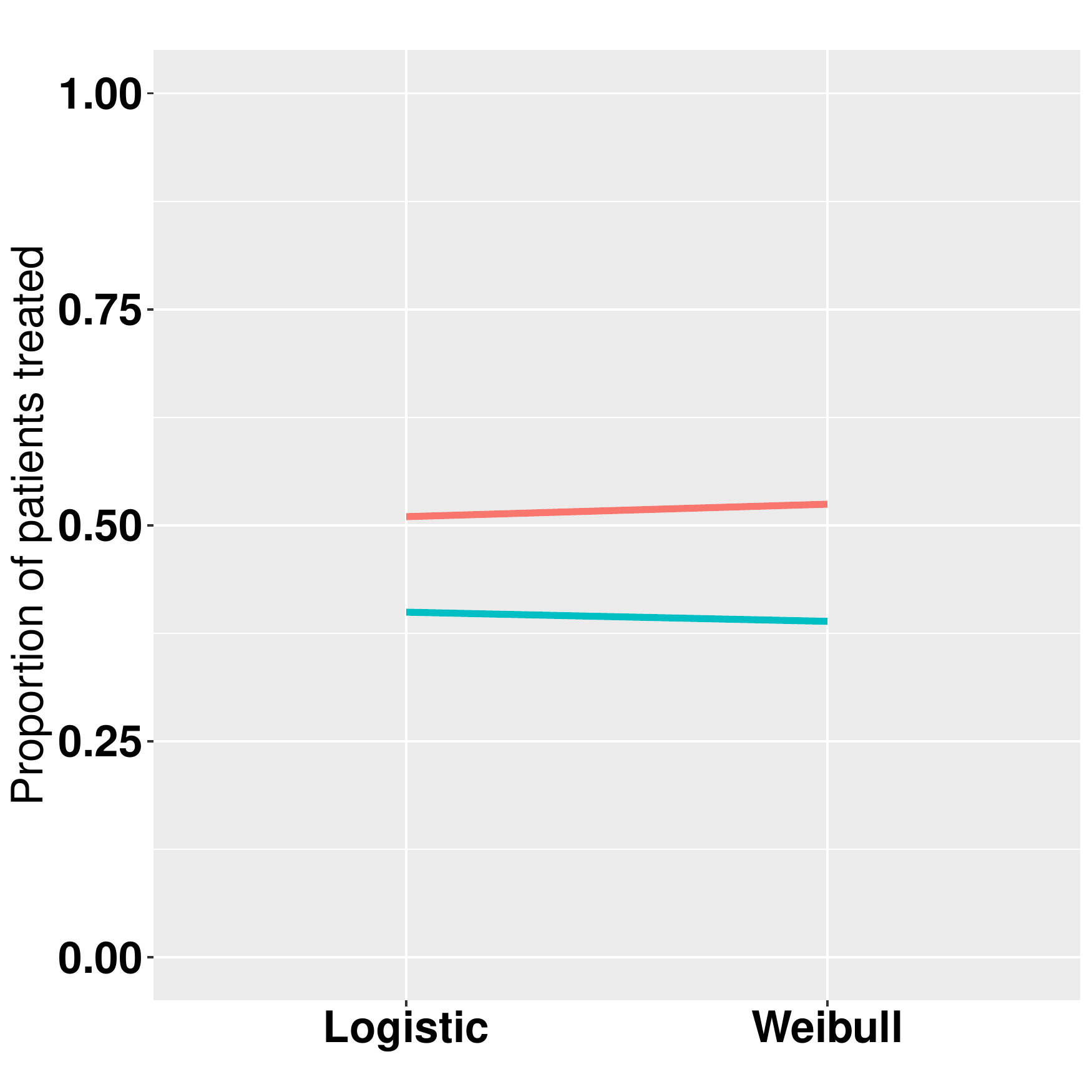}
			& \includegraphics[width=0.22\textwidth]{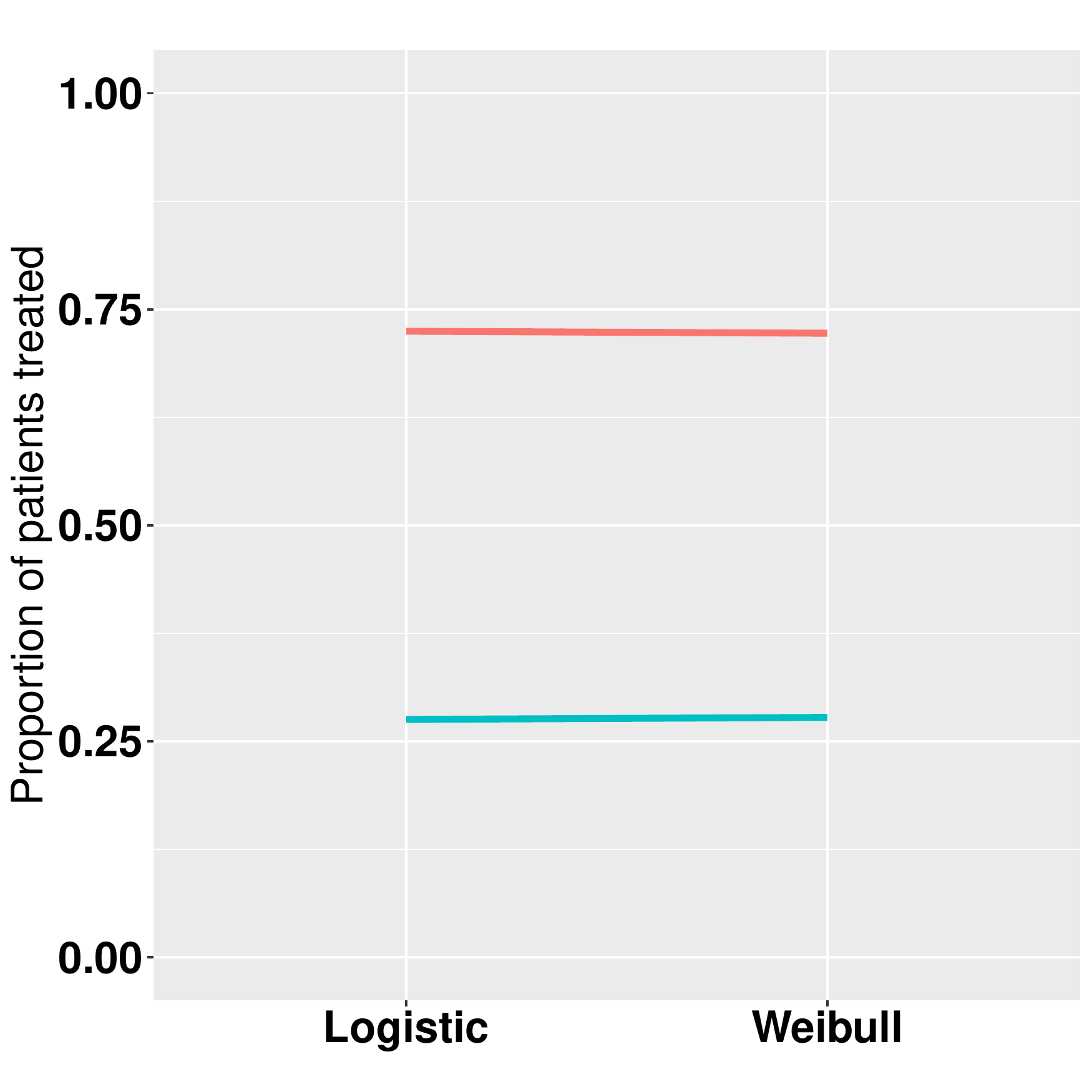}\\
			Utility function & \includegraphics[width=0.22\textwidth]{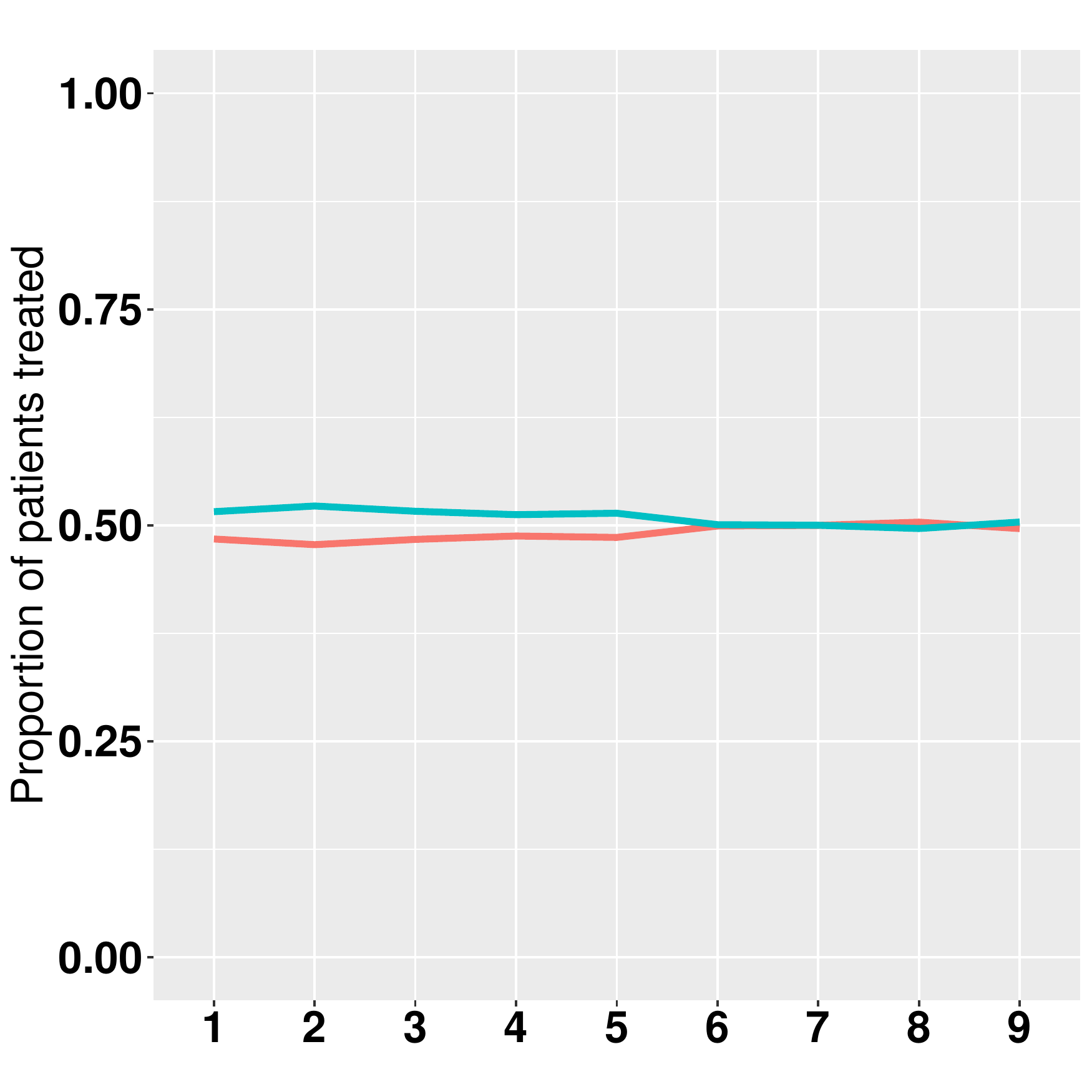}
			& \includegraphics[width=0.22\textwidth]{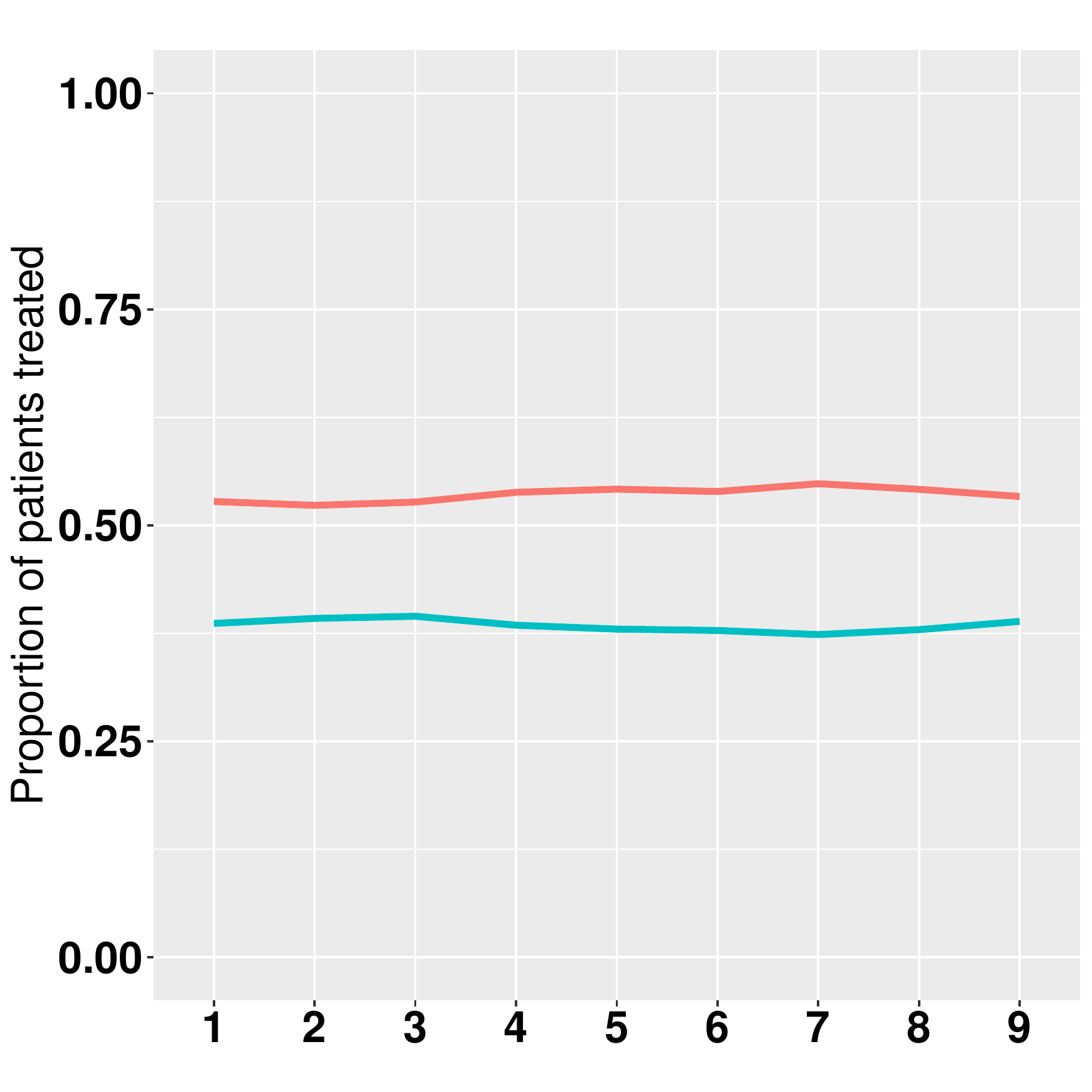}
			& \includegraphics[width=0.22\textwidth]{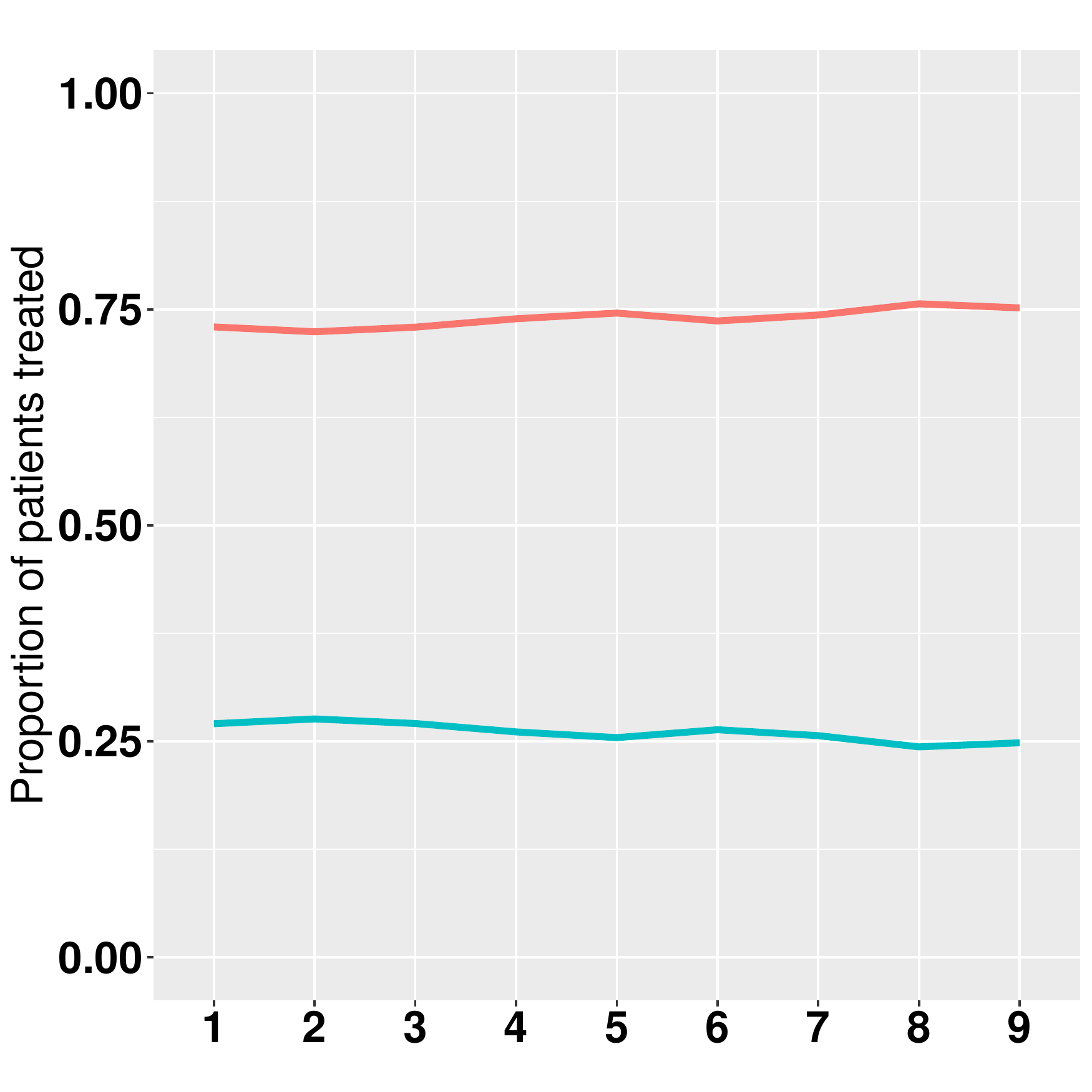}  
		\end{tabular}
		\caption{The sensitivity analysis of the proportion of the SE patients treated with each RT dose. The blue and red lines represent low and standard RT doses, respectively.}
		\label{tab:sensitivity_result_setp}
	\end{figure}

	\newpage 
	
	\section{Conclusion}
	
	This paper investigates phase II trial design using survival outcomes, focusing on the RT. Both the time to disease progression and time to normal tissue complications are considered as the co-primary outcomes, so the competing risk issue arises. We built a cause-specific hazard model to solve the competing risk problem and capture the association between the time-to-event and the RT dose and radiation susceptibility status. We propose to use a utility function method to tradeoff the risk-benefit of the RT dose on the cancer cell and normal tissue, which provides an overall measurement of the survival benefit of the different RT doses. Stratified by the radiation susceptibility status, we develop a Bayesian response-adaptive randomization scheme. More patients will be randomized to the RT dose reporting more favorable response outcomes in the posterior mean utility estimates. A subgroup-specific RT dose will be selected for SE and RE patients separately at the end of the trial. Numerical studies confirm the proposed design's desirable performances, compared with the conventional design ignoring the competing risk issue.   
	
	The proposed design considers one biomarker which stratifies the whole population into two sub-groups (RE or SE). An interesting extension is to consider multiple biomarker-induced sub-groups (e.g., the umbrella trial) and the ordinal relationship of the time-to-event exists for only part of the groups. We also assume that the biomarker can be accurately measured without missing. A practical extension of the proposed design is to consider prone to error and missing biomarker measurement \citep{Zang2015, Zang2016, Zang2018}. Besides competing risk outcomes, the time to disease progression and time to normal tissue complications may be considered as semi-competing risk outcomes in some clinical trials \citep{Murray2017, Zhang2022}. That is, although both events are still of primary interest, a subject will be treated off the protocol only if a specific adverse event has been observed for the subject. The proposed design cannot handle the semi-competing risk scenario, and a new design is required to address this problem.  
	
	\section*{Acknowledgement}
	
	Ick Hoon Jin's research was partially supported by the Yonsei University Research Fund 2019-22-0210 and by Basic Science Research Program through the National Research Foundation of Korea (NRF 2020R1A2C1A01009881). Yong Zang's research was partially supported by NIH grants P30 CA082709, R21 CA264257, and the Ralph W. and Grace M. Showalter Research Trust award. Correspondence should be addressed to Dr. Yong Zang (zangy@iu.edu) and Dr. Ick Hoon Jin (ijin@yonsei.ac.kr).

	\bibliographystyle{Chicago}
	\bibliography{referencev2}

\begin{thebibliography}{}

\bibitem[\protect\citeauthoryear{Barnett, West, Dunning, and etc.}{Barnett
  et~al.}{2009}]{Barnett2009}
Barnett, G., C.~West, A.~Dunning, and etc. (2009).
\newblock Normal tissue reactions to radiotherapy: towards tailoring treatment
  dose by genotype.
\newblock {\em Nature Reviews Cancer\/}~{\em 9}, 134--142.

\bibitem[\protect\citeauthoryear{Biard, Lee, and Cheng}{Biard
  et~al.}{2021}]{Biard2021}
Biard, L., S.~Lee, and B.~Cheng (2021).
\newblock Seamless phase {I}/{II} design for novel anticancer agents with
  competing disease progression.
\newblock {\em Statistics in Medicine\/}~{\em 40}, 4568--4581.

\bibitem[\protect\citeauthoryear{Bradley, Paulus, Komaki, and etc.}{Bradley
  et~al.}{2015}]{Bradley2015}
Bradley, J., R.~Paulus, R.~Komaki, and etc. (2015).
\newblock Standard-dose versus high-dose conformal radiotherapy with concurrent
  and consolidation carboplatin plus paclitaxel with or without cetuximab for
  patients with stage {IIIA} or {IIIB} non-small-cell lung cancer ({RTOG}
  0617): a randomised, two-by-two factorial phase 3 study.
\newblock {\em Lancet Oncology\/}~{\em 16}, 187--199.

\bibitem[\protect\citeauthoryear{Busch}{Busch}{1994}]{Busch1994}
Busch, D. (1994).
\newblock Genetic susceptibility to radiation and chemotherapy injury:
  diagnosis and management.
\newblock {\em International Journal of Radiation Oncology Biology
  Physics\/}~{\em 30}, 997--1002.

\bibitem[\protect\citeauthoryear{Carvalho, Leijenaar, ER, and etc.}{Carvalho
  et~al.}{2013}]{Carvalho2013}
Carvalho, S., R.~Leijenaar, E.~V. ER, and etc. (2013).
\newblock Prognostic value of metabolic metrics extracted from baseline
  positron emission tomography images in non-small cell lung cancer.
\newblock {\em Acta Oncologica\/}~{\em 52}, 1398--1404.

\bibitem[\protect\citeauthoryear{Chen}{Chen}{1997}]{Chen1997}
Chen, T. (1997).
\newblock Optimal three-stage designs for phase {II} cancer clinical trials.
\newblock {\em Statistics in Medicine\/}~{\em 16}, 2701--2711.

\bibitem[\protect\citeauthoryear{Chistiakov, Voronova, and
  Chistiakov}{Chistiakov et~al.}{2008}]{Chistiakov2008}
Chistiakov, D., N.~Voronova, and P.~Chistiakov (2008).
\newblock Genetic variations in {DNA} repair genes, radiosensitivity to cancer
  and susceptibility to acute tissue reactions in radiotherapy-treated cancer
  patients.
\newblock {\em Acta Oncologica\/}~{\em 47}, 809--824.

\bibitem[\protect\citeauthoryear{Dutton and Holmes}{Dutton and
  Holmes}{2018}]{Dutton2018}
Dutton, P. and J.~Holmes (2018).
\newblock Single arm two-stage studies: Improved designs for molecularly
  targeted agents.
\newblock {\em Pharmaceutical Statistics\/}~{\em 17}, 761--769.

\bibitem[\protect\citeauthoryear{Ensign, Gehan, Kamen, and Thall}{Ensign
  et~al.}{1994}]{Ensign1994}
Ensign, L., E.~Gehan, D.~Kamen, and P.~Thall (1994).
\newblock An optimal three-stage design for phase {II} clinical trials.
\newblock {\em Statistics in Medicine\/}~{\em 13}, 1727--1736.

\bibitem[\protect\citeauthoryear{Guo and Zang}{Guo and Zang}{2019}]{Guo2019}
Guo, B. and Y.~Zang (2019).
\newblock A {B}ayesian adaptive phase {II} clinical trial design accounting for
  spatial variation.
\newblock {\em Statistical Method in Medical Research\/}~{\em 28}, 3187--3204.

\bibitem[\protect\citeauthoryear{Guo and Zang}{Guo and Zang}{2020}]{Guo2020}
Guo, B. and Y.~Zang (2020).
\newblock {BILITE}: {A} {B}ayesian randomized phase {II} design for
  immunotherapy by jointly modeling the longitudinal immune response and
  time-to-event efficacy.
\newblock {\em Statistics in Medicine\/}~{\em 39}, 4439--4451.

\bibitem[\protect\citeauthoryear{Hanfelt, Slack, and Gehan}{Hanfelt
  et~al.}{1999}]{Hanfelt1999}
Hanfelt, J., R.~Slack, and E.~Gehan (1999).
\newblock A modification of {S}imon’s optimal design for phase {II} trials
  when the criterion is median sample size.
\newblock {\em Controlled Clincial Trials\/}~{\em 20}, 555--566.

\bibitem[\protect\citeauthoryear{Heitjan}{Heitjan}{1997}]{Heitjan1997}
Heitjan, D. (1997).
\newblock Bayesian interim analysis of phase {II} cancer clinical trials.
\newblock {\em Statistics in Medicine\/}~{\em 16}, 1791--1802.

\bibitem[\protect\citeauthoryear{Hendry, Jeremic, and Zubizarreta}{Hendry
  et~al.}{2006}]{Hendry2006}
Hendry, J., B.~Jeremic, and E.~Zubizarreta (2006).
\newblock Normal tissue complications after radiation therapy.
\newblock {\em Pan American Journal of Public Health\/}~{\em 20}, 151--160.

\bibitem[\protect\citeauthoryear{Huang, Ning, Li, Estey, Issa, and Berry}{Huang
  et~al.}{2009}]{Huang2009}
Huang, X., J.~Ning, Y.~Li, E.~Estey, J.~Issa, and D.~Berry (2009).
\newblock Using short-term response information to facilitate adaptive
  randomization for survival clinical trials.
\newblock {\em Statistics in Medicine\/}~{\em 28}, 1680--1689.

\bibitem[\protect\citeauthoryear{Iten, Muller, Schindle, and etc.}{Iten
  et~al.}{2007}]{Iten2007}
Iten, F., B.~Muller, C.~Schindle, and etc. (2007).
\newblock Response to yttrium-dota-toc treatment is associated with long-term
  survival benefit in metastasized medullary thyroid cancer: A phase {I}{I}
  clinical trial.
\newblock {\em Clincial Cancer Research\/}~{\em 13}, 6696--6702.

\bibitem[\protect\citeauthoryear{Jarnagin, Schwartz, Gultekin, and
  etc.}{Jarnagin et~al.}{2009}]{Jarnagin2009}
Jarnagin, W., L.~Schwartz, D.~Gultekin, and etc. (2009).
\newblock Regional chemotherapy for unresectable primary liver cancer: results
  of a phase {II} clinical trial and assessment of {DCE-MRI} as a biomarker of
  survival.
\newblock {\em Annals of Oncology\/}~{\em 20}, 1589--1595.

\bibitem[\protect\citeauthoryear{Johnson and Cook}{Johnson and
  Cook}{2009}]{Johnson2009}
Johnson, V. and J.~Cook (2009).
\newblock Bayesian design of single-arm phase {II} clinical trials with
  continuous monitoring.
\newblock {\em Clinical Trials\/}~{\em 6}, 217--226.

\bibitem[\protect\citeauthoryear{Jones and Holmgren}{Jones and
  Holmgren}{2007}]{Jones2007}
Jones, C. and E.~Holmgren (2007).
\newblock An adaptive {S}imon two-stage design for phase 2 studies of targeted
  therapies.
\newblock {\em Contemporary Clinical Trials\/}~{\em 28}, 654--661.

\bibitem[\protect\citeauthoryear{Jung, Carey, and Kim}{Jung
  et~al.}{2001}]{Jung2001}
Jung, S., M.~Carey, and K.~Kim (2001).
\newblock Graphical search for two-stage designs for phase {II} clinical
  trials.
\newblock {\em Controlled Clincial Trials\/}~{\em 22}, 367--372.

\bibitem[\protect\citeauthoryear{Kleinerman}{Kleinerman}{2009}]{Kleinerman2009}
Kleinerman, R. (2009).
\newblock Radiation-sensitive genetically susceptible pediatric
  sub-populations.
\newblock {\em Pediatric Radiology\/}~{\em 39}, S37--S31.

\bibitem[\protect\citeauthoryear{Kola and Landis}{Kola and
  Landis}{2004}]{Kola2004}
Kola, I. and J.~Landis (2004).
\newblock Can the pharmaceutical industry reduce attrition rates?
\newblock {\em Nature Reviews Drug Discovery\/}~{\em 3}, 711--715.

\bibitem[\protect\citeauthoryear{Kudo, Ueshima, Ikeda, and etc.}{Kudo
  et~al.}{2021}]{Kudo2021}
Kudo, M., K.~Ueshima, M.~Ikeda, and etc. (2021).
\newblock {TACTICS}: Final overall survival (os) data from a randomized, open
  label, multicenter, phase {I}{I} trial of transcatheter arterial
  chemoembolization (tace) therapy in combination with sorafenib as compared
  with tace alone in patients (pts) with hepatocellular carcinoma (hcc).
\newblock {\em Journal of Clincial Oncology\/}~{\em 39}, 270--270.

\bibitem[\protect\citeauthoryear{Lee and Liu}{Lee and Liu}{2008}]{Lee2008}
Lee, J.~J. and D.~D. Liu (2008).
\newblock A predictive probability design for phase {II} cancer clinical
  trials.
\newblock {\em Clinical Trials\/}~{\em 5\/}(2), 93--106.

\bibitem[\protect\citeauthoryear{Lin and Shih}{Lin and Shih}{2004}]{Lin2004}
Lin, Y. and W.~Shih (2004).
\newblock Adaptive two-stage designs for single-arm phase {IIA} cancer clinical
  trials.
\newblock {\em Biometrics\/}~{\em 60}, 482--490.

\bibitem[\protect\citeauthoryear{Liu, Barry, Birrer, and etc.}{Liu
  et~al.}{2019}]{Liu2019}
Liu, J., W.~Barry, M.~Birrer, and etc. (2019).
\newblock Overall survival and updated progression-free survival outcomes in a
  randomized phase ii study of combination cediranib and olaparib versus
  olaparib in relapsed platinum-sensitive ovarian cancer.
\newblock {\em Annals of Oncology\/}~{\em 30}, 551--557.

\bibitem[\protect\citeauthoryear{Murray, Thall, Yuan, McAvoy, and Gomez}{Murray
  et~al.}{2017}]{Murray2017}
Murray, T., P.~Thall, Y.~Yuan, S.~McAvoy, and D.~Gomez (2017).
\newblock Robust treatment comparison based on utilities of semi-competing
  risks in non-small-cell lung cancer.
\newblock {\em Journal of the American Statistical Association\/}~{\em 112},
  11--23.

\bibitem[\protect\citeauthoryear{Parashar, Bowden, Starr, Wernisch, and
  Mander}{Parashar et~al.}{2016}]{Parashar2016}
Parashar, D., J.~Bowden, C.~Starr, L.~Wernisch, and A.~Mander (2016).
\newblock An optimal stratified {S}imon two-stage design.
\newblock {\em Pharmaceutical Statistics\/}~{\em 15}, 333--340.

\bibitem[\protect\citeauthoryear{Pimentel, Lohmann, Ennis, and etc.}{Pimentel
  et~al.}{2019}]{Pimentel2019}
Pimentel, I., A.~Lohmann, M.~Ennis, and etc. (2019).
\newblock A phase {I}{I} randomized clinical trial of the effect of metformin
  versus placebo on progression-free survival in women with metastatic breast
  cancer receiving standard chemotherapy.
\newblock {\em The Breast\/}~{\em 48}, 17--23.

\bibitem[\protect\citeauthoryear{Pusztai, Anderson, and Hess}{Pusztai
  et~al.}{2007}]{Pusztai2007}
Pusztai, L., K.~Anderson, and K.~Hess (2007).
\newblock Pharmacogenomic predictor discovery in phase {II} clinical trials for
  breast cancer.
\newblock {\em American Association for Cancer Research\/}~{\em 13},
  6080--6086.

\bibitem[\protect\citeauthoryear{Santivasi and Xia}{Santivasi and
  Xia}{2014}]{Santivasi2014}
Santivasi, W. and F.~Xia (2014).
\newblock Ionizing radiation-induced {DNA} damage, response, and repair.
\newblock {\em Antioxidants Redox Signaling\/}~{\em 21}, 251--259.

\bibitem[\protect\citeauthoryear{Schipper, Taylor, TenHaken, Matuzak, Kong, and
  Lawrence}{Schipper et~al.}{2014}]{Schipper2014}
Schipper, M.~J., J.~M.~G. Taylor, R.~TenHaken, M.~M. Matuzak, F.-M. Kong, and
  T.~S. Lawrence (2014).
\newblock Personalized dose selection in radiation therapy using statistical
  models for toxicity and efficacy with dose and biomarkers as covariates.
\newblock {\em Statistics in Medicine\/}~{\em 33\/}(30), 5330--5339.

\bibitem[\protect\citeauthoryear{Shuster}{Shuster}{2002}]{Shuster2002}
Shuster, J. (2002).
\newblock Optimal two-stage designs for single arm phase {II} cancer trials.
\newblock {\em Journal of Biopharmaceutical Statistics\/}~{\em 12}, 39--51.

\bibitem[\protect\citeauthoryear{Simon}{Simon}{1989}]{Simon1989}
Simon, R. (1989).
\newblock Optimal two-stage designs for phase {II} clinical trials.
\newblock {\em Controlled Clincial Trials\/}~{\em 10}, 1--10.

\bibitem[\protect\citeauthoryear{Sinha and Hader}{Sinha and
  Hader}{2002}]{Sinha2002}
Sinha, R. and D.~Hader (2002).
\newblock {UV}-induced {DNA} damage and repair: a review.
\newblock {\em Photochemical and Photobiological Sciences\/}~{\em 1}, 225--236.

\bibitem[\protect\citeauthoryear{Thall and Simon}{Thall and
  Simon}{1994}]{Thall1994}
Thall, P.~F. and R.~Simon (1994).
\newblock A bayesian approach to establishang sample size and monitoring
  criteria for phase {II} clinical trials.
\newblock {\em Controlled Clinical Trials\/}~{\em 15\/}(6), 463--481.

\bibitem[\protect\citeauthoryear{Thall, Simon, and Estey}{Thall
  et~al.}{1995}]{Thall1995}
Thall, P.~F., R.~M. Simon, and E.~H. Estey (1995).
\newblock Bayesian sequential monitoring designs for single-arm clinical trials
  with multiple outcomes.
\newblock {\em Statistics in Medicine\/}~{\em 14\/}(4), 357--379.

\bibitem[\protect\citeauthoryear{Wang, Matuszak, and Kong}{Wang
  et~al.}{2012}]{Wang2012}
Wang, W., M.~Matuszak, and F.~Kong (2012).
\newblock Single nucleotide polymorphisms in {DNA} repair genes may be
  associated with radiation pneumonitis in patients with non-small cell lung
  cancer treated with definitive radiotherapy.
\newblock {\em Journal of Thoracic Oncology\/}~{\em 7}, S218.

\bibitem[\protect\citeauthoryear{Yin, Chen, and Lee}{Yin
  et~al.}{2012}]{Yin2012}
Yin, G., N.~Chen, and J.~Lee (2012).
\newblock Phase {II} trial design with {B}ayesian adaptive randomization and
  predictive probability.
\newblock {\em Journal of the Royal Statistical Society: Series C\/}~{\em 61},
  219--235.

\bibitem[\protect\citeauthoryear{Yuan, Guo, Munsell, Lu, and Jazaeri}{Yuan
  et~al.}{2016}]{Yuan2016}
Yuan, Y., B.~Guo, M.~Munsell, K.~Lu, and A.~Jazaeri (2016).
\newblock {MIDAS}: a practical {B}ayesian design for platform trials with
  molecularly targeted agents.
\newblock {\em Statistics in Medicine\/}~{\em 35}, 3892--3906.

\bibitem[\protect\citeauthoryear{Zang and Guo}{Zang and Guo}{2018}]{Zang2018}
Zang, Y. and B.~Guo (2018).
\newblock Optimal two-stage enrichment design correcting for biomarker
  misclassification.
\newblock {\em Statistical Method in Medical Research\/}~{\em 27}, 35--47.

\bibitem[\protect\citeauthoryear{Zang, Lee, and Yuan}{Zang
  et~al.}{2016}]{Zang2016}
Zang, Y., J.~Lee, and Y.~Yuan (2016).
\newblock Two-stage marker stratified clinical trial design in the presence of
  biomarker misclassification.
\newblock {\em Journal of the Royal Statistical Society: Series C\/}~{\em 65},
  585--601.

\bibitem[\protect\citeauthoryear{Zang, Liu, and Yuan}{Zang
  et~al.}{2015}]{Zang2015}
Zang, Y., S.~Liu, and Y.~Yuan (2015).
\newblock Optimal marker-adaptive designs for targeted therapy based on
  imperfectly measured biomarkers.
\newblock {\em Journal of the Royal Statistical Society: Series C\/}~{\em 64},
  635--650.

\bibitem[\protect\citeauthoryear{Zang and Yuan}{Zang and Yuan}{2017}]{Zang2017}
Zang, Y. and Y.~Yuan (2017).
\newblock Optimal sequential enrichment designs for phase {II} clinical trials.
\newblock {\em Statistics in Medicine\/}~{\em 36}, 54--66.

\bibitem[\protect\citeauthoryear{Zhang, Cao, Zhang, Jin, and Zang}{Zhang
  et~al.}{2021}]{Zhang2021}
Zhang, Y., S.~Cao, C.~Zhang, I.~Jin, and Y.~Zang (2021).
\newblock A {B}ayesian adaptive phase {I}/{II} clinical trial design with
  late-onset competing risk outcomes.
\newblock {\em Biometrics\/}~{\em 77}, 796--808.

\bibitem[\protect\citeauthoryear{Zhang, Guo, Cao, Zhang, and Zang}{Zhang
  et~al.}{2022}]{Zhang2022}
Zhang, Y., B.~Guo, S.~Cao, C.~Zhang, and Y.~Zang (2022).
\newblock {SCI}: A {B}ayesian adaptive phase {I}/{II} dose-finding design
  accounting for semi-competing risks outcomes for immunotherapy trials.
\newblock {\em Pharmaceutical Statistics\/}~{\em in press}.

\end{thebibliography}

\end{document}